\documentclass[final,2p,times,twocolumn]{elsarticle}

\usepackage{amssymb}
\usepackage{amsmath}
\usepackage{upgreek}

\usepackage{color}
\usepackage{orcidlink}
\usepackage{lineno}
\usepackage{geometry}
\let\oldthebibliography\thebibliography
\renewcommand{\thebibliography}[1]{%
  \oldthebibliography{#1}%
  \setlength{\itemsep}{5pt}  
}

\newcommand{\feii}{$\rm Fe\,\textsc{ii}$}
\newcommand{\oiii}{$\rm [O\,\textsc{iii}]$}
\newcommand{\Bowenoiii}{$\rm O\,\textsc{iii}$}
\newcommand{\niii}{$\rm N\,\textsc{iii}$}

\journal{Nuclear Physics B}

\newcommand{\received}{05-Nov-2025}
\newcommand{\revised}{07-Mar-2026}
\newcommand{\accepted}{03-Aug-2026}
\newcommand{\paperhistory}{%
  Received \received; revised \revised; accepted \accepted.%
}

\makeatletter
\let\ps@pprintTitle@orig\ps@pprintTitle
\def\ps@pprintTitle{%
  \ps@pprintTitle@orig
  \def\@oddhead{\@paperhistoryhead}%
  \def\@evenhead{\@paperhistoryhead}%
}
\def\@paperhistoryhead{%
  \hbox to \textwidth{\footnotesize\itshape\paperhistory\hfil}%
}
\makeatother

\begin{document}

\begin{frontmatter}

\title{A Star's Death by a Thousand Cuts: The Runaway Periodic Eruptions of AT2023uqm}


\author[USTC1,USTC2]{Yibo Wang\orcidlink{0000-0003-4225-5442}\corref{cor1}} 
\ead{wybustc@ustc.edu.cn}
\cortext[cor1]{Corresponding author.}
\author[USTC1,USTC2,DSS,GZU]{Tinggui Wang\orcidlink{0000-0002-1517-6792}\corref{cor1}}
\ead{twang@ustc.edu.cn}
\author[USTC1,USTC2]{Shifeng Huang\orcidlink{0000-0001-7689-6382}}
\author[USTC1,USTC2]{Jiazheng Zhu\orcidlink{0000-0003-3824-9496}}
\author[USTC1,USTC2]{Ning Jiang\orcidlink{0000-0002-7152-3621}} 
\author[UCB]{Wenbin Lu} 
\author[SYU,CSST]{Rongfeng Shen\orcidlink{0000-0001-5012-2362} }
\author[YNU]{Shiyan Zhong\orcidlink{0000-0003-4121-5684}}
\author[TDLI]{Dong Lai}
\author[THU]{Yi Yang\orcidlink{0000-0002-6535-8500}}
\author[ANU]{Xinwen Shu\orcidlink{0000-0002-7020-4290}}
\author[USTC1,USTC2]{Tianyu Xia\orcidlink{0000-0002-8438-8529}}
\author[USTC1,USTC2]{Di Luo\orcidlink{0009-0007-1153-8112}}
\author[UOA]{Jianwei Lyu\orcidlink{0000-0002-6221-1829}}
\author[UCB]{Alexei V. Filippenko\orcidlink{0000-0003-3460-0103}}
\author[UCB]{Thomas G. Brink\orcidlink{0000-0001-5955-2502}} 
\author[UCB]{Weikang Zheng\orcidlink{0000-0002-2636-6508}}
\author[USTC1,USTC2]{Minxuan Cai\orcidlink{0000-0003-4721-6477}} 
\author[USTC1,USTC2]{Zelin   Xu} 
\author[USTC1,USTC2]{Mingxin Wu}
\author[SAO,UCAS]{Xiaer Zhang\orcidlink{0000-0002-1542-8080}}
\author[USTC1,USTC2]{Weiyu Wu}
\author[USTC1,USTC2,DSS,GZU]{Lulu Fan\orcidlink{0000-0003-4200-4432}}
\author[USTC1,USTC2,NAOJ]{Ji-an Jiang\orcidlink{0000-0002-9092-0593}} 
\author[USTC1,USTC2,DSS]{Xu Kong\orcidlink{0000-0002-7660-2273}}
\author[PMO]{Bin Li} 
\author[USTC3]{Feng Li}
\author[NOAO]{Ming Liang}
\author[DSS]{Wentao Luo\orcidlink{0000-0003-1297-6142}}
\author[CIOE]{Jinlong Tang}
\author[USTC1,USTC2]{Zhen Wan\orcidlink{0000-0002-3105-3821}}
\author[PMO]{Hairen Wang} 
\author[USTC3,DSS]{Jian Wang\orcidlink{0000-0003-1617-2002}}
\author[USTC1,USTC2]{Yongquan Xue\orcidlink{0000-0002-1935-8104}} 
\author[PMO]{Dazhi Yao}
\author[USTC3]{Hongfei Zhang\orcidlink{0000-0002-1463-9070}}
\author[USTC1,USTC2,GZU]{Wen Zhao\orcidlink{0000-0002-1330-2329}}
\author[TDLI]{Xianzhong Zheng\orcidlink{0000-0003-3728-9912}}
\author[USTC1, USTC2,DSS]{Qingfeng Zhu\orcidlink{0000-0003-0694-8946}}
\author[PMO]{Yingxi Zuo}

\affiliation[USTC1]{organization={Department of Astronomy, University of Science and Technology of China},
            city={Hefei 230026},
            country={China}} 
\affiliation[USTC2]{organization={School of Astronomy and Space Sciences,University of Science and Technology of China},
            city={Hefei 230026},
            country={China}}
\affiliation[YNU]{organization={South-Western Institute for Astronomy Research, Yunnan University},
            city={ Kunming 650500},
            country={China}}   
\affiliation[SYU]{organization={School of Physics and Astronomy, Sun Yat-Sen University},
            city={Zhuhai 519000},
            country={China}} 
\affiliation[CSST]{organization={CSST Science Center for the Guangdong-Hongkong-Macau Greater Bay Area, Sun Yat-Sen University},
            city={Zhuhai 519082},
            country={China}} 
\affiliation[UCB]{organization={Department of Astronomy, University of California,  Berkeley},
            city={CA 94720-3411},
            country={USA}} 
\affiliation[SAO]{organization={Shanghai Astronomical Observatory, Chinese Academy of Sciences},
            city={Shanghai 200030},
            country={China}} 
\affiliation[UCAS]{organization={School of Astronomy and Space Sciences, University of Chinese Academy of Sciences},
            city={Beijing 100049},
            country={China}} 
\affiliation[DSS]{organization={Institute of Deep Space Sciences, Deep Space Exploration Laboratory},
            city={Hefei 230026},
            country={China}} 
\affiliation[PMO]{organization={Purple Mountain Observatory, Chinese Academy of Sciences},
            city={Nanjing 210023},
            country={China}}
\affiliation[CIOE]{organization={Institute of Optics and Electronics, Chinese Academy of Sciences},
            city={Chengdu 610209},
            country={China}} 
\affiliation[TDLI]{organization={Tsung-Dao Lee Institute and Key Laboratory for Particle Physics, Astrophysics and Cosmology, Ministry of Education, Shanghai Jiao Tong University},
            city={Shanghai 201210},
            country={China}} 
\affiliation[NAOJ]{organization={National Astronomical Observatory of Japan, National Institutes of Natural Sciences},
            city={Tokyo 181-8588},
            country={Japan}} 
\affiliation[USTC3]{organization={State Key Laboratory of Particle Detection and Electronics, University of Science and Technology of China},
            city={Hefei 230026},
            country={China}} 
\affiliation[NOAO]{organization={National Optical Astronomy Observatory (NSF’s National Optical-Infrared Astronomy Research Laboratory)},
            city={Tucson 85726},
            country={USA}} 
\affiliation[ANU]{organization={Department of Physics, Anhui Normal University},
            city={Wuhu 241002},
            country={China}} 
\affiliation[THU]{organization={Physics Department and Tsinghua Center for Astrophysics, Tsinghua University},
            city={Beijing 100084},
            country={China}} 
            
\affiliation[GZU]{organization={Department of Physics and Astronomy, College of Physics, Guizhou University},
            city={Guiyang 550025},
            country={China}}    

\affiliation[UOA]{organization={Steward Observatory, University of Arizona,933 North Cherry Avenue},
            city={Tucson 85721},
            country={USA}}  



\begin{abstract}
Stars on bound orbits around a supermassive black hole may undergo repeating partial tidal disruption events (rpTDEs), producing periodic flares. While several candidates have been suggested, definitive confirmation of these events remains elusive. We report the discovery of AT2023uqm, a nuclear transient that has exhibited at least five periodic optical flares, making it only the second confirmed case of periodicity after ASASSN-14ko. Uniquely, the flares from AT2023uqm show a nearly exponential increase in energy --- a ``runaway'' phenomenon signaling the star's progressive destruction. This behavior is consistent with rpTDEs of low-mass main-sequence stars or evolved giant stars. Multiwavelength observations and spectroscopic analysis of the two most recent flares reinforce its interpretation as an rpTDE. Intriguingly, each flare displays a similar double-peaked structure, potentially originating from a double-peaked mass fallback rate or two discrete collisions per orbit. The extreme ratio of peak separation to orbital period draws attention to the possibility of a giant star being disrupted, which could be distinguished from a low-mass main-sequence star by its future mass-loss evolution. Our analysis demonstrates the power of rpTDEs to probe the properties of disrupted stars and the physical processes of tidal disruption, though it is currently limited by our knowledge of these events. AT2023uqm emerges as the most compelling rpTDE thus far, serving as a crucial framework for modeling and understanding these phenomena. 
\end{abstract}

\begin{keyword}
tidal disruption events \sep optical periodic eruptions \sep supermassive black hole \sep time-domain astronomy
\end{keyword}
\end{frontmatter}



\section{Introduction}
\label{intro}
When a star in the vicinity of a galactic nucleus approaches the central supermassive black hole (SMBH) too closely, near the so-called tidal disruption radius, it can be torn apart by the BH’s tidal forces. A portion of the resulting stellar debris is bound to the SMBH and may eventually be accreted onto it, producing a luminous flare in the optical/ultraviolet (UV) or soft X-ray bands. These phenomena, known as tidal disruption events (TDEs; \cite{Rees1988, Gezari2021}), offer a unique opportunity to study the SMBH itself, the physics of accretion, and the nuclear star clusters. 

Depending on the internal structure of the star and its proximity to the SMBH, a partial tidal disruption event (pTDE) may occur, where the core of the star stays intact while its outer layers are stripped
\cite{Diener1997, Ivanov2001, MacLeod2012GaiantTDE, Guillochon2013}. In contrast to full TDEs, where the fallback rate declines as $t^{-5/3}$ over time, pTDEs can show a steeper decline~\citep{Coughlin2019}. More intriguingly, the surviving remnant can return to the pericenter again and undergo repeated partial TDEs (rpTDEs), resulting in periodic or recurring nuclear flares~\citep{MacLeod2013}. While most stars involved in TDEs are assumed to follow nearly parabolic orbits, in some cases the star may instead be on a bound, eccentric orbit, which can lead to recurring flares on timescales observable by current surveys. These rpTDEs provide a valuable opportunity to explore the tidal disruption process in detail and to constrain the properties of the disrupted star.

The rapid increase of TDEs detected by wide-field sky surveys has recently led to the identification of several rpTDE candidates. These include AT2018fyk~\citep{Wevers2023}, AT2020vdq~\citep{Somalwar2025}, AT2022dbl~\citep{LinZheyu2024}, eRASSt J0456~\citep{LiuZhu2024}, IRAS~F01004-2237~\citep{SunLuming2024}, and ASASSN-14ko~\citep{Payne2021, HuangShifeng2023ASASN14ko,HuangShifeng2025ASASN14ko}. Among them, most sources show only two flares, making it difficult to confirm whether they result from repeated partial disruptions of the same star. Only ASASSN-14ko and eRASSt J0456 exhibited multiple flares, with the latter mainly in X-rays. ASASSN-14ko, the most extensively studied, has shown over 30 flares of consistent brightness, which are thought to originate from a star in a grazing orbit around the SMBH. The discovery of these events has sparked intense interest in understanding the evolution of stars undergoing rpTDEs. However, it remains challenging to draw firm conclusions owing to the complexity of the process and the limited number of confirmed rpTDEs with multiple flares.

Here we report the discovery of AT2023uqm, which exhibits at least five periodic optical flares, making it the second known case of periodic optical flares following ASASSN-14ko. Unlike ASASSN-14ko, AT2023uqm exhibits flares increasing in brightness over time. We demonstrate that this unique property can be well explained within the framework of rpTDEs. 

\section{Discovery and Observations}
AT2023uqm was initially reported as a nuclear outburst by the Asteroid Terrestrial-impact Last Alert System (ATLAS~\cite{Tonry2018}) under the name ATLAS23txi on October 10, 2023 (UTC). It was classified as an active galactic nucleus (AGN)-related outburst at a redshift $z = 0.238$ in November 2023~\cite{Ramsden2023}. In September 2024, we noticed three earlier, weaker outbursts in the forced-photometry light curve from the Zwicky Transient Facility (ZTF; \cite{Bellm2019}). The earliest one can be traced back to June 2019.
These outbursts shared similar light-curve morphologies and a constant interval, indicating a common origin of repeated partial TDEs. Consequently, we began monitoring the source using the UltraViolet/Optical Telescope (UVOT; \cite{Roming2005}) and the X-Ray Telescope (XRT; \cite{Burrows2005}) aboard the {\it Neil Gehrels Swift Observatory} ({\it Swift}; \cite{Gehrels2004}) from December 2024. 

\begin{figure*}[htb]
\centering
\begin{minipage}{1.0\textwidth}
\centering{\includegraphics[angle=0,width=1.0\textwidth]{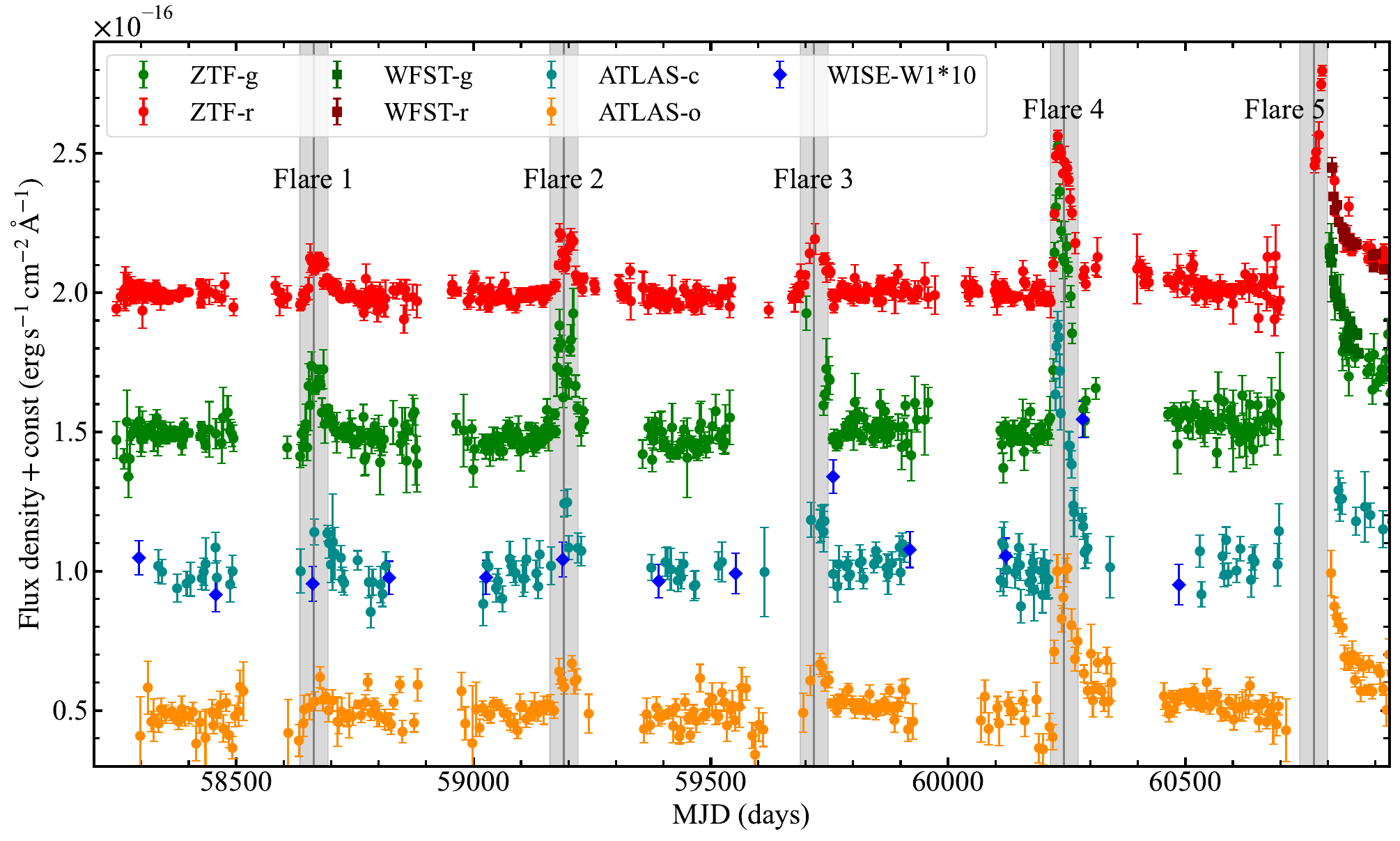}}
\end{minipage}
\caption{\label{fig:all-lc} Multiband light curves of AT2023uqm. Facilities and filters are indicated in the legend. For the purpose of presentation, we shifted the light curves vertically and scaled the {\it WISE} photometry by a factor of 10. Vertical gray lines indicate the time of the dip of the double-peaked flare, calculated using the period derived in Section~\ref{sec:period}. Gray shaded regions spanning $\pm 30$  from each vertical line highlight the flare events. }
\end{figure*}

The fifth flare occurred in April 2025, roughly at the predicted time, which was reported by the ALeRCE broker using ZTF data~\cite{Munoz2025}. We subsequently launched a multiwavelength follow-up campaign across optical/UV (WFST \citep{WangTinggui2023} and {\it Swift}/UVOT),  X-ray ({\it Swift}/XRT, {\it Einstein Probe (EP)}~\citep{Yuan2022}, and {\it XMM-Newton}~\citep{Jasen2001}), and radio (e-MERLIN~\citep{eMCP}) wavelengths. Notably, the 2.5~m Wide Field Survey Telescope (WFST), with its deeper limiting magnitude compared to ZTF, enabled a more precise characterization of AT2023uqm’s optical light curve. Archival photometry from ATLAS, ZTF, Catalina Real-Time Transient Survey (CRTS; \cite{Drake2009}) and the {\it Wide Infrared Survey Explorer} ({\it WISE}; \citep{Wright2010}) was also retrieved. Fig.~\ref{fig:all-lc} presents the multiband light curves, clearly demonstrating the periodic outbursts. Optical spectroscopic observations were also triggered following the fifth flare, using the Next Generation Palomar Spectrograph (NGPS; \citep{Jiang2018NGPS}) on the Palomar 200-inch (P200) Hale Telescope, the Low Resolution Imaging Spectrometer (LRIS; \citep{1995PASP..107..375O}) on the Keck I 10\,m telescope, the Kast double spectrograph~\citep{LickShane} on the 3~m Shane telescope at Lick Observatory, and BINOSPEC spectrograph~\citep{Fabricant2019} on the 6.5~m MMT. Including one archival spectrum obtained during the fourth flare, eight spectra in total were acquired and are displayed in Fig.~\ref{fig:all-spec}. 

Details of the data-reduction process are provided in Section S1 (online) of the supplementary materials\footnote{Hereafter, section, table, and figure numbers prefixed with ``S'' (e.g., Section S1, Table S1, Fig. S1) refer to items in the online supplementary materials.}. In Fig. S1 (online), we show that the position of the transient lies in the nucleus of the host galaxy.



\begin{figure*}[htb]
\centering
\begin{minipage}{0.9\textwidth}
\centering{\includegraphics[angle=0,width=1.0\textwidth]{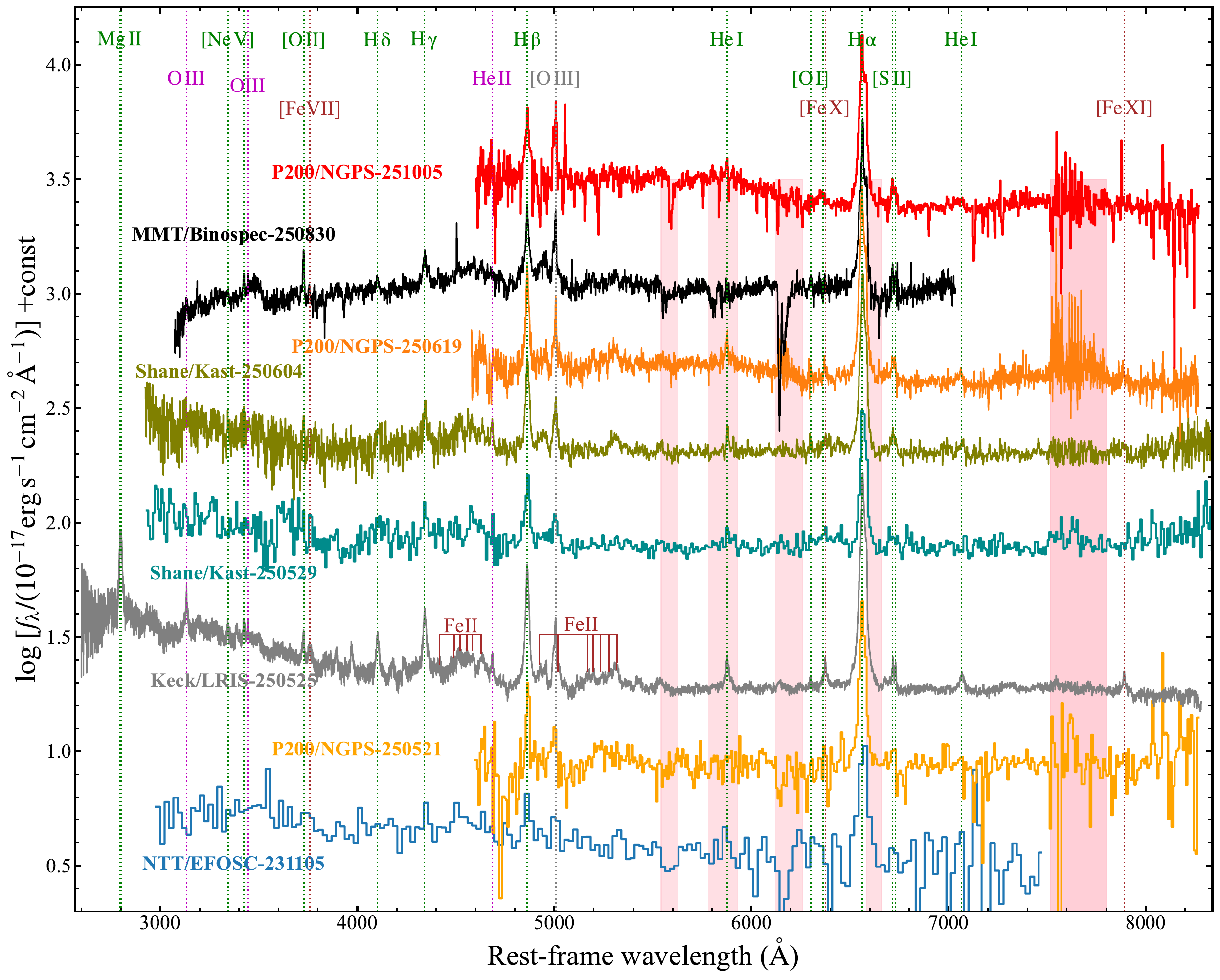}}
\end{minipage}
\caption{\label{fig:all-spec} Optical spectra of AT2023uqm. The blue curve represents a spectrum taken during the fourth flare, while the others were obtained during the fifth flare. Vertical lines mark prominent emission features, including the Balmer series, Bowen fluorescence \Bowenoiii,  $\rm Fe\, \textsc{ii}$, and high-ionization coronal lines. The pink shaded regions highlight prominent telluric bands. For clarity, we rebinned the NTT spectrum, the two P200/NGPS spectra (taken in May and October), and one Shane/Kast spectrum (taken in May). } 
\end{figure*}

\section{Data Analysis}  
\subsection{Host properties} 
\label{sec:Host}

In order to assess the AGN activity prior to the outburst, we examined the CRTS light curve and found no significant variability.  The pre-outburst {\it WISE} $W1-W2$ color was around 0.5 mag, falling below the 0.8 mag threshold for strong AGNs \citep{Stern2012,Yan2013}. Furthermore, emission-line intensity ratios measured from the post-outburst Keck spectrum place the source in the nonactive regime of the Baldwin-Phillips-Terlevich (BPT) diagram~\citep{Baldwin1981,Veilleux1987}. Spectral energy distribution (SED) fitting with multiband archival pre-outburst photometry suggests the presence of a weak AGN component, likely owing to contamination in the {\it WISE} photometry from a nearby source (see Section S2.1 online). Therefore, there is no significant AGN activity prior to the outburst, although weak AGN activity cannot be completely ruled out.  


Moreover, the stellar mass inferred from the SED fitting is around $10^{10.74}\, \mathrm{M_\odot}$.  
Using the relation for S0 galaxies with classical bulges or ellipticals from Kormendy et al.~\citep{Kormendy2013} (their Eq. (10)), we estimate a BH mass of $\rm 10^{8.39}\, M_\odot$ (root-mean-square (rms) scatter $\approx 0.3$ dex). However, since pre-outburst AGN activity cannot be ruled out for AT2023uqm, we also apply Eq. (4) of Reines et al.~\citep{Reines2015}, yielding a lower mass estimate of $\rm 10^{7.18}\, M_\odot$ (rms scatter $\approx 0.55$ dex). Given the large discrepancy, we further estimate the BH mass using the line's full width at half-maximum intensity ($\rm FWHM\approx 2000\,\rm km\,s^{-1}$) of $\rm Fe\,\textsc{ii}$ (or $\rm [Fe\,\textsc{x}]$) detected in the post-outburst spectra (see Section~\ref{sec:spectrumAnalysis}). These iron lines most likely originate from virialized, pre-existing ambient gas located in the inner region of the dusty torus (i.e., $R_{\rm line}\lesssim R_{\rm torus}$; \citep{HeZhicheng2021}). Under this assumption, the BH mass is $M_{\rm BH}\approx {\rm FWHM}^2 R_{\rm line}/\mathrm{G} \lesssim {\rm FWHM}^2 R_{\rm torus}/\mathrm{G} \approx 10^{7.4}\,\rm M_\odot$, where the torus scale is inferred from infrared echo modeling (Section S2.2 online). Hence, in this work we adopt a fiducial value of $\rm 10^{7}\,M_\odot$ for the BH mass. However, it is important to note the significant uncertainty in the BH mass. 



\subsection{Optical/UV light curves}
\subsubsection{Light-curve morphology of the outburst} 
\label{sec:lcshape}
\begin{figure*}[htb]
\centering
\begin{minipage}{0.9\textwidth}
\centering{\includegraphics[angle=0,width=1.0\textwidth]{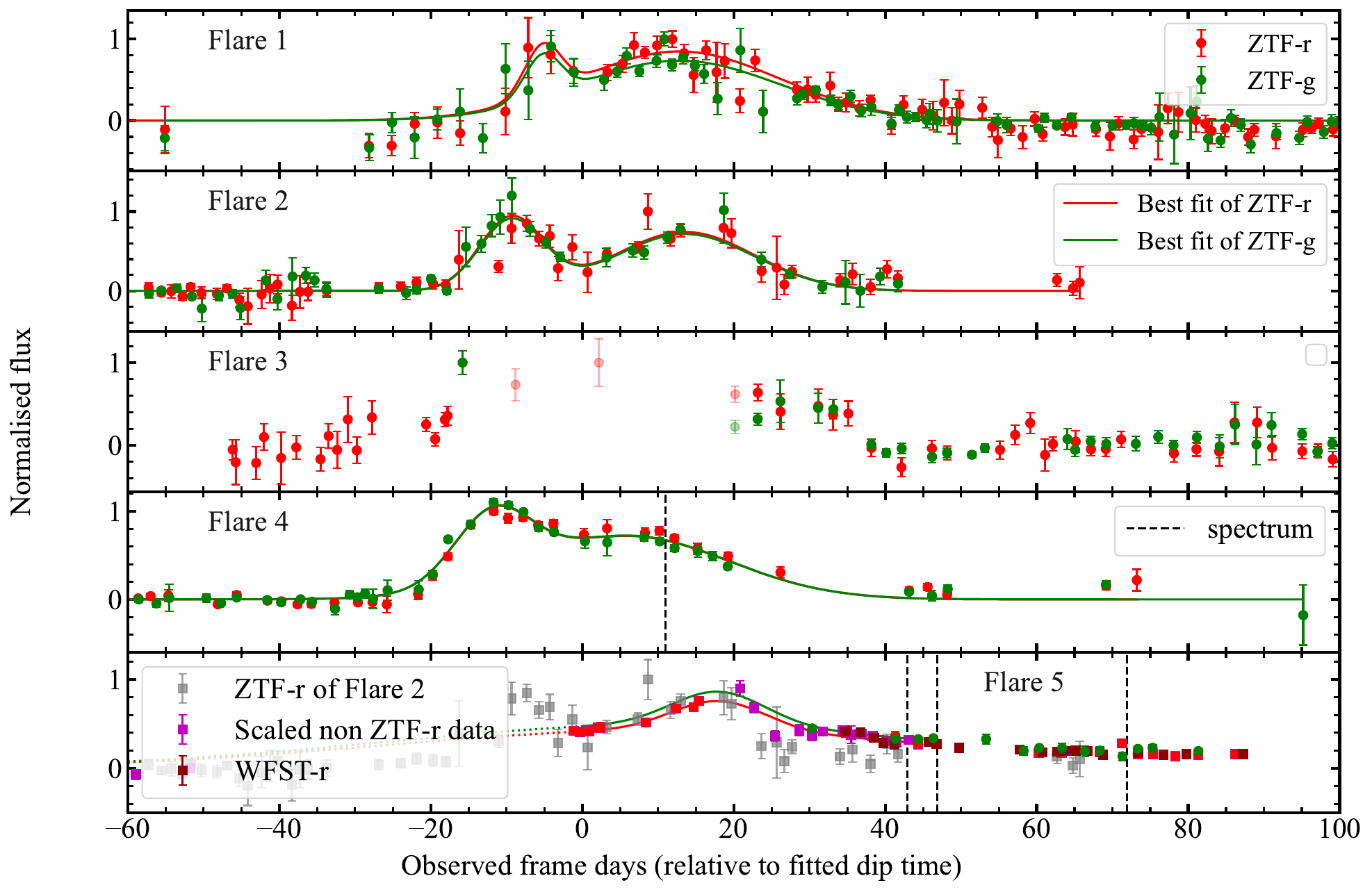}}
\end{minipage}
\caption{\label{fig:all-flares} The zoomed-in ZTF light curve for each flare. The green and red points depict the ZTF photometric data, while curves of the same color indicate the best double-Gaussian fits (see Section~\ref{sec:lcshape}). The dotted curves represent the unreliable fitting results. The fitting results for the third flare are not shown because the sparse sampling and poor photometric quality (points plotted with lower opacity) make the fit highly unreliable. Accordingly, this flare is excluded from the analysis presented in this paper. 
Vertical dashed lines mark the time of spectroscopic observations. The timing of each flare is relative to the corresponding dip in the Gaussian fit. For the latest flare, WFST $r$ (dark red square) and scaled photometry (magenta square) from all bands except ZTF $r$ are also included. The ZTF $r$ light curve of the second flare is overlaid on the latest flare, and their similarity suggests that the observed latest flare corresponds to the second peak of a double-peaked flare. } 
\end{figure*}

First, we examine the light-curve behavior of each outburst. As shown in Fig.~\ref{fig:all-flares}, the first four flares all exhibit double-peaked light curves. Flare 3 is excluded from most subsequent analyses owing to sparse sampling and unreliable photometric measurements caused by poor image subtraction. 
The latest outburst showed just one peak, likely the second in a double-peaked eruption, with the first peak missed owing to a seasonal gap. This is consistent with the timing of previous flares.
Otherwise, the interval between the two most recent outbursts would be significantly longer. 
Furthermore, the normalized shape\footnote{To compensate for the gap in the decline phase of the ZTF and WFST $r$-band light curve during the fifth flare, we scaled photometry from other bands (particularly the {\it Swift}/UVOT data) to align with the ZTF $r$-band observations.} of this flare resembles the second peak of other flares, with an apparent dip in the first few data points (see Fig.~\ref{fig:LC-recent}), supporting this interpretation. The morphological match and temporal consistency make this the most parsimonious explanation.


We adopt a two-component Gaussian model 
to characterize the observational properties of each flare episode. This model is purely phenomenological rather than physical; its primary purpose is to provide a smooth representation of the light curve from which the relevant timing features can be measured. Light curves obtained in all filters except for those with sparse sampling or low signal-to-noise ratio (S/N) were fitted simultaneously to improve the reliability of the results. This fitting procedure assumes that each flare would manifest a universal pattern across all photometric bands, differing only in flux amplitude. The model provides a reasonably good fit for all flares (see Fig.~\ref{fig:all-flares}). To estimate the uncertainties in the fitting parameters, we generated 10,000 perturbed realizations of the original light curves. For the most recent flare, sparse data coverage prevented robust constraints on both the first peak and the  dip between peaks. 
Therefore, we applied a simple second-order polynomial fit to the first few data points (see Fig.~\ref{fig:LC-recent}) of the flare to estimate the time of the dip. All fitting results are summarized in Table~S1 (online). In general, we find that the rise time to the first peak is $\sim 10$ days, and the separation between the two peaks is about 17--20 days. These timescales are given in the observer's frame, as are all other values hereafter unless otherwise noted.

Two single-peak models, a single Gaussian function and a Gaussian rise plus power-law decay, were also tested to fit the light-curve morphology. An F-test at the 0.99 significance level shows that the two-Gaussian model is preferred over the single-Gaussian model for all flares except the first, owing to its low S/N, and over the Gaussian-plus-power-law model for all flares except the latest flare, for which the first peak was likely missed because of an observational gap.

Compared to typical TDEs, each flare of AT2023uqm evolves on a shorter timescale, but is comparable to that of ASASSN-14ko (see Figs. S2 and S5 online). Focusing on the decline phase following the second peak of each flare, AT2023uqm clearly exhibits a faster decline in the $L_{\rm peak}$–$\Delta L_{40}$ plane (see Fig. S19 online and Hinkle et al.~\citep{Hinkle2020Declinerate}). Actually, in the early flares, the emission nearly fades within 40 days of the second peak. This is consistent with rpTDE scenarios, in which the gravitational influence of a surviving stellar core \citep{Coughlin2019} or bound stellar orbital energy ~\citep{Park2020} can induce a fast decline. Given the double-peaked morphology (see Section~\ref{sec:doublepeakOrigin}) and the overall short timescale, the latter may play a dominant role.

We note that double-peaked structures have also been observed in typical TDEs (e.g., AT2021uvz; Fig. S2 online), but it is uncertain whether they share a similar origin with those in AT2023uqm. Additionally, Huang et al.~\citep{HuangShifeng2023ASASN14ko} reported that several successive flares of the rpTDE candidate ASASSN-14ko exhibit an early bump preceding the main peak, followed by a rebrightening phase. A few flares with a weak early bump may mimic a double-peaked morphology (see Fig. S2 online). However, such features may arise from physical processes~\citep{HuangShifeng2023ASASN14ko} distinct from those discussed in Section~\ref{sec:doublepeakOrigin} for AT2023uqm.

\subsubsection{Period of the outburts} 
\label{sec:period}
We examine three features for pinpointing flare timing: the first peak time ($t_{\rm peak1}$), the interpeak dip time ($t_{\rm dip}$),\footnote{The $t_{\rm dip}$ is defined as the interpeak minimum point, as determined from a two-Gaussian fit.} and the second peak time ($t_{\rm peak2}$). To obtain the period, we utilize two models based on these timing features: one assumes a static period (func1), while the other incorporates a period-evolution rate $\dot{P}$ (func2, Payne et al.~\citep{Payne2021}). Specifically,
\begin{align}
    & {\rm func1:}  & t= t_0+ n\times P_0\, , \hspace{8.6em}  \\
    & {\rm func2:}  & t= t_0+ n\times P_0 + \frac{1}{2}n^2P_0\dot{P} +\frac{1}{6}n^3P_0\dot{P}^2\, ,
\end{align}
where $n$ represents the number of the flare. Using \textit{scipy.optimize.curve\_fit} from Python, we performed the fitting, excluding the third flare's timing features and the first peak of the fifth flare owing to inadequate sampling. Results are in Table S2 (online) and Fig.~S3 (online). Including $\dot{P}$ in the model shows no significant improvement, with $\dot{P}$ values being negligible or within their uncertainties. Thus, we use the constant-period model throughout this paper. Among the three features, $t_{\rm peak1}$ agrees best with a constant period, while $t_{\rm peak2}$ varies the most. However, all period estimates concur within uncertainties. We simply adopt the $t_{\rm dip}$-derived period of $526.75 \pm 0.87$\,days in the subsequent analysis, given consistency of the three timing features.

\subsubsection{The 2025 outburst} 
\label{sec:2025outburst}
\begin{figure*}[htb]
\centering
\begin{minipage}{0.9\textwidth}
\centering{\includegraphics[angle=0,width=1.0\textwidth]{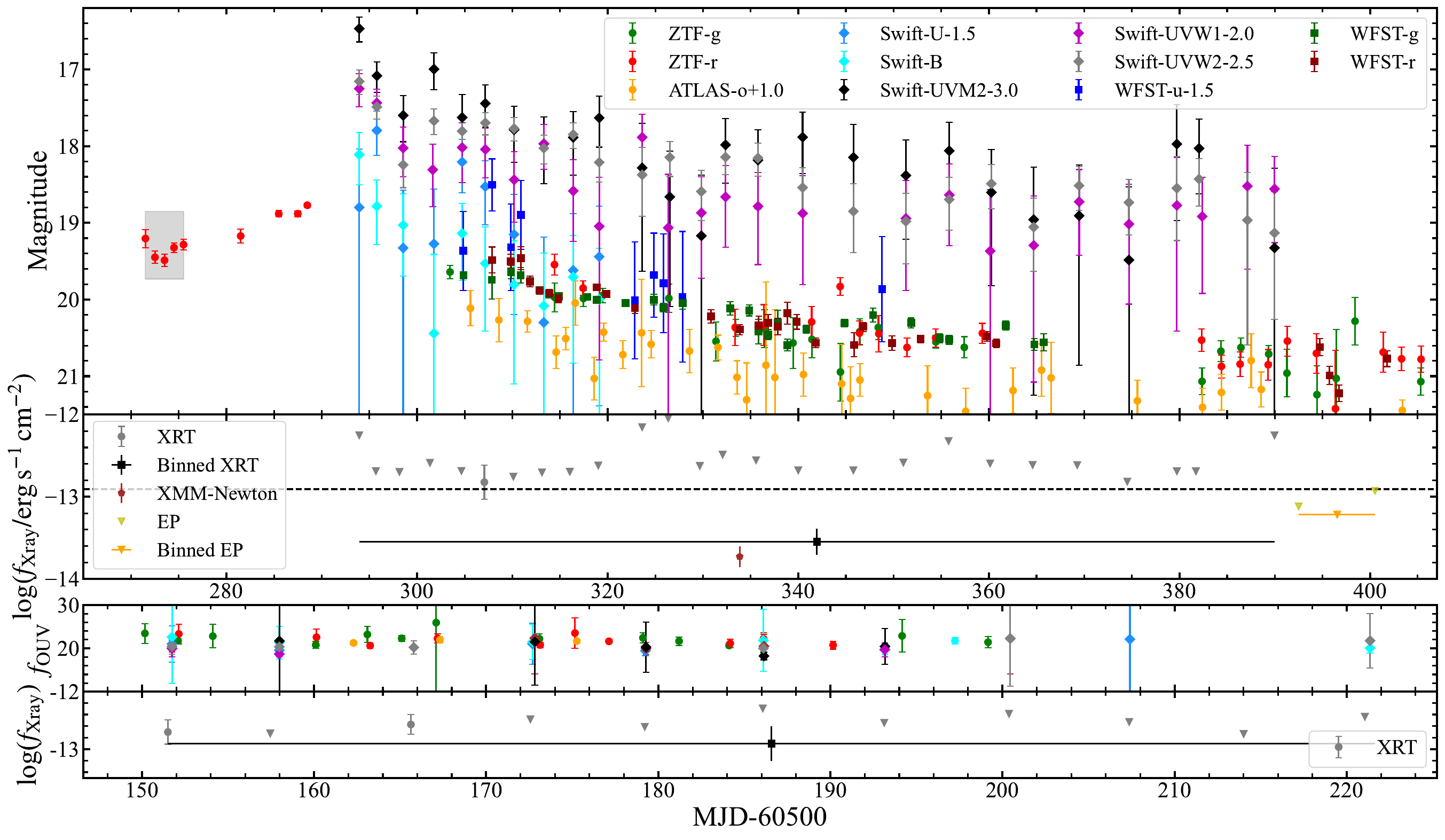}}
\end{minipage}
\caption{\label{fig:LC-recent} 
Multiband light curves for the quiescent phase preceding the recent outburst (lower two panels) and during the outburst (upper panels). In the top panel, multiband optical/UV photometry --- identified in the legend --- has been vertically offset for clarity. The ZTF $r$ data points highlighted by the gray shadow were used to determine $t_{\rm dip}$ for the fifth flare (see Section~\ref{sec:lcshape}). In the X-ray panels, gray points (downward triangles) represent marginal detections (upper limits) from {\it Swift}/XRT observations. Black squares indicate binned results, with the horizontal error bars reflecting the bin width in Modified Julian Date (MJD). For upper limits obtained from {\it EP} observations (yellow downward triangles), those with horizontal error bars correspond to binned data (orange downward triangles ).} 
\end{figure*}

After the fourth flare, we initiated multiband photometric monitoring with {\it Swift}, WFST~\citep{WangTinggui2023}, and additional facilities, leading to decent coverage for the latest flare (see Fig.~\ref{fig:LC-recent}). After correcting for Galactic extinction, we fit the multiband photometry assuming a blackbody SED. All six {\it Swift}/UVOT bands were used, except for the $V$ band owing to poor data quality. Photometry from other facilities was interpolated onto the MJD grid of the {\it Swift} observations during the fitting. The results are shown in Fig. S4 (online), alongside a comparison with ASASSN-14ko~\citep{HuangShifeng2023ASASN14ko,HuangShifeng2025ASASN14ko}. The peak blackbody luminosity of the recent outburst is $\sim 10^{44.5} \, \rm erg\,s^{-1}$, with a nearly constant temperature around $18{,}000\,\rm K$ and a blackbody radius on the order of $10^{15}\, \rm cm$. These values and their temporal evolution are broadly consistent with those observed in typical TDEs~\citep{Hammerstein2023,Yao2023} (see also Fig. S5 online). 

X-ray observations were conducted before and after the most recent outburst (see Fig.~\ref{fig:LC-recent}).
Weak X-ray emission was deteted in both phases, with X-ray luminosities near $10^{43}\, \rm erg\,s^{-1}$ (see Sections S1.3 and S2.3). The post-outburst X-ray emission appears to be fainter, although the significance of this change is limited by the large uncertainties. 
In  optical TDEs, X-ray emission is rarely detected near optical peak brightness, though delayed X-ray onset has often been observed~\citep{Guolo2024}, potentially due to delayed accretion~\citep{Piran2015} or thinning of the reprocessing layer~\citep{Thomsen2022}. In the rpTDE candidate ASASSN-14ko, the X-ray flux drops during the optical/UV rise and then recovers several days later~\citep{Payne2022ASASSN14ko, Payne2023ASASSN14ko}. The UV absorption lines in this system suggest X-ray absorption by outflows~\citep{Payne2023ASASSN14ko}. In AT2023uqm, optical spectra reveal outflows (Section ~\ref{sec:spectrumAnalysis} ), indicating potential X-ray absorption. Furthermore, the presence of strong optical coronal lines implies intense X-ray activity, raising the question of whether the currently observed faint X-ray flux is sufficient to produce them.

Radio observations were also conducted with e-MERLIN following the recent outburst (see Section S1.4 online), but no source was detected, with the last epoch observed on 2025-09-20 UTC.

\subsubsection{Evolution of the flare energy} 
\label{sec:mloss}

\begin{figure}[htb]
\centering
\begin{minipage}{0.5\textwidth}
\centering{\includegraphics[angle=0,width=1.0\textwidth]{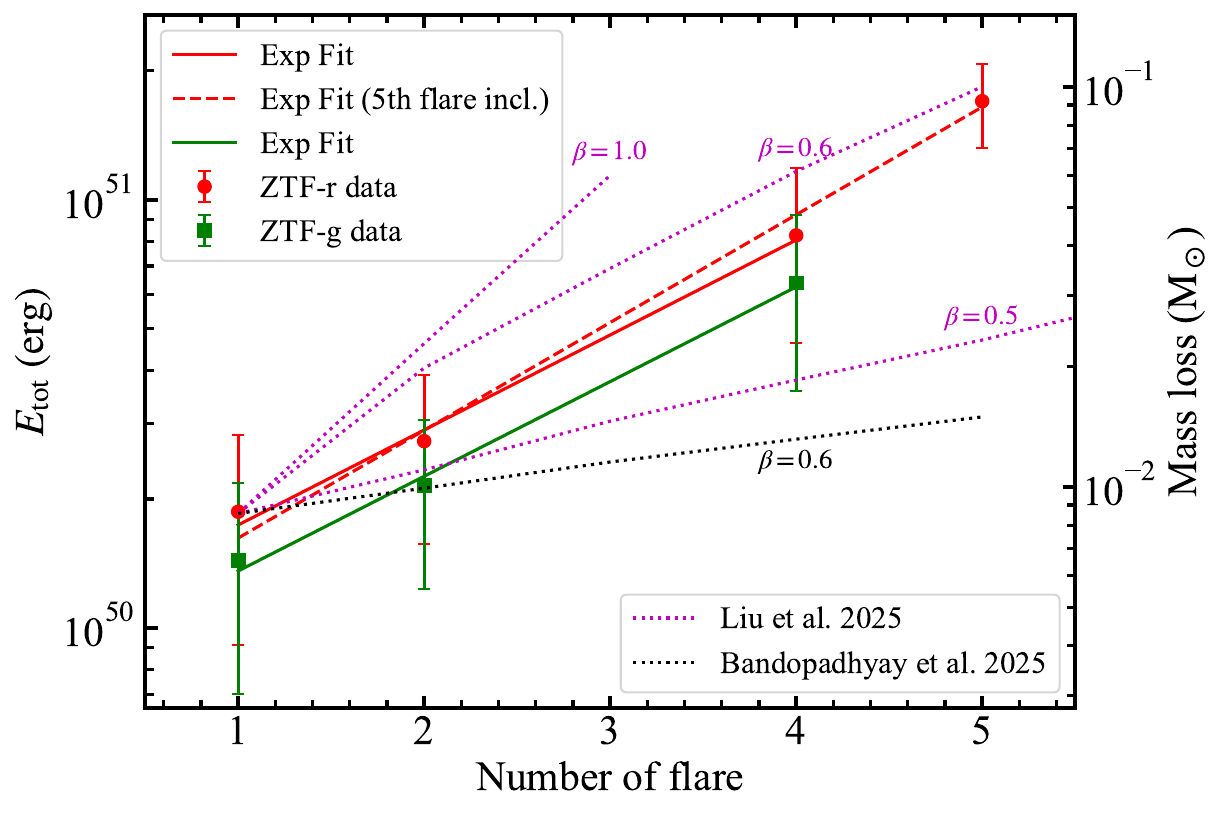}}
\end{minipage}
\caption{\label{fig:mloss} The evolution of integrated energy (left ordinate axis) and mass loss (right ordinate axis) as a function of flare number. The red points and green squares represent the integrated energy for each flare converted from ZTF $r$ and ZTF $g$ monochromatic fluxes, respectively (see details in the text).  Solid lines of the corresponding colors show exponential function fits to the data. The red dashed line represents the fit to the ZTF $r$ data when the fifth flare is included. The magenta dotted lines indicate the mass-loss evolution for a Sun-like star, based on hydrodynamic simulations by~Liu et al.~\citep{LiuChang2025}, with different values of $\beta$ (impact parameter, defined as the ratio of orbital pericenter to the classical tidal disruption radius) labeled on each line. The black dotted lines represent the mass-loss evolution for a $1\,\rm M_\odot$ zero-age main-sequence (ZAMS) star with $\beta=0.6$, given by the simulations of Bandopadhyay et al.~\citep{Bandopadhyay2024}. These simulation results have been scaled to facilitate comparison of the evolutionary behavior. }  
\end{figure}

The flares in AT2023uqm display a persistent increasing trend of peak flux, which we quantify through integrated energy calculations. Like typical optical TDEs and rpTDE candidates (e.g., AT2022dbl \citep{LinZheyu2024}), AT2023uqm maintains a nearly constant blackbody temperature during flare decay phases.  We therefore adopt a single temperature for all flares, derived from the decline phase of the most recent flare  (with variations representing uncertainty). While this assumption introduces some systematic error, it enables bolometric luminosity estimates from monochromatic fluxes without requiring complete multiband coverage. The  energy of each flare was derived by integrating its luminosity light curve. For the most recent flare, where only the second peak was observed, we applied a correction factor of 1.67 to estimate the total energy\footnote{This value was derived from the second flare, which has a similar second peak to that of the fifth flare.}. Fig.~\ref{fig:mloss} presents the flare energy evolution based on ZTF $g$ or $r$ data, indicating an approximately exponential growth pattern. Successive flares show energy increases by a factor of 1.79 (or 1.67 when excluding the last flare), with both bands yielding consistent results within the uncertainties.

We then try to convert the flare energy into mass loss, assuming a radiative efficiency of $\eta \approx 0.01$ (see Fig.~\ref{fig:mloss})\footnote{Given the ``missing energy'' puzzle for optical TDEs~\citep{LuWenbin2018}, we adopt a relatively conservative efficiency of 0.01.} . For the first flare observed by ZTF, the estimated mass loss is $\sim 0.01\,\rm M_\odot$, increasing to about $0.1\,\rm M_\odot$ for the most recent outburst.


\subsubsection{Earlier outbursts?}
\label{sec:earlier}

\begin{figure}[tb]
\centering
\begin{minipage}{0.5\textwidth}
\centering{\includegraphics[angle=0,width=1.0\textwidth]{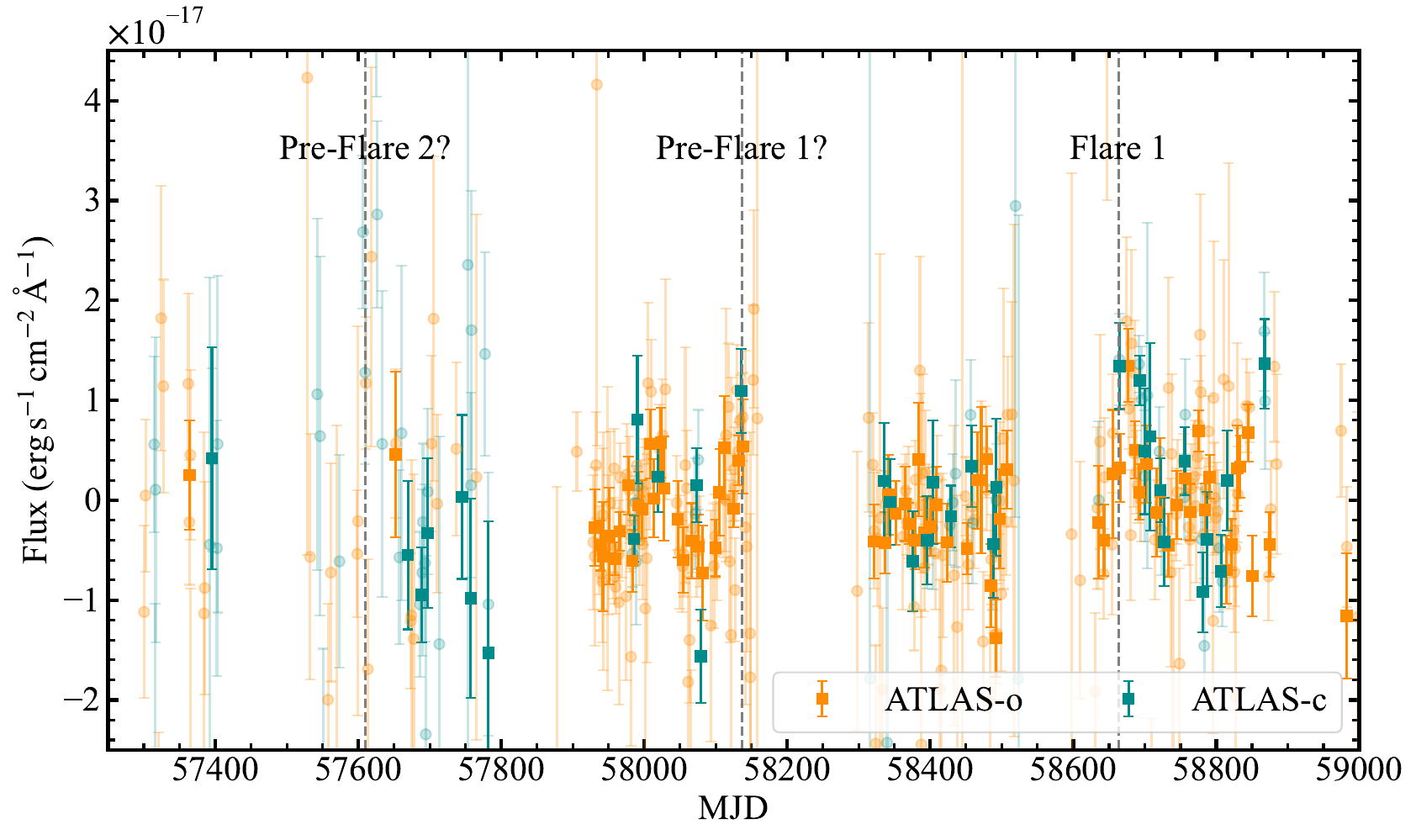}}
\end{minipage}
\caption{\label{fig:pre-flare} The ATLAS light curves at early times. The light-colored points 
depict daily-grouped light curves, while darker squares mark more rigorously filtered curves, binned every 5 days with a minimum of 3 data points per bin.} 
\end{figure}

We search for possible flares before the ZTF survey starting around MJD 58200 using archiveal ATLAS and CRTS data. While the CRTS data show no significant flares, the ATLAS light curves exhibit tentative flaring activity at the expected times before ZTF's first detection (see Fig.~\ref{fig:pre-flare}). Notably, the flare around MJD 58146 is particularly robust.
Although these pre-ZTF flares have a low S/N, they may have important implications.
In the rpTDE scenario, such early weak flares could have occurred many times~\citep{LiuChang2025,Bandopadhyay2024}, and they  influence accretion-disk formation or stellar heating via tidal interactions~\citep{Li2013} or collision processes~\citep{YaoPhilippe2025}. 

\subsection{Spectroscopic analysis}
\label{sec:spectrumAnalysis}
Eight spectra of AT2023uqm were obtained (see Fig.~\ref{fig:all-spec}). The NTT/EFOSC spectrum, taken during the fourth flare, reveals intermediate-width Balmer lines (IWBLs), similar to the spectra obtained during the fifth flare. The higher S/N Keck/LRIS spectrum shows, in addition to IWBLs, numerous emission features including prominent $\rm Fe\,\textsc{ii}$ lines, high-ionization iron coronal lines (e.g., $\rm [Fe\, \textsc{x}]\, \lambda 6374$ with luminosity $\approx \rm 10^{41} \, erg \, s^{-1}$), $\rm He\,\textsc{ii}$, blueshifted \oiii, and Bowen fluorescence lines such as $\rm O\,\textsc{iii}\, \lambda3133$. These emission lines appear to weaken in the late-time spectra (see also Table S6 online), indicating their relation to the outburst. Detailed spectral decomposition and analysis is presented in Section S2.4 (online).

Compared to typical TDEs, which are characterized by broad emission lines~\citep{vV2020SSR, Charalampopoulos2022}, the spectra of AT2023uqm are dominated by IWBLs and strong Fe$\,$\textsc{ii} emission. These spectral features resemble those of narrow-line Seyfert 1 (NLSy1) galaxies and rapid transients discovered in faint AGNs (see Fig. S6 and Section S2.4.2 online). The latter include Bowen fluorescence flares (BFFs; \cite{Tadhunter2017}), which are thought to be promising TDE candidates in AGNs, as well as other well-known TDE-in-AGN candidates such as PS16dtm~\citep{Blanchard2017} and the candidate rpTDE, ASASSN-14ko. Considering the absence of significant AGN activity prior to the outburst (see Section~\ref{sec:Host}) and the presence of strong Bowen fluorescence emission lines, AT2023uqm argues against a classification as a normal NLSy1 and instead favors an interpretation as a TDE-related event. Furthermore, the detection of high-luminosity iron coronal lines provides additional support for a TDE origin, as such lines are considered a possible indicator of TDEs occurring in dust-rich environments~\citep{WangTinggui2012}.
The observed IWBLs might originate from distant gas responding to the outburst, where the emitting gas could either be pre-existing nuclear gas or material expelled by the event itself through outflows launched by previous outbursts.

The absence of the very broad Balmer lines (FWHM $\sim 10,000\ \rm km\,s^{-1}$) typically observed in TDEs may be attributed to several factors: the lack of high-S/N spectra obtained near or before peak brightness, given the observed narrowing of line widths in TDEs over time~\citep{vV2020SSR,Charalampopoulos2022}; the event’s possible membership in the class of featureless TDEs~\citep{Hammerstein2023}; or the blending of very broad Balmer components with other emission features and/or the continuum, making them difficult to decompose. Given that BFFs and several other promising TDE candidates in gas-rich environments also generally lack prominent, very broad emission features, intrinsic physical processes may be responsible for this behavior. A detailed investigation of these possibilities, however, is beyond the scope of this paper.

Unlike in BFFs and Bowen emission TDEs --- where the $\rm N\,\textsc{iii}$-to-$\rm O\,\textsc{iii}$ line-intensity ratio is typically high, an effect attributed to nitrogen abundance enhancement following the disruption of a massive main-sequence star~\citep{Kochanek2016} --- this ratio is low in AT2023uqm, although $\rm N\,\textsc{iii}\, \lambda4640$ is detected (see Section S2.4 online). This result is nevertheless consistent with an rpTDE interpretation, as a partial disruption would initially strip only the stellar envelope, which lacks the nitrogen-enriched material produced by the CNO cycle in the stellar core. Specifically, for AT2023uqm, the favored low-mass main-sequence progenitor (see Section~\ref{sec:rpTDEscenario}) is unlikely to show strong nitrogen abundance enhancement owing to its slow evolutionary timescale relative to the age of the universe. An alternative giant-star scenario is also consistent, as such an event would only strip the stellar envelope. Dilution from a potentially gas-rich environment could also contribute to the observed low line-intensity ratio.

In summary, the spectral characteristics of AT2023uqm are consistent with an rpTDE candidate. However, the dominance of IWBLs and the presence of strong infrared echoes suggest that the host may harbor a faint or recently faded AGN. A pre-existing AGN could in fact enhance the TDE rate and facilitate the formation of low-eccentricity stellar orbits~\citep{JiangNing2025QPE}.



\section{The rpTDE scenario}
\label{sec:rpTDEscenario}


Based on the light curve and spectroscopic features analyzed above, AT2023uqm is undoubtedly a highly confident candidate for rare rpTDEs. In particular, we observe, for the first time, an approximately exponential increase in flare energy across multiple periodic outbursts. This suggests we are witnessing a star being disrupted by the BH's tidal forces in a runaway manner. In this section, we aim to quantitatively explain the observed phenomena within the rpTDE framework and simultaneously constrain the properties of the disrupted star. A detailed discussion disfavoring alternative scenarios, involving an SMBH with a companion or intrinsic AGN variability, is presented in Section S3.4 (online). We denote the star’s mass, radius, orbital pericenter, orbital energy, orbital semimajor axis, and orbital period as $M_*$, $R_*$, $r_{\rm p}$, $E_{\rm orb}$, $a_{\rm orb}$, and $P$, respectively. Changes in the orbital period and pericenter over multiple passages are negligible (see Section S3.1 online). We note that a Newtonian framework is adopted in this work. However, for the disruption of a main-sequence star by a massive BH ($M_{\rm BH} \gtrsim 10^{7}\,\rm M_\odot$), the orbital dynamics may become strongly relativistic. In addition, current BH mass estimates remain highly uncertain. For more massive BHs ($\gtrsim 10^{8}\,\rm M_\odot$), the tidal disruption of main-sequence stars may be suppressed, as stars are likely to be directly captured by the BH event horizon~\citep{Bardeen1972,Hills1975}.

\subsection{Orbital parameters}
First, the orbital period of the star can be robustly determined as $P \approx 425.5$\,days in the rest frame. The corresponding orbital semimajor axis and orbital energy are
\begin{align}
   &  a_{\rm orb} = \left(\frac{\mathrm{G}M_{\rm BH}P^2}{4 \uppi ^2}\right)^{1/3} \approx 3.57\times 10^{15} \left(\frac{M_{\rm BH}}{10^7\, {\rm M}_\odot}\right)^{1/3}\, \rm cm, \\
   &  E_{\rm orb} = -\frac{\mathrm{G}M_{\rm BH}}{2a_{\rm orb}}\approx -1.86\times 10^{17}\left(\frac{M_{\rm BH}}{10^7\, {\rm M}_\odot}\right)^{2/3}\,\rm erg\,g^{-1}.
\end{align}
The orbital eccentricity of the star is then
\begin{align}
    e = 1- \frac{r_\mathrm{p}}{a_{\rm orb}} \approx 1- 0.007 \left(\frac{0.6}{\beta}\right)\left(\frac{R_*}{ R_\odot}\right) \left( \frac{M_*}{ {\rm M}_\odot}\right)^{-1/3},
\end{align}
where $\beta$ is the impact parameter, defined as the ratio of the orbital pericenter to the classical tidal disruption radius,
\begin{align}
r_\mathrm{t}=\left(\frac{M_{\rm BH}}{M_*}\right)^{1/3}R_*.  
\end{align}
For typical main-sequence stars, the eccentricity is about 0.99, which can be naturally produced via the tidal breakup of binary systems (e.g.,~\citep{Cufari2022,Yu2024}). 

\subsection{Mass-loss evolution} 
\label{sec:mloss-ms}
Unlike ASASSN-14ko, which exhibits flares of similar strength, AT2023uqm displays novel behavior: a nearly exponential increase in integrated energy and thus inferred mass loss across its flare series. This phenomenon has been observed for the first time, with the exception of AT2020vdq~\citep{Somalwar2025}, for which only two flares were observed and the second was significantly brighter than the first by a factor of $\sim 6$.

Recent simulations and analytical works have begun to explore mass-loss evolution in rpTDEs with multiple passages. While a unified picture has yet to emerge, a common trend is identified for main-sequence stars: low-mass stars with convective envelopes experience runaway mass loss across encounters~\citep{LiuChang2025,Bandopadhyay2024,LuQuataert2023,Bandopadhyay2025}, while  high-mass stars with radiative envelopes exhibit the opposite behavior. Specifically, Liu et al.~\citep{LiuChang2025} reported that solar-like stars undergoing multiple passages can experience an exponential increase in mass loss, eventually leading to complete disintegration, with enhancement factors per passage ranging from 1.3 to 2.6, depending on the initial impact parameter $\beta$. Notably, the observed mass-loss enhancement for AT2023uqm falls within this range (see Fig.~\ref{fig:mloss}). Hence, based on our current understanding, the runaway behavior of AT2023uqm indicates that the disrupted star was likely of low mass ($\lesssim 1\, \rm M_\odot$). More precise determinations of its mass and initial impact parameter will become possible with a clearer understanding of rpTDE evolution. In this context, AT2023uqm can serve as a benchmark system for studying these processes.

\subsection{The future flares} 
The observed period predicts the next outburst (if it occurs) will peak (the first peak) around MJD 61280 (2026-08-28 UTC). Assuming a radiative efficiency of $\eta \approx 0.01$, the recent outburst corresponds to a mass loss of $\sim 0.1 \, \rm M_\odot$. Given the inferred potential low mass of the progenitor star and an estimated mass-loss enhancement factor of $\sim 1.7$ per passage, the star may be completely disrupted after only one or two more flares. This conclusion could change if the actual efficiency is higher, as this would imply a lower mass loss per flare and potentially allow for more events before complete disruption.

The predictable timing of the next flare provides an opportunity to capture the detailed evolution of the rising phase, which is crucial for understanding the origin of the radiation.


\section{On the origin of double-peaked light curve}
\label{sec:doublepeakOrigin}
Beyond the constraints imposed by the orbital period and mass-loss evolution, the light-curve structure offers additional insights into the system's origin. Specifically, Section~\ref{sec:lcshape} shows that all outbursts exhibit similar double-peaked structures, with a separation of $\sim 20$ days, two peaks of comparable strength, and a longer duration for the second peak. However, considerable uncertainty remains regarding the physical origin of the observed optical/UV radiation. In the following, we explore several emission scenarios and their implications based on the light-curve data. Owing to the sparse sampling of the third and fifth flares, we primarily focus on interpreting the similarity of the double-peaked properties among flares, while deferring any investigation of their potential evolution to future studies with improved temporal coverage of subsequent flares.
Notably, our analysis suggests that giant stars, in addition to main-sequence stars, have a significant likelihood of being the disrupted objects. 



\subsection{Double-peaked fallback rate}
\label{sec:fallbackrate}
To understand the double-peaked structure, we first consider the fallback rate in partial TDEs. In such an event, only the outer layers of the star are stripped away, leaving a surviving core. This naturally results in a double-peaked distribution of debris mass as a function of specific binding energy ($\mathrm{d}m/\mathrm{d}\epsilon$; see Fig. 2 of Ryu et al.~\citep{Ryu2020pTDE}), with a dip near the orbital energy of the star.
This contrasts sharply with the nearly uniform distribution characteristic of complete disruptions. For rpTDEs on elliptical orbits, a double-peaked fallback rate arises naturally when the orbital energy is sufficiently low to bind the most energetic debris to the BH. Although the precise relationship between fallback rate and observed emission remains unclear (e.g., \citep{Gezari2021,Roth2020, Bonnerot2021}), models generally predict light curves that trace the fallback rate. This provides a natural explanation for the double-peaked flares in AT2023uqm within the rpTDE framework.

In detail, the fallback rate can be expressed as\footnote{The Keplerian orbital energy-period relation adopted here is not strictly valid for pTDEs because of the gravitational influence of the surviving core~\citep{Coughlin2019}. However, this effect has a negligible impact on the peak fallback time and thus on the peak separation concerned (e.g., Fig. 10 of Guillochon et al.~\citep{Guillochon2013}). }
\begin{equation}
    \dot{m}= \frac{\mathrm{d}m}{\mathrm{d}\epsilon} \left(\frac{1}{3}\right) (2\uppi \mathrm{G} M_{\rm BH})^{2/3} t^{-5/3}\, ,
\end{equation}
in which $t$ represents the fallback time since pericenter passage.  The equation indicates that a  symmetric, double-peaked $\mathrm{d}m/\mathrm{d}\epsilon$ will cause the second peak to be lower and more temporally extended than the first, owing to the stretching term $t^{-5/3}$. This agrees with our observations (see Fig.~\ref{fig:all-flares}). We estimate the peak ratio in the fallback rate\footnote{Here we assume the $\mathrm{d}m/\mathrm{d}\epsilon$ is identical at both peaks, though this may not hold owing to the $t^{-5/3}$ stretching. A simple test using the mock fallback rates (see Footnote 8 below) yields similar results when $\Delta t_{\rm peak}/P$ approaches the observed value.
 } as $[P_{\rm orb}/(P_{\rm orb} + \Delta t_{\rm peak})]^{-5/3} \approx 1.06$, which is comparable to the observed peak strength ratio (see Table S1 online). Hence, under the assumption that radiation completely follows the fallback rates, the observed double-peaked light curves can be roughly reproduced. 

To assess the feasibility of this scenario, we suggest that half of the energy spread between the two fallback peaks is $\mathrm{d}E = f_\mathrm{E}(\mathrm{G}M_{\rm BH}/r_\mathrm{p}^2)R_*$, where $f_\mathrm{E}$ is a factor of order unity. The ratio of this energy spread to the orbital energy is 
\begin{align}
    \frac{\mathrm{d}E}{|E_{\rm orb}|}  & \approx 0.8 \left(\frac{f_\mathrm{E}}{1}\right) \left(\frac{10^7\, M_\odot}{M_{\rm BH}}\right)^{1/3} \left(\frac{M_*}{{\rm M}_\odot}\right)^{2/3}\left(\frac{R_\odot}{R_*}\right)\left(\frac{\beta}{0.6}\right)^2.  \label{eq:Eratio} 
\end{align}
This ratio indicates that it is plausible for solar-like stars to capture both the inner and outer debris streams. However, the time separation between the two peaks can be estimated as\footnote{Here, we assume $\mathrm{d}E \ll E_{\rm orb}$, which is inadequate for main-sequence stars. However, the actual value is even larger, worsening the noted discrepancy. Additionally, we tested this simple equation against mock fallback rates for eccentric orbits constructed from the simulation results of Law-Smith et al.~\citep{LawSmith2020} using a method similar to that of Zhong et al.~\citep{ZhongShiyan2025}, and found that it reproduces $\Delta t_{\rm peak}/P$ reasonably well, with discrepancies of only a factor around unity when adopting $f_{\rm E} = 1$.}
\begin{align}
 \frac{\Delta t_{\rm peak}}{P}\approx 3\frac{\mathrm{d}E}{|E_{\rm orb}|}\, . \label{eq:Pratio}
\end{align}

For main-sequence stars, the predicted ratio exceeds the observed value of $\sim 0.038$ by more than an order of magnitude. Eq.~\eqref{eq:Eratio} suggests that increasing the stellar radius by tens to hundreds of times, making it a giant star, can solve the discrepancy (see Section S3.2 online for a detailed illustration). We discuss the giant-star scenario in detail in Section S3.3 (online), including its origin, mass-loss evolution, and orbital properties.  In particular, we suggest the demanded giant star most likely formed through normal stellar evolution, rather than via mechanisms such as star-disk collisions~\citep{YaoPhilippe2025} or tidal heating~\citep{Li2013,YaoPhilippe2025,LuWenbin2021}, though these alternatives cannot be entirely ruled out. Furthermore, while such a giant star can undergo runaway mass-loss evolution as observed initially,  this rate is expected to  eventually decline as the gravity of the stellar core becomes dominant~\citep{MacLeod2013}. Detecting this transition would provide a key discriminant from the main-sequence scenario, which ultimately leads to complete tidal disruption (see Section~\ref{sec:mloss-ms}).
\subsection{The presence of an accretion disk} 
One issue with the simple scenario outlined above is that the circularization process may not occur rapidly enough, particularly for the required giant-star progenitor. To accelerate circularization, one could introduce an accretion disk, which could also shorten the viscous timescale if the debris collides with the disk in a retrograde manner. In reality, an accretion disk is indeed very likely for AT2023uqm, either coming from pre-event weak AGN activity or forming through multiple previous weak pTDE process before the observed flares. However, the interaction between stripped debris and an accretion disk influences more than just the acceleration of the circularization process. Specifically, there are two key points related to our results: (i) the fallback material may not collide with or inject material at the pericenter, and (ii) the debris-disk collision may become the dominant radiation process. For the second point, the double-peak structure might not be related to the fallback rate at all, but rather arise from two separate collisions within one orbit.

\subsubsection{Position of debris-disk collision }
\label{sec:collisionPosition}
As mentioned above, the presence of an accretion disk can accelerate the circularization process through debris-disk collisions. In particular, if the disk is dense enough relative to the debris, the stripped material may be directly injected into the disk and immediately contribute to the accretion process after collision. Such a collision may occur either during the first or second crossing within one orbit, depending on the properties of both the debris and the disk at the collision point. Moreover, the mass-injection point is not necessarily at periapsis but rather at the collision point, which depends on the relative geometry of the disk and orbit. Therefore, a factor $f_{\rm P}\in$ [0,1] should be introduced into Eq.~\eqref{eq:Pratio}, such that
\begin{align}
    \frac{\Delta t_{\rm peak}}{P}\approx 3f_{\rm P} \frac{\mathrm{d}E}{E_{\rm orb}}\, . \label{eq:Pratio-fp}
\end{align}
As a result, the requirement for a giant star, based on the observed ratio of peak separation to period, can be relaxed to a main-sequence star. However, to achieve this, the orbit's semimajor axis must be nearly parallel to the disk plane (see Figs. S17 and S18 online), and the collision point should occur just before apocenter, but far from pericenter, considering the orbit's extreme eccentricity\footnote{We assessed this with the typical case, a star of $1\,\rm M_\odot$.}. 
\subsubsection{Dominance of collision-driven emission} 
Besides accelerating the circularization process, debris-disk collisions would also produce luminous radiation, which could be the dominant source of optical/UV light flares. In this scenario, part of the debris's and disk's kinetic energy would be converted into radiation. Such a collision-powered model has also been proposed for X-ray quasiperiodic eruptions (QPEs) ~\citep{YaoPhilippe2025,LuQuataert2023,Linial2025}, although in that case, the debris is typically thought to originate from star-disk collisions rather than tidal disruption processes.  In this sense, AT2023uqm could be considered an optical analog to QPEs, but the underlying reason for this wavelength difference is not yet understood.

Generally, there are two opportunities per orbit for debris to collide with the disk. In some cases, only one of these collisions produces luminous emission. This occurs in two possible ways: (i) most of the debris is dissipated during the first collision before reaching apocenter; and (ii) the debris passes through the disk during the first collision without significant emission but produces luminous radiation during the second collision after falling back from apocenter\footnote{This occurs when the first collision happens immediately after disruption, where the debris has not yet significantly expanded and therefore retains a small cross-section and elongation, and a high density. Such a collision is likely a star-disk interaction generating negligible optical emission. By the time of the second collision, gravitational evolution combined with heating from the first collision causes the debris stream to attain a larger cross-section and elongation. This increase in size enables more efficient dissipation of the kinetic energy of both the debris and the disk material, leading to a brighter flare.  }. 
The outcome depends on the properties of the disk and debris, where the star's orbital geometry relative to the disk also plays a critical role by determining the degree of expansion and stretching of the debris at the collision point after being stripped from the star~\citep{Bonnerot2022}. In this single-collision-dominated scenario, the double-peaked structure in observed flares should therefore originate from the double-peaked fallback rate mentioned above. Assuming the presence of a spherical outflow or envelope from which radiation can escape after collision, the corresponding photon diffusion timescale can be estimated as $t_{\text{\rm diff}}\approx (3\Delta M \kappa)/(4\uppi R_{\text{ph}} \mathrm{c}) \approx 0.7\,(\Delta M)/(0.01\, M_\odot)$\, day. Here, we adopt a photospheric radius of $R_{\text{ph}}\approx 10^{15}$\,cm based on blackbody fits, and an electron scattering opacity $\kappa$. For an inferred mass loss of 0.01--0.1\,M$_\odot$ from the observed flares, the resulting diffusion time is shorter than the flare duration, supporting the assumption that the light curve closely tracks the fallback rate. This scenario imposes the same constraints on the progenitor star as derived in Sections~\ref{sec:fallbackrate} and~\ref{sec:collisionPosition} --- namely, a giant star or a main-sequence star with a collision point far from pericenter. However, in both cases, the efficiency is typically lower than the assumed 0.01 value\footnote{In fact, for the case of a giant star disrupted by a more massive BH ($\gtrsim10^{8}\,\rm M_\odot$), such an efficiency can be achieved if the majority of the debris's kinematic energy is converted to radiation.}, raising issues such as a longer diffusion timescale and an unusually massive progenitor star.

Interestingly, two luminous collisions per orbit are also plausible. In this model, the double peaks may represent these collisions rather than the fallback rate. 
Specifically, the first peak would arise from the collision of fallback debris from apocenter with the disk. After the pericenter passage, newly stripped debris would then collide with the disk on the opposite side, creating the second peak (see Fig. S17). This scenario can accommodate a main-sequence progenitor and also explain the peak separation. However, for solar-like stars, the second collision needs to happen far from pericenter. Although this promotes debris spread, enhancing collision cross-section, low efficiency remains a challenge. The resulting higher mass loss increases the diffusion timescale, potentially wiping out light-curve details linked to fallback-rate variations, but necessitating an unusually massive star. 
Additionally, Eq.~\eqref{eq:Pratio-fp} shows that the first collision would produce a longer optical peak, contradicting the observations, unless the kinematic distribution of the debris was significantly modified by the collision before apocenter --- a plausible scenario. Finally, the relative strength of the two collisions is uncertain, hinging on disk properties, collision sites, and the evolution of mass loss across different rpTDE passages.

\section{Conclusion} 
In this paper, we report the discovery of the periodically outbursting nuclear transient AT2023uqm, with at least five distinct flares observed. Through detailed analysis of the light curves, we find that AT2023uqm exhibits a period of $\sim 526.75$ days (observer's frame), a characteristic double-peaked flare profile, and a nearly exponential increase in flare strength. Multiwavelength follow-up observations were conducted for the most recent outbursts, covering optical/UV, X-ray, and radio bands. No significant emission (or brightening) was detected in the radio (or X-ray). The optical/UV data indicate a nearly constant blackbody temperature around $18,000\,\mathrm{K}$. Spectra during the last two flares reveal similar features dominated by intermediate-width Balmer lines, strong \feii\ emission, high-ionization coronal lines, and Bowen fluorescence lines (especially \Bowenoiii\ ). 

AT2023uqm represents the first known case of periodic nuclear flares with exponentially rising strength. This novel phenomenon suggests that we are witnessing a star being gradually destroyed by an SMBH. The characteristics of the light curves strongly suggest an rpTDE, and the  lower \niii-to-\Bowenoiii\ Bowen fluorescence line ratio, in comparison to typical TDEs or BFFs, further supports this. Nevertheless, uncertainties about the radiation mechanism limit precise system characterization. Our study reveals that the observed double-peaked light curve may be produced by a giant star. A main-sequence star could only account for this if an accretion disk exists, requiring specific stellar orbit and disk properties. Future monitoring could distinguish these scenarios: a giant star's dense core might cause the flare strength to peak and then decline, while a main-sequence star would lead to a continual increase in flare strength until its complete disintegration after up to two more events. 
Several challenges, including timescales and emission efficiency, remain for the scenarios discussed and should be addressed in dedicated future work.

In summary, AT2023uqm is a compelling rpTDE candidate owing to its distinct temporal evolution, potentially serving as a template for studying these phenomena. Prior to this, even the most promising candidate, ASASSN-14ko, was suggested to have alternative origins, such as recurrent star-disk collisions~\citep{Linial2024}, with observed flares powered by debris from these collisions rather than tidal disruption. 
Our analysis demonstrates the power of rpTDEs in constraining the progenitor star, the disruption conditions, and the origin of the star, thus shedding light on the processes underlying tidal disruption. Moreover, AT2023uqm offers a unique opportunity to study the early-time multiwavelength evolution of TDEs, thanks to its predictable recurrence.  The next outburst is expected to peak (the first peak) around MJD$\,$61280 (2026-08-28 UTC). Given the estimated mass loss, only a few more outbursts are anticipated for AT2023uqm. With the advent of more sensitive wide-field surveys such as WFST \citep{WangTinggui2023} and LSST \citep{Ivezic2019}, the discovery of similar events is highly anticipated.



\section*{Conflict of interest} 
The authors declare that they have no conflict of interest.

\section*{Acknowledgments}
\footnotesize

This work is supported by the National Natural Science Foundation of China (grant 12393814), 
the National Key Research and Development Program of China (grant 2023YFA1608100), the Strategic Priority Research Program of the Chinese Academy of Sciences (grant XDB0550200), the National Natural Science Foundation of China (grants 12192221, 12522303, 12025303, 12503017), the Postdoctoral Fellowship Program of CPSF under grant GZC20252096, the Fundamental Research Funds for Central Universities (grants WK2030000097, WK2030250127), the China Postdoctoral Science Foundation (2025M773198), Anhui Provincial Natural Science Foundation (2508085QA006) and the China Manned Space Project. The authors appreciate the support of the Cyrus Chun Ying Tang Foundations. 
Yi Yang's research is partially supported by the Tsinghua University Dushi Program.
Alexei V. Filippenko’s research group at UC Berkeley acknowledges financial assistance from the Christopher R. Redlich Fund, as well as donations from Gary and Cynthia Bengier, Clark and Sharon Winslow, Alan Eustace and Kathy Kwan, Timothy and Melissa Draper, Briggs and Kathleen Wood, Ellyn and Alan Seelenfreund, Sanford Robertson (Weikang Zheng is a Bengier-Winslow-Eustace Specialist in Astronomy, Thomas G. Brink is a Draper-Wood-Seelenfreund Specialist in Astronomy, Yi Yang is a Bengier-Winslow-Robertson Fellow in Astronomy), and many other donors.

We appreciate the expert assistance of the staff at the various observatories where data were obtained. 
The Wide Field Survey Telescope (WFST) is a joint facility of the University of Science and Technology of China, Purple Mountain Observatory. 
%
This research uses data obtained through the Telescope Access Program (TAP), which has been funded by the TAP member institutes. Observations with the Hale Telescope at Palomar Observatory were obtained as part of an agreement between the National Astronomical Observatories, the Chinese Academy of Sciences, and the California Institute of Technology. 
Some of the data presented herein were obtained at the W. M. Keck
Observatory, which is operated as a scientific partnership among the
California Institute of Technology, the University of California, and NASA; the observatory was made possible by the generous financial support of the W. M. Keck Foundation.
A major upgrade of the Kast spectrograph on the Shane 3\,m telescope at Lick Observatory, led by Brad Holden, was made possible through gifts from the Heising-Simons Foundation, William and Marina Kast, and the University of California Observatories. Research at Lick Observatory is partially supported by a gift from Google.

This work based on observations obtained with {\it XMM-Newton}, an ESA science mission with instruments and contributions directly funded by ESA Member States and NASA. e-MERLIN is a National Facility operated by the University of Manchester at Jodrell Bank Observatory on behalf of STFC, part of UK Research and Innovation.  
The authors acknowledge the use of public data from the {\it Swift} data archive. We thank the {\it Swift} ToO team for accepting our proposal and executing the observations.

This work has made use of data from the Asteroid Terrestrial-impact Last Alert System (ATLAS) project. The ATLAS project is primarily funded to search for near-Earth objects (NEOs) through NASA grants NN12AR55G, 80NSSC18K0284, and 80NSSC18K1575; byproducts of the NEO search include images and catalogs from the survey area.  The ATLAS science products have been made possible through the contributions of the University of Hawaii Institute for Astronomy, the Queen’s University Belfast, the Space Telescope Science Institute, the South African Astronomical Observatory, and The Millennium Institute of Astrophysics (MAS), Chile.

\section*{Author Contributions}
Yibo Wang and Tinggui Wang led the project. Tinggui Wang discovered the event, and Yibo Wang carried out the data analysis and wrote the manuscript. Yibo Wang and Tinggui Wang led the theoretical interpretation, with insightful discussions from Wenbin Lu, Dong Lai, Rongfeng Shen and Shiyan Zhong.
Shifeng Huang applied for the X-ray observations and reduced the corresponding data. Jiazheng Zhu retrieved the ZTF light curves and performed the host-galaxy SED fitting. Yi Yang, Ning Jiang, Alexei V. Filippenko, Thomas G. Brink, and Weikang Zheng contributed to the Keck spectrum. Alexei V. Filippenko and Thomas G. Brink contributed the Shane/Kast spectra. Ning Jiang, Weiyu Wu, Yibo Wang, and Shifeng Huang contributed to the P200/NGPS spectra. Jianwei Lyu and Ning Jiang contributed the MMT spectrum. Yibo Wang performed the spectroscopic analysis, and Xiaer Zhang suggested the use of FANTASY for $\rm Fe\,\textsc{ii}$ modeling. 
Tinggui Wang and Xinwen Shu proposed the eMERLIN radio observations, and Xinwen Shu reduced the radio data. Tianyu Xia performed the initial light-curve fitting, which was not included in the final manuscript. Di Luo carried out the infrared echo modeling. Alexei V. Filippenko edited the manuscript and improved the language quality.
Tinggui Wang, Wenbin Lu, Ning Jiang, Shifeng Huang, Alexei V. Filippenko, Yi Yang, Xu Kong, and other coauthors read the manuscript and provided valuable suggestions. Minxuan Cai, Zelin Xu, Mingxin Wu, Lulu Fan, Ji-an Jiang, Xu Kong, Bin Li, Feng Li, Ming Liang, Wentao Luo, Jinlong Tang, Zhen Wan, Hairen Wang, Jian Wang, Yongquan Xue, Dazhi Yao, Hongfei Zhang, Wen Zhao, Xianzhong Zheng, Qingfeng Zhu, and Yingxi Zuo contributed to the WFST data.

\footnotesize
\bibliographystyle{scibull} 
\bibliography{Intro}

@ARTICLE{Thomsen2022,
       author = {{Thomsen}, Lars L. and {Kwan}, Tom M. and {Dai}, Li-Xin and {Wu}, Samantha C. and {Roth}, Nathaniel and {Ramirez-Ruiz}, Enrico},
        title = "{Dynamical unification of tidal disruption events}",
      journal = {\apjl},
         year = 2022,
        month = oct,
       volume = {937},
       number = {2},
          eid = {L28},
        pages = {L28},
          doi = {10.3847/2041-8213/ac911f},
archivePrefix = {arXiv},
       eprint = {2206.02804},
 primaryClass = {astro-ph.HE},
       adsurl = {https://ui.adsabs.harvard.edu/abs/2022ApJ...937L..28T}
}

@ARTICLE{Frederick2019,
       author = {{Frederick}, Sara and {Gezari}, Suvi and {Graham}, Matthew J. and {Cenko}, S. Bradley and {van Velzen}, Sjoert and {Stern}, Daniel and {Blagorodnova}, Nadejda and {Kulkarni}, Shrinivas R. and {Yan}, Lin and {De}, Kishalay and {Fremling}, U. Christoffer and {Hung}, Tiara and {Kara}, Erin and {Shupe}, David L. and {Ward}, Charlotte and {Bellm}, Eric C. and {Dekany}, Richard and {Duev}, Dmitry A. and {Feindt}, Ulrich and {Giomi}, Matteo and {Kupfer}, Thomas and {Laher}, Russ R. and {Masci}, Frank J. and {Miller}, Adam A. and {Neill}, James D. and {Ngeow}, Chow-Choong and {Patterson}, Maria T. and {Porter}, Michael and {Rusholme}, Ben and {Sollerman}, Jesper and {Walters}, Richard},
        title = "{A New Class of Changing-look LINERs}",
      journal = {\apj},
         year = 2019,
        month = sep,
       volume = {883},
       number = {1},
          eid = {31},
        pages = {31},
          doi = {10.3847/1538-4357/ab3a38},
archivePrefix = {arXiv},
       eprint = {1904.10973},
 primaryClass = {astro-ph.HE},
       adsurl = {https://ui.adsabs.harvard.edu/abs/2019ApJ...883...31F}
}

@ARTICLE{Sniegowska2020,
       author = {{Sniegowska}, M. and {Czerny}, B. and {Bon}, E. and {Bon}, N.},
        title = "{Possible mechanism for multiple changing-look phenomena in active galactic nuclei}",
      journal = {\aap},
         year = 2020,
        month = sep,
       volume = {641},
          eid = {A167},
        pages = {A167},
          doi = {10.1051/0004-6361/202038575},
archivePrefix = {arXiv},
       eprint = {2007.06441},
 primaryClass = {astro-ph.GA},
       adsurl = {https://ui.adsabs.harvard.edu/abs/2020A&A...641A.167S}
}

@ARTICLE{Bellm2019,
       author = {{Bellm}, Eric C. and {Kulkarni}, Shrinivas R. and {Graham}, Matthew J. and {Dekany}, Richard and {Smith}, Roger M. and {Riddle}, Reed and {Masci}, Frank J. and {Helou}, George and {Prince}, Thomas A. and {Adams}, Scott M. and {Barbarino}, C. and {Barlow}, Tom and {Bauer}, James and {Beck}, Ron and {Belicki}, Justin and {Biswas}, Rahul and {Blagorodnova}, Nadejda and {Bodewits}, Dennis and {Bolin}, Bryce and {Brinnel}, Valery and {Brooke}, Tim and {Bue}, Brian and {Bulla}, Mattia and {Burruss}, Rick and {Cenko}, S. Bradley and {Chang}, Chan-Kao and {Connolly}, Andrew and {Coughlin}, Michael and {Cromer}, John and {Cunningham}, Virginia and {De}, Kishalay and {Delacroix}, Alex and {Desai}, Vandana and {Duev}, Dmitry A. and {Eadie}, Gwendolyn and {Farnham}, Tony L. and {Feeney}, Michael and {Feindt}, Ulrich and {Flynn}, David and {Franckowiak}, Anna and {Frederick}, S. and {Fremling}, C. and {Gal-Yam}, Avishay and {Gezari}, Suvi and {Giomi}, Matteo and {Goldstein}, Daniel A. and {Golkhou}, V. Zach and {Goobar}, Ariel and {Groom}, Steven and {Hacopians}, Eugean and {Hale}, David and {Henning}, John and {Ho}, Anna Y.~Q. and {Hover}, David and {Howell}, Justin and {Hung}, Tiara and {Huppenkothen}, Daniela and {Imel}, David and {Ip}, Wing-Huen and {Ivezi{\'c}}, {\v{Z}}eljko and {Jackson}, Edward and {Jones}, Lynne and {Juric}, Mario and {Kasliwal}, Mansi M. and {Kaspi}, S. and {Kaye}, Stephen and {Kelley}, Michael S.~P. and {Kowalski}, Marek and {Kramer}, Emily and {Kupfer}, Thomas and {Landry}, Walter and {Laher}, Russ R. and {Lee}, Chien-De and {Lin}, Hsing Wen and {Lin}, Zhong-Yi and {Lunnan}, Ragnhild and {Giomi}, Matteo and {Mahabal}, Ashish and {Mao}, Peter and {Miller}, Adam A. and {Monkewitz}, Serge and {Murphy}, Patrick and {Ngeow}, Chow-Choong and {Nordin}, Jakob and {Nugent}, Peter and {Ofek}, Eran and {Patterson}, Maria T. and {Penprase}, Bryan and {Porter}, Michael and {Rauch}, Ludwig and {Rebbapragada}, Umaa and {Reiley}, Dan and {Rigault}, Mickael and {Rodriguez}, Hector and {van Roestel}, Jan and {Rusholme}, Ben and {van Santen}, Jakob and {Schulze}, S. and {Shupe}, David L. and {Singer}, Leo P. and {Soumagnac}, Maayane T. and {Stein}, Robert and {Surace}, Jason and {Sollerman}, Jesper and {Szkody}, Paula and {Taddia}, F. and {Terek}, Scott and {Van Sistine}, Angela and {van Velzen}, Sjoert and {Vestrand}, W. Thomas and {Walters}, Richard and {Ward}, Charlotte and {Ye}, Quan-Zhi and {Yu}, Po-Chieh and {Yan}, Lin and {Zolkower}, Jeffry},
        title = "{The zwicky transient facility: system overview, performance, and first results}",
      journal = {\pasp},
         year = 2019,
        month = jan,
       volume = {131},
       number = {995},
        pages = {018002},
          doi = {10.1088/1538-3873/aaecbe},
archivePrefix = {arXiv},
       eprint = {1902.01932},
 primaryClass = {astro-ph.IM},
       adsurl = {https://ui.adsabs.harvard.edu/abs/2019PASP..131a8002B}
}

@ARTICLE{Fitzpatrick1999,
       author = {{Fitzpatrick}, Edward L.},
        title = "{Correcting for the Effects of Interstellar Extinction}",
      journal = {\pasp},
         year = 1999,
        month = jan,
       volume = {111},
       number = {755},
        pages = {63-75},
          doi = {10.1086/316293},
archivePrefix = {arXiv},
       eprint = {astro-ph/9809387},
 primaryClass = {astro-ph},
       adsurl = {https://ui.adsabs.harvard.edu/abs/1999PASP..111...63F}
}

@ARTICLE{Ivezic2019,
       author = {{Ivezi{\'c}}, {\v{Z}}eljko and {Kahn}, Steven M. and {Tyson}, J. Anthony and {Abel}, Bob and {Acosta}, Emily and {Allsman}, Robyn and {Alonso}, David and {AlSayyad}, Yusra and {Anderson}, Scott F. and {Andrew}, John and {Angel}, James Roger P. and {Angeli}, George Z. and {Ansari}, Reza and {Antilogus}, Pierre and {Araujo}, Constanza and {Armstrong}, Robert and {Arndt}, Kirk T. and {Astier}, Pierre and {Aubourg}, {\'E}ric and {Auza}, Nicole and {Axelrod}, Tim S. and {Bard}, Deborah J. and {Barr}, Jeff D. and {Barrau}, Aurelian and {Bartlett}, James G. and {Bauer}, Amanda E. and {Bauman}, Brian J. and {Baumont}, Sylvain and {Bechtol}, Ellen and {Bechtol}, Keith and {Becker}, Andrew C. and {Becla}, Jacek and {Beldica}, Cristina and {Bellavia}, Steve and {Bianco}, Federica B. and {Biswas}, Rahul and {Blanc}, Guillaume and {Blazek}, Jonathan and {Blandford}, Roger D. and {Bloom}, Josh S. and {Bogart}, Joanne and {Bond}, Tim W. and {Booth}, Michael T. and {Borgland}, Anders W. and {Borne}, Kirk and {Bosch}, James F. and {Boutigny}, Dominique and {Brackett}, Craig A. and {Bradshaw}, Andrew and {Brandt}, William Nielsen and {Brown}, Michael E. and {Bullock}, James S. and {Burchat}, Patricia and {Burke}, David L. and {Cagnoli}, Gianpietro and {Calabrese}, Daniel and {Callahan}, Shawn and {Callen}, Alice L. and {Carlin}, Jeffrey L. and {Carlson}, Erin L. and {Chandrasekharan}, Srinivasan and {Charles-Emerson}, Glenaver and {Chesley}, Steve and {Cheu}, Elliott C. and {Chiang}, Hsin-Fang and {Chiang}, James and {Chirino}, Carol and {Chow}, Derek and {Ciardi}, David R. and {Claver}, Charles F. and {Cohen-Tanugi}, Johann and {Cockrum}, Joseph J. and {Coles}, Rebecca and {Connolly}, Andrew J. and {Cook}, Kem H. and {Cooray}, Asantha and {Covey}, Kevin R. and {Cribbs}, Chris and {Cui}, Wei and {Cutri}, Roc and {Daly}, Philip N. and {Daniel}, Scott F. and {Daruich}, Felipe and {Daubard}, Guillaume and {Daues}, Greg and {Dawson}, William and {Delgado}, Francisco and {Dellapenna}, Alfred and {de Peyster}, Robert and {de Val-Borro}, Miguel and {Digel}, Seth W. and {Doherty}, Peter and {Dubois}, Richard and {Dubois-Felsmann}, Gregory P. and {Durech}, Josef and {Economou}, Frossie and {Eifler}, Tim and {Eracleous}, Michael and {Emmons}, Benjamin L. and {Fausti Neto}, Angelo and {Ferguson}, Henry and {Figueroa}, Enrique and {Fisher-Levine}, Merlin and {Focke}, Warren and {Foss}, Michael D. and {Frank}, James and {Freemon}, Michael D. and {Gangler}, Emmanuel and {Gawiser}, Eric and {Geary}, John C. and {Gee}, Perry and {Geha}, Marla and {Gessner}, Charles J.~B. and {Gibson}, Robert R. and {Gilmore}, D. Kirk and {Glanzman}, Thomas and {Glick}, William and {Goldina}, Tatiana and {Goldstein}, Daniel A. and {Goodenow}, Iain and {Graham}, Melissa L. and {Gressler}, William J. and {Gris}, Philippe and {Guy}, Leanne P. and {Guyonnet}, Augustin and {Haller}, Gunther and {Harris}, Ron and {Hascall}, Patrick A. and {Haupt}, Justine and {Hernandez}, Fabio and {Herrmann}, Sven and {Hileman}, Edward and {Hoblitt}, Joshua and {Hodgson}, John A. and {Hogan}, Craig and {Howard}, James D. and {Huang}, Dajun and {Huffer}, Michael E. and {Ingraham}, Patrick and {Innes}, Walter R. and {Jacoby}, Suzanne H. and {Jain}, Bhuvnesh and {Jammes}, Fabrice and {Jee}, M. James and {Jenness}, Tim and {Jernigan}, Garrett and {Jevremovi{\'c}}, Darko and {Johns}, Kenneth and {Johnson}, Anthony S. and {Johnson}, Margaret W.~G. and {Jones}, R. Lynne and {Juramy-Gilles}, Claire and {Juri{\'c}}, Mario and {Kalirai}, Jason S. and {Kallivayalil}, Nitya J. and {Kalmbach}, Bryce and {Kantor}, Jeffrey P. and {Karst}, Pierre and {Kasliwal}, Mansi M. and {Kelly}, Heather and {Kessler}, Richard and {Kinnison}, Veronica and {Kirkby}, David and {Knox}, Lloyd and {Kotov}, Ivan V. and {Krabbendam}, Victor L. and {Krughoff}, K. Simon and {Kub{\'a}nek}, Petr and {Kuczewski}, John and {Kulkarni}, Shri and {Ku}, John and {Kurita}, Nadine R. and {Lage}, Craig S. and {Lambert}, Ron and {Lange}, Travis and {Langton}, J. Brian and {Le Guillou}, Laurent and {Levine}, Deborah and {Liang}, Ming and {Lim}, Kian-Tat and {Lintott}, Chris J. and {Long}, Kevin E. and {Lopez}, Margaux and {Lotz}, Paul J. and {Lupton}, Robert H. and {Lust}, Nate B. and {MacArthur}, Lauren A. and {Mahabal}, Ashish and {Mandelbaum}, Rachel and {Markiewicz}, Thomas W. and {Marsh}, Darren S. and {Marshall}, Philip J. and {Marshall}, Stuart and {May}, Morgan and {McKercher}, Robert and {McQueen}, Michelle and {Meyers}, Joshua and {Migliore}, Myriam and {Miller}, Michelle and {Mills}, David J. and {Miraval}, Connor and {Moeyens}, Joachim and {Moolekamp}, Fred E. and {Monet}, David G. and {Moniez}, Marc and {Monkewitz}, Serge and {Montgomery}, Christopher and {Morrison}, Christopher B. and {Mueller}, Fritz and {Muller}, Gary P. and {Mu{\~n}oz Arancibia}, Freddy and {Neill}, Douglas R. and {Newbry}, Scott P. and {Nief}, Jean-Yves and {Nomerotski}, Andrei and {Nordby}, Martin and {O'Connor}, Paul and {Oliver}, John and {Olivier}, Scot S. and {Olsen}, Knut and {O'Mullane}, William and {Ortiz}, Sandra and {Osier}, Shawn and {Owen}, Russell E. and {Pain}, Reynald and {Palecek}, Paul E. and {Parejko}, John K. and {Parsons}, James B. and {Pease}, Nathan M. and {Peterson}, J. Matt and {Peterson}, John R. and {Petravick}, Donald L. and {Libby Petrick}, M.~E. and {Petry}, Cathy E. and {Pierfederici}, Francesco and {Pietrowicz}, Stephen and {Pike}, Rob and {Pinto}, Philip A. and {Plante}, Raymond and {Plate}, Stephen and {Plutchak}, Joel P. and {Price}, Paul A. and {Prouza}, Michael and {Radeka}, Veljko and {Rajagopal}, Jayadev and {Rasmussen}, Andrew P. and {Regnault}, Nicolas and {Reil}, Kevin A. and {Reiss}, David J. and {Reuter}, Michael A. and {Ridgway}, Stephen T. and {Riot}, Vincent J. and {Ritz}, Steve and {Robinson}, Sean and {Roby}, William and {Roodman}, Aaron and {Rosing}, Wayne and {Roucelle}, Cecille and {Rumore}, Matthew R. and {Russo}, Stefano and {Saha}, Abhijit and {Sassolas}, Benoit and {Schalk}, Terry L. and {Schellart}, Pim and {Schindler}, Rafe H. and {Schmidt}, Samuel and {Schneider}, Donald P. and {Schneider}, Michael D. and {Schoening}, William and {Schumacher}, German and {Schwamb}, Megan E. and {Sebag}, Jacques and {Selvy}, Brian and {Sembroski}, Glenn H. and {Seppala}, Lynn G. and {Serio}, Andrew and {Serrano}, Eduardo and {Shaw}, Richard A. and {Shipsey}, Ian and {Sick}, Jonathan and {Silvestri}, Nicole and {Slater}, Colin T. and {Smith}, J. Allyn and {Smith}, R. Chris and {Sobhani}, Shahram and {Soldahl}, Christine and {Storrie-Lombardi}, Lisa and {Stover}, Edward and {Strauss}, Michael A. and {Street}, Rachel A. and {Stubbs}, Christopher W. and {Sullivan}, Ian S. and {Sweeney}, Donald and {Swinbank}, John D. and {Szalay}, Alexander and {Takacs}, Peter and {Tether}, Stephen A. and {Thaler}, Jon J. and {Thayer}, John Gregg and {Thomas}, Sandrine and {Thornton}, Adam J. and {Thukral}, Vaikunth and {Tice}, Jeffrey and {Trilling}, David E. and {Turri}, Max and {Van Berg}, Richard and {Vanden Berk}, Daniel and {Vetter}, Kurt and {Virieux}, Francoise and {Vucina}, Tomislav and {Wahl}, William and {Walkowicz}, Lucianne and {Walsh}, Brian and {Walter}, Christopher W. and {Wang}, Daniel L. and {Wang}, Shin-Yawn and {Warner}, Michael and {Wiecha}, Oliver and {Willman}, Beth and {Winters}, Scott E. and {Wittman}, David and {Wolff}, Sidney C. and {Wood-Vasey}, W. Michael and {Wu}, Xiuqin and {Xin}, Bo and {Yoachim}, Peter and {Zhan}, Hu},
        title = "{LSST: from science drivers to reference design and anticipated data products}",
      journal = {\apj},
         year = 2019,
        month = mar,
       volume = {873},
       number = {2},
          eid = {111},
        pages = {111},
          doi = {10.3847/1538-4357/ab042c},
archivePrefix = {arXiv},
       eprint = {0805.2366},
 primaryClass = {astro-ph},
       adsurl = {https://ui.adsabs.harvard.edu/abs/2019ApJ...873..111I}
}

@ARTICLE{JiangNing2016,
       author = {{Jiang}, Ning and {Dou}, Liming and {Wang}, Tinggui and {Yang}, Chenwei and {Lyu}, Jianwei and {Zhou}, Hongyan},
        title = "{The WISE Detection of an Infrared Echo in Tidal Disruption Event ASASSN-14li}",
      journal = {\apjl},
         year = 2016,
        month = sep,
       volume = {828},
       number = {1},
          eid = {L14},
        pages = {L14},
          doi = {10.3847/2041-8205/828/1/L14},
archivePrefix = {arXiv},
       eprint = {1605.04640},
 primaryClass = {astro-ph.HE},
       adsurl = {https://ui.adsabs.harvard.edu/abs/2016ApJ...828L..14J}
}

@ARTICLE{Kankare2017,
       author = {{Kankare}, E. and {Kotak}, R. and {Mattila}, S. and {Lundqvist}, P. and {Ward}, M.~J. and {Fraser}, M. and {Lawrence}, A. and {Smartt}, S.~J. and {Meikle}, W.~P.~S. and {Bruce}, A. and {Harmanen}, J. and {Hutton}, S.~J. and {Inserra}, C. and {Kangas}, T. and {Pastorello}, A. and {Reynolds}, T. and {Romero-Ca{\~n}izales}, C. and {Smith}, K.~W. and {Valenti}, S. and {Chambers}, K.~C. and {Hodapp}, K.~W. and {Huber}, M.~E. and {Kaiser}, N. and {Kudritzki}, R. -P. and {Magnier}, E.~A. and {Tonry}, J.~L. and {Wainscoat}, R.~J. and {Waters}, C.},
        title = "{A population of highly energetic transient events in the centres of active galaxies}",
      journal = {\NatAs},
         year = 2017,
        month = nov,
       volume = {1},
        pages = {865-871},
          doi = {10.1038/s41550-017-0290-2},
archivePrefix = {arXiv},
       eprint = {1711.04577},
 primaryClass = {astro-ph.HE},
       adsurl = {https://ui.adsabs.harvard.edu/abs/2017NatAs...1..865K}
}

@ARTICLE{Komossa2008,
       author = {{Komossa}, S. and {Zhou}, H. and {Wang}, T. and {Ajello}, M. and {Ge}, J. and {Greiner}, J. and {Lu}, H. and {Salvato}, M. and {Saxton}, R. and {Shan}, H. and {Xu}, D. and {Yuan}, W.},
        title = "{Discovery of Superstrong, Fading, Iron Line Emission and Double-peaked Balmer Lines of the Galaxy SDSS J095209.56+214313.3: The Light Echo of a Huge Flare}",
      journal = {\apjl},
         year = 2008,
        month = may,
       volume = {678},
       number = {1},
        pages = {L13},
          doi = {10.1086/588281},
archivePrefix = {arXiv},
       eprint = {0804.2670},
 primaryClass = {astro-ph},
       adsurl = {https://ui.adsabs.harvard.edu/abs/2008ApJ...678L..13K}
}

@ARTICLE{Kormendy2013,
       author = {{Kormendy}, John and {Ho}, Luis C.},
        title = "{Coevolution (or not) of supermassive black holes and host galaxies}",
      journal = {\araa},
         year = 2013,
        month = aug,
       volume = {51},
       number = {1},
        pages = {511-653},
          doi = {10.1146/annurev-astro-082708-101811},
archivePrefix = {arXiv},
       eprint = {1304.7762},
 primaryClass = {astro-ph.CO},
       adsurl = {https://ui.adsabs.harvard.edu/abs/2013ARA&A..51..511K}
}

@ARTICLE{Oke1982,
       author = {{Oke}, J.~B. and {Gunn}, J.~E.},
        title = "{An Efficient Low Resolution and Moderate Resolution Spectrograph for the Hale Telescope}",
      journal = {\pasp},
         year = 1982,
        month = jun,
       volume = {94},
        pages = {586},
          doi = {10.1086/131027},
       adsurl = {https://ui.adsabs.harvard.edu/abs/1982PASP...94..586O}
}

@ARTICLE{Rees1988,
       author = {{Rees}, Martin J.},
        title = "{Tidal disruption of stars by black holes of {}10$^{6}$-{}10$^{8}$ solar masses in nearby galaxies}",
      journal = {\nat},
         year = 1988,
        month = jun,
       volume = {333},
       number = {6173},
        pages = {523-528},
          doi = {10.1038/333523a0},
       adsurl = {https://ui.adsabs.harvard.edu/abs/1988Natur.333..523R}
}

@ARTICLE{Trakhtenbrot2019NewClassFlare,
       author = {{Trakhtenbrot}, Benny and {Arcavi}, Iair and {Ricci}, Claudio and {Tacchella}, Sandro and {Stern}, Daniel and {Netzer}, Hagai and {Jonker}, Peter G. and {Horesh}, Assaf and {Mej{\'\i}a-Restrepo}, Juli{\'a}n Esteban and {Hosseinzadeh}, Griffin and {Hallefors}, Valentina and {Howell}, D. Andrew and {McCully}, Curtis and {Balokovi{\'c}}, Mislav and {Heida}, Marianne and {Kamraj}, Nikita and {Lansbury}, George Benjamin and {Wyrzykowski}, {\L}ukasz and {Gromadzki}, Mariusz and {Hamanowicz}, Aleksandra and {Cenko}, S. Bradley and {Sand}, David J. and {Hsiao}, Eric Y. and {Phillips}, Mark M. and {Diamond}, Tiara R. and {Kara}, Erin and {Gendreau}, Keith C. and {Arzoumanian}, Zaven and {Remillard}, Ron},
        title = "{A new class of flares from accreting supermassive black holes}",
      journal = {\NatAs},
         year = 2019,
        month = jan,
       volume = {3},
        pages = {242-250},
          doi = {10.1038/s41550-018-0661-3},
archivePrefix = {arXiv},
       eprint = {1901.03731},
 primaryClass = {astro-ph.GA},
       adsurl = {https://ui.adsabs.harvard.edu/abs/2019NatAs...3..242T}
}

@ARTICLE{vV2016,
       author = {{van Velzen}, S. and {Mendez}, A.~J. and {Krolik}, J.~H. and {Gorjian}, V.},
        title = "{Discovery of Transient Infrared Emission from Dust Heated by Stellar Tidal Disruption Flares}",
      journal = {\apj},
         year = 2016,
        month = sep,
       volume = {829},
       number = {1},
          eid = {19},
        pages = {19},
          doi = {10.3847/0004-637X/829/1/19},
archivePrefix = {arXiv},
       eprint = {1605.04304},
 primaryClass = {astro-ph.HE},
       adsurl = {https://ui.adsabs.harvard.edu/abs/2016ApJ...829...19V}
}

@ARTICLE{vV2020SSR,
       author = {{van Velzen}, Sjoert and {Holoien}, Thomas W. -S. and {Onori}, Francesca and {Hung}, Tiara and {Arcavi}, Iair},
        title = "{Optical-ultraviolet tidal disruption events}",
      journal = {\ssr},
         year = 2020,
        month = oct,
       volume = {216},
       number = {8},
          eid = {124},
        pages = {124},
          doi = {10.1007/s11214-020-00753-z},
archivePrefix = {arXiv},
       eprint = {2008.05461},
 primaryClass = {astro-ph.HE},
       adsurl = {https://ui.adsabs.harvard.edu/abs/2020SSRv..216..124V}
}

@ARTICLE{vV2021ApJ,
       author = {{van Velzen}, Sjoert and {Gezari}, Suvi and {Hammerstein}, Erica and {Roth}, Nathaniel and {Frederick}, Sara and {Ward}, Charlotte and {Hung}, Tiara and {Cenko}, S. Bradley and {Stein}, Robert and {Perley}, Daniel A. and {Taggart}, Kirsty and {Foley}, Ryan J. and {Sollerman}, Jesper and {Blagorodnova}, Nadejda and {Andreoni}, Igor and {Bellm}, Eric C. and {Brinnel}, Valery and {De}, Kishalay and {Dekany}, Richard and {Feeney}, Michael and {Fremling}, Christoffer and {Giomi}, Matteo and {Golkhou}, V. Zach and {Graham}, Matthew J. and {Ho}, Anna. Y.~Q. and {Kasliwal}, Mansi M. and {Kilpatrick}, Charles D. and {Kulkarni}, Shrinivas R. and {Kupfer}, Thomas and {Laher}, Russ R. and {Mahabal}, Ashish and {Masci}, Frank J. and {Miller}, Adam A. and {Nordin}, Jakob and {Riddle}, Reed and {Rusholme}, Ben and {van Santen}, Jakob and {Sharma}, Yashvi and {Shupe}, David L. and {Soumagnac}, Maayane T.},
        title = "{Seventeen Tidal Disruption Events from the First Half of ZTF Survey Observations: Entering a New Era of Population Studies}",
      journal = {\apj},
         year = 2021,
        month = feb,
       volume = {908},
       number = {1},
          eid = {4},
        pages = {4},
          doi = {10.3847/1538-4357/abc258},
archivePrefix = {arXiv},
       eprint = {2001.01409},
 primaryClass = {astro-ph.HE},
       adsurl = {https://ui.adsabs.harvard.edu/abs/2021ApJ...908....4V}
}

@ARTICLE{WangTinggui2012,
       author = {{Wang}, Ting-Gui and {Zhou}, Hong-Yan and {Komossa}, S. and {Wang}, Hui-Yuan and {Yuan}, Weimin and {Yang}, Chenwei},
        title = "{Extreme coronal line emitters: tidal disruption of stars by massive black holes in galactic nuclei?}",
      journal = {\apj},
         year = 2012,
        month = apr,
       volume = {749},
       number = {2},
          eid = {115},
        pages = {115},
          doi = {10.1088/0004-637X/749/2/115},
archivePrefix = {arXiv},
       eprint = {1202.1064},
 primaryClass = {astro-ph.HE},
       adsurl = {https://ui.adsabs.harvard.edu/abs/2012ApJ...749..115W}
}

@ARTICLE{Guolo2024,
       author = {{Guolo}, Muryel and {Gezari}, Suvi and {Yao}, Yu-Han and {van Velzen}, Sjoert and {Hammerstein}, Erica and {Cenko}, S. Bradley and {Tokayer}, Yarone M.},
        title = "{A systematic analysis of the x-ray emission in optically selected tidal disruption events: observational evidence for the unification of the optically and X-ray-selected populations}",
      journal = {\apj},
         year = 2024,
        month = may,
       volume = {966},
       number = {2},
          eid = {160},
        pages = {160},
          doi = {10.3847/1538-4357/ad2f9f},
archivePrefix = {arXiv},
       eprint = {2308.13019},
 primaryClass = {astro-ph.HE},
       adsurl = {https://ui.adsabs.harvard.edu/abs/2024ApJ...966..160G}
}

@ARTICLE{Short2023,
       author = {{Short}, P. and {Lawrence}, A. and {Nicholl}, M. and {Ward}, M. and {Reynolds}, T.~M. and {Mattila}, S. and {Yin}, C. and {Arcavi}, I. and {Carnall}, A. and {Charalampopoulos}, P. and {Gromadzki}, M. and {Jonker}, P.~G. and {Kim}, S. and {Leloudas}, G. and {Mandel}, I. and {Onori}, F. and {Pursiainen}, M. and {Schulze}, S. and {Villforth}, C. and {Wevers}, T.},
        title = "{Delayed appearance and evolution of coronal lines in the TDE AT2019qiz}",
      journal = {\mnras},
         year = 2023,
        month = oct,
       volume = {525},
       number = {1},
        pages = {1568-1587},
          doi = {10.1093/mnras/stad2270},
archivePrefix = {arXiv},
       eprint = {2307.13674},
 primaryClass = {astro-ph.HE},
       adsurl = {https://ui.adsabs.harvard.edu/abs/2023MNRAS.525.1568S}
}

@ARTICLE{WangTinggui2023,
       author = {{Wang}, Ting-Gui and {Liu}, Gui-Lin and {Cai}, Zhen-Yi and {Geng}, Jin-Jun and {Fang}, Min and {He}, Haoning and {Jiang}, Ji-an and {Jiang}, Ning and {Kong}, Xu and {Li}, Bin and {Li}, Ye and {Luo}, Wentao and {Pan}, Zhizheng and {Wu}, Xuefeng and {Yang}, Ji and {Yu}, Jiming and {Zheng}, Xianzhong and {Zhu}, Qingfeng and {Cai}, Yi-Fu and {Chen}, Yuanyuan and {Chen}, Zhiwei and {Dai}, Zigao and {Fan}, Lulu and {Fan}, Yizhong and {Fang}, Wenjuan and {He}, Zhicheng and {Hu}, Lei and {Hu}, Maokai and {Jin}, Zhiping and {Jiang}, Zhibo and {Li}, Guoliang and {Li}, Fan and {Li}, Xuzhi and {Liang}, Runduo and {Lin}, Zheyu and {Liu}, Qingzhong and {Liu}, Wenhao and {Liu}, Zhengyan and {Liu}, Wei and {Liu}, Yao and {Lou}, Zheng and {Qu}, Han and {Sheng}, Zhenfeng and {Shi}, Jianchun and {Shu}, Yiping and {Su}, Zhenbo and {Sun}, Tianrui and {Wang}, Hongchi and {Wang}, Huiyuan and {Wang}, Jian and {Wang}, Junxian and {Wei}, Daming and {Wei}, Junjie and {Xue}, Yongquan and {Yan}, Jingzhi and {Yang}, Chao and {Yuan}, Ye and {Yuan}, Yefei and {Zhang}, Hongxin and {Zhang}, Miaomiao and {Zhao}, Haibin and {Zhao}, Wen},
        title = "{Science with the 2.5-meter wide field survey telescope (WFST)}",
      journal = {\SCPMA},
         year = 2023,
        month = oct,
       volume = {66},
       number = {10},
          eid = {109512},
        pages = {109512},
          doi = {10.1007/s11433-023-2197-5},
archivePrefix = {arXiv},
       eprint = {2306.07590},
 primaryClass = {astro-ph.IM},
       adsurl = {https://ui.adsabs.harvard.edu/abs/2023SCPMA..6609512W}
}

@ARTICLE{Yao2023,
       author = {{Yao}, Yu-Han and {Ravi}, Vikram and {Gezari}, Suvi and {van Velzen}, Sjoert and {Lu}, Wenbin and {Schulze}, Steve and {Somalwar}, Jean J. and {Kulkarni}, S.~R. and {Hammerstein}, Erica and {Nicholl}, Matt and {Graham}, Matthew J. and {Perley}, Daniel A. and {Cenko}, S. Bradley and {Stein}, Robert and {Ricarte}, Angelo and {Chadayammuri}, Urmila and {Quataert}, Eliot and {Bellm}, Eric C. and {Bloom}, Joshua S. and {Dekany}, Richard and {Drake}, Andrew J. and {Groom}, Steven L. and {Mahabal}, Ashish A. and {Prince}, Thomas A. and {Riddle}, Reed and {Rusholme}, Ben and {Sharma}, Yashvi and {Sollerman}, Jesper and {Yan}, Lin},
        title = "{Tidal disruption event demographics with the zwicky transient facility: volumetric rates, luminosity function, and implications for the local black hole mass function}",
      journal = {\apjl},
         year = 2023,
        month = sep,
       volume = {955},
       number = {1},
          eid = {L6},
        pages = {L6},
          doi = {10.3847/2041-8213/acf216},
archivePrefix = {arXiv},
       eprint = {2303.06523},
 primaryClass = {astro-ph.HE},
       adsurl = {https://ui.adsabs.harvard.edu/abs/2023ApJ...955L...6Y}
}

@ARTICLE{Hammerstein2023,
       author = {{Hammerstein}, Erica and {van Velzen}, Sjoert and {Gezari}, Suvi and {Cenko}, S. Bradley and {Yao}, Yuhan and {Ward}, Charlotte and {Frederick}, Sara and {Villanueva}, Natalia and {Somalwar}, Jean J. and {Graham}, Matthew J. and {Kulkarni}, Shrinivas R. and {Stern}, Daniel and {Andreoni}, Igor and {Bellm}, Eric C. and {Dekany}, Richard and {Dhawan}, Suhail and {Drake}, Andrew J. and {Fremling}, Christoffer and {Gatkine}, Pradip and {Groom}, Steven L. and {Ho}, Anna Y.~Q. and {Kasliwal}, Mansi M. and {Karambelkar}, Viraj and {Kool}, Erik C. and {Masci}, Frank J. and {Medford}, Michael S. and {Perley}, Daniel A. and {Purdum}, Josiah and {van Roestel}, Jan and {Sharma}, Yashvi and {Sollerman}, Jesper and {Taggart}, Kirsty and {Yan}, Lin},
        title = "{The final season reimagined: 30 tidal disruption events from the ZTF-I survey}",
      journal = {\apj},
         year = 2023,
        month = jan,
       volume = {942},
       number = {1},
          eid = {9},
        pages = {9},
          doi = {10.3847/1538-4357/aca283},
archivePrefix = {arXiv},
       eprint = {2203.01461},
 primaryClass = {astro-ph.HE},
       adsurl = {https://ui.adsabs.harvard.edu/abs/2023ApJ...942....9H}
}

@ARTICLE{Petrushevska2023,
       author = {{Petrushevska}, T. and {Leloudas}, G. and {Ili{\'c}}, D. and {Bronikowski}, M. and {Charalampopoulos}, P. and {Jaisawal}, G.~K. and {Paraskeva}, E. and {Pursiainen}, M. and {Raki{\'c}}, N. and {Schulze}, S. and {Taggart}, K. and {Wedderkopp}, C.~K. and {Anderson}, J.~P. and {de Boer}, T. and {Chambers}, K. and {Chen}, T.~W. and {Damljanovi{\'c}}, G. and {Fraser}, M. and {Gao}, H. and {Gomboc}, A. and {Gromadzki}, M. and {Ihanec}, N. and {Maguire}, K. and {Mar{\v{c}}un}, B. and {M{\"u}ller-Bravo}, T.~E. and {Nicholl}, M. and {Onori}, F. and {Reynolds}, T.~M. and {Smartt}, S.~J. and {Sollerman}, J. and {Smith}, K.~W. and {Wevers}, T. and {Wyrzykowski}, {\L}.},
        title = "{The rise and fall of the iron-strong nuclear transient PS16dtm}",
      journal = {\aap},
         year = 2023,
        month = jan,
       volume = {669},
          eid = {A140},
        pages = {A140},
          doi = {10.1051/0004-6361/202244623},
archivePrefix = {arXiv},
       eprint = {2211.17097},
 primaryClass = {astro-ph.HE},
       adsurl = {https://ui.adsabs.harvard.edu/abs/2023A&A...669A.140P}
}

@ARTICLE{Onori2022,
       author = {{Onori}, F. and {Cannizzaro}, G. and {Jonker}, P.~G. and {Kim}, M. and {Nicholl}, M. and {Mattila}, S. and {Reynolds}, T.~M. and {Fraser}, M. and {Wevers}, T. and {Brocato}, E. and {Anderson}, J.~P. and {Carini}, R. and {Charalampopoulos}, P. and {Clark}, P. and {Gromadzki}, M. and {Guti{\'e}rrez}, C.~P. and {Ihanec}, N. and {Inserra}, C. and {Lawrence}, A. and {Leloudas}, G. and {Lundqvist}, P. and {M{\"u}ller-Bravo}, T.~E. and {Piranomonte}, S. and {Pursiainen}, M. and {Rybicki}, K.~A. and {Somero}, A. and {Young}, D.~R. and {Chambers}, K.~C. and {Gao}, H. and {de Boer}, T.~J.~L. and {Magnier}, E.~A.},
        title = "{The nuclear transient AT 2017gge: a tidal disruption event in a dusty and gas-rich environment and the awakening of a dormant SMBH}",
      journal = {\mnras},
         year = 2022,
        month = nov,
       volume = {517},
       number = {1},
        pages = {76-98},
          doi = {10.1093/mnras/stac2673},
archivePrefix = {arXiv},
       eprint = {2206.00049},
 primaryClass = {astro-ph.HE},
       adsurl = {https://ui.adsabs.harvard.edu/abs/2022MNRAS.517...76O}
}

@ARTICLE{ATLAS17jrp,
       author = {{Wang}, Yibo and {Jiang}, Ning and {Wang}, Tinggui and {Zhu}, Jiazheng and {Dou}, Liming and {Lin}, Zheyu and {Sun}, Luming and {Liu}, Hui and {Sheng}, Zhenfeng},
        title = "{Discovery of ATLAS17jrp as an Optical-, X-Ray-, and Infrared-bright Tidal Disruption Event in a Star-forming Galaxy}",
      journal = {\apjl},
         year = 2022,
        month = may,
       volume = {930},
       number = {1},
          eid = {L4},
        pages = {L4},
          doi = {10.3847/2041-8213/ac6670},
archivePrefix = {arXiv},
       eprint = {2204.05461},
 primaryClass = {astro-ph.GA},
       adsurl = {https://ui.adsabs.harvard.edu/abs/2022ApJ...930L...4W}
}

@ARTICLE{Charalampopoulos2022,
       author = {{Charalampopoulos}, P. and {Leloudas}, G. and {Malesani}, D.~B. and {Wevers}, T. and {Arcavi}, I. and {Nicholl}, M. and {Pursiainen}, M. and {Lawrence}, A. and {Anderson}, J.~P. and {Benetti}, S. and {Cannizzaro}, G. and {Chen}, T. -W. and {Galbany}, L. and {Gromadzki}, M. and {Guti{\'e}rrez}, C.~P. and {Inserra}, C. and {Jonker}, P.~G. and {M{\"u}ller-Bravo}, T.~E. and {Onori}, F. and {Short}, P. and {Sollerman}, J. and {Young}, D.~R.},
        title = "{A detailed spectroscopic study of tidal disruption events}",
      journal = {\aap},
         year = 2022,
        month = mar,
       volume = {659},
          eid = {A34},
        pages = {A34},
          doi = {10.1051/0004-6361/202142122},
archivePrefix = {arXiv},
       eprint = {2109.00016},
 primaryClass = {astro-ph.HE},
       adsurl = {https://ui.adsabs.harvard.edu/abs/2022A&A...659A..34C}
}

@ARTICLE{Frederick2021,
       author = {{Frederick}, Sara and {Gezari}, Suvi and {Graham}, Matthew J. and {Sollerman}, Jesper and {van Velzen}, Sjoert and {Perley}, Daniel A. and {Stern}, Daniel and {Ward}, Charlotte and {Hammerstein}, Erica and {Hung}, Tiara and {Yan}, Lin and {Andreoni}, Igor and {Bellm}, Eric C. and {Duev}, Dmitry A. and {Kowalski}, Marek and {Mahabal}, Ashish A. and {Masci}, Frank J. and {Medford}, Michael and {Rusholme}, Ben and {Smith}, Roger and {Walters}, Richard},
        title = "{A Family Tree of Optical Transients from Narrow-line Seyfert 1 Galaxies}",
      journal = {\apj},
         year = 2021,
        month = oct,
       volume = {920},
       number = {1},
          eid = {56},
        pages = {56},
          doi = {10.3847/1538-4357/ac110f},
archivePrefix = {arXiv},
       eprint = {2010.08554},
 primaryClass = {astro-ph.HE},
       adsurl = {https://ui.adsabs.harvard.edu/abs/2021ApJ...920...56F}
}

@ARTICLE{Gezari2021,
       author = {{Gezari}, Suvi},
        title = "{Tidal disruption events}",
      journal = {\araa},
         year = 2021,
        month = sep,
       volume = {59},
        pages = {21-58},
          doi = {10.1146/annurev-astro-111720-030029},
archivePrefix = {arXiv},
       eprint = {2104.14580},
 primaryClass = {astro-ph.HE},
       adsurl = {https://ui.adsabs.harvard.edu/abs/2021ARA&A..59...21G}
}

@ARTICLE{MIRONGI,
       author = {{Jiang}, Ning and {Wang}, Tinggui and {Dou}, Liming and {Shu}, Xinwen and {Hu}, Xueyang and {Liu}, Hui and {Wang}, Yibo and {Yan}, Lin and {Sheng}, Zhenfeng and {Yang}, Chenwei and {Sun}, Luming and {Zhou}, Hongyan},
        title = "{Mid-infrared Outbursts in Nearby Galaxies (MIRONG). I. Sample Selection and Characterization}",
      journal = {\apjs},
         year = 2021,
        month = feb,
       volume = {252},
       number = {2},
          eid = {32},
        pages = {32},
          doi = {10.3847/1538-4365/abd1dc},
archivePrefix = {arXiv},
       eprint = {2012.06806},
 primaryClass = {astro-ph.GA},
       adsurl = {https://ui.adsabs.harvard.edu/abs/2021ApJS..252...32J}
}

@ARTICLE{Roth2020,
       author = {{Roth}, Nathaniel and {Rossi}, Elena Maria and {Krolik}, Julian and {Piran}, Tsvi and {Mockler}, Brenna and {Kasen}, Daniel},
        title = "{Radiative emission mechanisms}",
      journal = {\ssr},
         year = 2020,
        month = oct,
       volume = {216},
       number = {7},
          eid = {114},
        pages = {114},
          doi = {10.1007/s11214-020-00735-1},
archivePrefix = {arXiv},
       eprint = {2008.01117},
 primaryClass = {astro-ph.HE},
       adsurl = {https://ui.adsabs.harvard.edu/abs/2020SSRv..216..114R}
}

@ARTICLE{Leloudas2019,
       author = {{Leloudas}, Giorgos and {Dai}, Lixin and {Arcavi}, Iair and {Vreeswijk}, Paul M. and {Mockler}, Brenna and {Roy}, Rupak and {Malesani}, Daniele B. and {Schulze}, Steve and {Wevers}, Thomas and {Fraser}, Morgan and {Ramirez-Ruiz}, Enrico and {Auchettl}, Katie and {Burke}, Jamison and {Cannizzaro}, Giacomo and {Charalampopoulos}, Panos and {Chen}, Ting-Wan and {Cikota}, Aleksandar and {Della Valle}, Massimo and {Galbany}, Lluis and {Gromadzki}, Mariusz and {Heintz}, Kasper E. and {Hiramatsu}, Daichi and {Jonker}, Peter G. and {Kostrzewa-Rutkowska}, Zuzanna and {Maguire}, Kate and {Mandel}, Ilya and {Nicholl}, Matt and {Onori}, Francesca and {Roth}, Nathaniel and {Smartt}, Stephen J. and {Wyrzykowski}, Lukasz and {Young}, Dave R.},
        title = "{The Spectral Evolution of AT 2018dyb and the Presence of Metal Lines in Tidal Disruption Events}",
      journal = {\apj},
         year = 2019,
        month = dec,
       volume = {887},
       number = {2},
          eid = {218},
        pages = {218},
          doi = {10.3847/1538-4357/ab5792},
archivePrefix = {arXiv},
       eprint = {1903.03120},
 primaryClass = {astro-ph.HE},
       adsurl = {https://ui.adsabs.harvard.edu/abs/2019ApJ...887..218L}
}

@ARTICLE{JiangNing2019,
       author = {{Jiang}, Ning and {Wang}, Tinggui and {Mou}, Guobin and {Liu}, Hui and {Dou}, Liming and {Sheng}, Zhenfeng and {Wang}, Yibo},
        title = "{Infrared Echo and Late-stage Rebrightening of Nuclear Transient Ps1-10adi: Exploring the Torus with Tidal Disruption Events in Active Galactic Nuclei}",
      journal = {\apj},
         year = 2019,
        month = jan,
       volume = {871},
       number = {1},
          eid = {15},
        pages = {15},
          doi = {10.3847/1538-4357/aaf6b2},
archivePrefix = {arXiv},
       eprint = {1812.01295},
 primaryClass = {astro-ph.GA},
       adsurl = {https://ui.adsabs.harvard.edu/abs/2019ApJ...871...15J}
}

@ARTICLE{Masci2019,
       author = {{Masci}, Frank J. and {Laher}, Russ R. and {Rusholme}, Ben and {Shupe}, David L. and {Groom}, Steven and {Surace}, Jason and {Jackson}, Edward and {Monkewitz}, Serge and {Beck}, Ron and {Flynn}, David and {Terek}, Scott and {Landry}, Walter and {Hacopians}, Eugean and {Desai}, Vandana and {Howell}, Justin and {Brooke}, Tim and {Imel}, David and {Wachter}, Stefanie and {Ye}, Quan-Zhi and {Lin}, Hsing-Wen and {Cenko}, S. Bradley and {Cunningham}, Virginia and {Rebbapragada}, Umaa and {Bue}, Brian and {Miller}, Adam A. and {Mahabal}, Ashish and {Bellm}, Eric C. and {Patterson}, Maria T. and {Juri{\'c}}, Mario and {Golkhou}, V. Zach and {Ofek}, Eran O. and {Walters}, Richard and {Graham}, Matthew and {Kasliwal}, Mansi M. and {Dekany}, Richard G. and {Kupfer}, Thomas and {Burdge}, Kevin and {Cannella}, Christopher B. and {Barlow}, Tom and {Van Sistine}, Angela and {Giomi}, Matteo and {Fremling}, Christoffer and {Blagorodnova}, Nadejda and {Levitan}, David and {Riddle}, Reed and {Smith}, Roger M. and {Helou}, George and {Prince}, Thomas A. and {Kulkarni}, Shrinivas R.},
        title = "{The Zwicky Transient Facility: Data Processing, Products, and Archive}",
      journal = {\pasp},
         year = 2019,
        month = jan,
       volume = {131},
       number = {995},
        pages = {018003},
          doi = {10.1088/1538-3873/aae8ac},
archivePrefix = {arXiv},
       eprint = {1902.01872},
 primaryClass = {astro-ph.IM},
       adsurl = {https://ui.adsabs.harvard.edu/abs/2019PASP..131a8003M}
}

@ARTICLE{LuWenbin2018,
       author = {{Lu}, Wen-Bin and {Kumar}, Pawan},
        title = "{On the missing energy puzzle of tidal disruption events}",
      journal = {\apj},
         year = 2018,
        month = oct,
       volume = {865},
       number = {2},
          eid = {128},
        pages = {128},
          doi = {10.3847/1538-4357/aad54a},
archivePrefix = {arXiv},
       eprint = {1802.02151},
 primaryClass = {astro-ph.HE},
       adsurl = {https://ui.adsabs.harvard.edu/abs/2018ApJ...865..128L}
}

@ARTICLE{Tonry2018,
       author = {{Tonry}, J.~L. and {Denneau}, L. and {Heinze}, A.~N. and {Stalder}, B. and {Smith}, K.~W. and {Smartt}, S.~J. and {Stubbs}, C.~W. and {Weiland}, H.~J. and {Rest}, A.},
        title = "{ATLAS: a high-cadence all-sky survey system}",
      journal = {\pasp},
         year = 2018,
        month = jun,
       volume = {130},
       number = {988},
        pages = {064505},
          doi = {10.1088/1538-3873/aabadf},
archivePrefix = {arXiv},
       eprint = {1802.00879},
 primaryClass = {astro-ph.IM},
       adsurl = {https://ui.adsabs.harvard.edu/abs/2018PASP..130f4505T}
}

@ARTICLE{Blanchard2017,
       author = {{Blanchard}, P.~K. and {Nicholl}, M. and {Berger}, E. and {Guillochon}, J. and {Margutti}, R. and {Chornock}, R. and {Alexander}, K.~D. and {Leja}, J. and {Drout}, M.~R.},
        title = "{PS16dtm: a tidal disruption event in a narrow-line seyfert 1 galaxy}",
      journal = {\apj},
         year = 2017,
        month = jul,
       volume = {843},
       number = {2},
          eid = {106},
        pages = {106},
          doi = {10.3847/1538-4357/aa77f7},
archivePrefix = {arXiv},
       eprint = {1703.07816},
 primaryClass = {astro-ph.HE},
       adsurl = {https://ui.adsabs.harvard.edu/abs/2017ApJ...843..106B}
}

@ARTICLE{Tadhunter2017,
       author = {{Tadhunter}, C. and {Spence}, R. and {Rose}, M. and {Mullaney}, J. and {Crowther}, P.},
        title = "{A tidal disruption event in the nearby ultra-luminous infrared galaxy F01004-2237}",
      journal = {\NatAs},
         year = 2017,
        month = mar,
       volume = {1},
          eid = {0061},
        pages = {0061},
          doi = {10.1038/s41550-017-0061},
archivePrefix = {arXiv},
       eprint = {1702.02573},
 primaryClass = {astro-ph.GA},
       adsurl = {https://ui.adsabs.harvard.edu/abs/2017NatAs...1E..61T}
}

@ARTICLE{Chambers2016,
       author = {{Chambers}, K.~C. and {Magnier}, E.~A. and {Metcalfe}, N. and {Flewelling}, H.~A. and {Huber}, M.~E. and {Waters}, C.~Z. and {Denneau}, L. and {Draper}, P.~W. and {Farrow}, D. and {Finkbeiner}, D.~P. and {Holmberg}, C. and {Koppenhoefer}, J. and {Price}, P.~A. and {Rest}, A. and {Saglia}, R.~P. and {Schlafly}, E.~F. and {Smartt}, S.~J. and {Sweeney}, W. and {Wainscoat}, R.~J. and {Burgett}, W.~S. and {Chastel}, S. and {Grav}, T. and {Heasley}, J.~N. and {Hodapp}, K.~W. and {Jedicke}, R. and {Kaiser}, N. and {Kudritzki}, R. -P. and {Luppino}, G.~A. and {Lupton}, R.~H. and {Monet}, D.~G. and {Morgan}, J.~S. and {Onaka}, P.~M. and {Shiao}, B. and {Stubbs}, C.~W. and {Tonry}, J.~L. and {White}, R. and {Ba{\~n}ados}, E. and {Bell}, E.~F. and {Bender}, R. and {Bernard}, E.~J. and {Boegner}, M. and {Boffi}, F. and {Botticella}, M.~T. and {Calamida}, A. and {Casertano}, S. and {Chen}, W. -P. and {Chen}, X. and {Cole}, S. and {Deacon}, N. and {Frenk}, C. and {Fitzsimmons}, A. and {Gezari}, S. and {Gibbs}, V. and {Goessl}, C. and {Goggia}, T. and {Gourgue}, R. and {Goldman}, B. and {Grant}, P. and {Grebel}, E.~K. and {Hambly}, N.~C. and {Hasinger}, G. and {Heavens}, A.~F. and {Heckman}, T.~M. and {Henderson}, R. and {Henning}, T. and {Holman}, M. and {Hopp}, U. and {Ip}, W. -H. and {Isani}, S. and {Jackson}, M. and {Keyes}, C.~D. and {Koekemoer}, A.~M. and {Kotak}, R. and {Le}, D. and {Liska}, D. and {Long}, K.~S. and {Lucey}, J.~R. and {Liu}, M. and {Martin}, N.~F. and {Masci}, G. and {McLean}, B. and {Mindel}, E. and {Misra}, P. and {Morganson}, E. and {Murphy}, D.~N.~A. and {Obaika}, A. and {Narayan}, G. and {Nieto-Santisteban}, M.~A. and {Norberg}, P. and {Peacock}, J.~A. and {Pier}, E.~A. and {Postman}, M. and {Primak}, N. and {Rae}, C. and {Rai}, A. and {Riess}, A. and {Riffeser}, A. and {Rix}, H.~W. and {R{\"o}ser}, S. and {Russel}, R. and {Rutz}, L. and {Schilbach}, E. and {Schultz}, A.~S.~B. and {Scolnic}, D. and {Strolger}, L. and {Szalay}, A. and {Seitz}, S. and {Small}, E. and {Smith}, K.~W. and {Soderblom}, D.~R. and {Taylor}, P. and {Thomson}, R. and {Taylor}, A.~N. and {Thakar}, A.~R. and {Thiel}, J. and {Thilker}, D. and {Unger}, D. and {Urata}, Y. and {Valenti}, J. and {Wagner}, J. and {Walder}, T. and {Walter}, F. and {Watters}, S.~P. and {Werner}, S. and {Wood-Vasey}, W.~M. and {Wyse}, R.},
        title = "{The Pan-STARRS1 Surveys}",
      journal = {arXiv e-prints},
         year = 2016,
        month = dec,
          eid = {arXiv:1612.05560},
          doi = {10.48550/arXiv.1612.05560},
archivePrefix = {arXiv},
       eprint = {1612.05560},
 primaryClass = {astro-ph.IM},
       adsurl = {https://ui.adsabs.harvard.edu/abs/2016arXiv161205560C}
}

@ARTICLE{HI4PI2016,
       author = {{HI4PI Collaboration} and {Ben Bekhti}, N. and {Fl{\"o}er}, L. and {Keller}, R. and {Kerp}, J. and {Lenz}, D. and {Winkel}, B. and {Bailin}, J. and {Calabretta}, M.~R. and {Dedes}, L. and {Ford}, H.~A. and {Gibson}, B.~K. and {Haud}, U. and {Janowiecki}, S. and {Kalberla}, P.~M.~W. and {Lockman}, F.~J. and {McClure-Griffiths}, N.~M. and {Murphy}, T. and {Nakanishi}, H. and {Pisano}, D.~J. and {Staveley-Smith}, L.},
        title = "{HI4PI: A full-sky H I survey based on EBHIS and GASS}",
      journal = {\aap},
         year = 2016,
        month = oct,
       volume = {594},
          eid = {A116},
        pages = {A116},
          doi = {10.1051/0004-6361/201629178},
archivePrefix = {arXiv},
       eprint = {1610.06175},
 primaryClass = {astro-ph.GA},
       adsurl = {https://ui.adsabs.harvard.edu/abs/2016A&A...594A.116H}
}

@ARTICLE{Reines2015,
       author = {{Reines}, Amy E. and {Volonteri}, Marta},
        title = "{Relations between central black hole mass and total galaxy stellar mass in the local universe}",
      journal = {\apj},
         year = 2015,
        month = nov,
       volume = {813},
       number = {2},
          eid = {82},
        pages = {82},
          doi = {10.1088/0004-637X/813/2/82},
archivePrefix = {arXiv},
       eprint = {1508.06274},
 primaryClass = {astro-ph.GA},
       adsurl = {https://ui.adsabs.harvard.edu/abs/2015ApJ...813...82R}
}

@ARTICLE{Piran2015,
       author = {{Piran}, Tsvi and {Svirski}, Gilad and {Krolik}, Julian and {Cheng}, Roseanne M. and {Shiokawa}, Hotaka},
        title = "{Disk formation versus disk accretion{\textemdash}what powers tidal disruption events?}",
      journal = {\apj},
         year = 2015,
        month = jun,
       volume = {806},
       number = {2},
          eid = {164},
        pages = {164},
          doi = {10.1088/0004-637X/806/2/164},
archivePrefix = {arXiv},
       eprint = {1502.05792},
 primaryClass = {astro-ph.HE},
       adsurl = {https://ui.adsabs.harvard.edu/abs/2015ApJ...806..164P}
}

@ARTICLE{Mainzer2014,
       author = {{Mainzer}, A. and {Bauer}, J. and {Cutri}, R.~M. and {Grav}, T. and {Masiero}, J. and {Beck}, R. and {Clarkson}, P. and {Conrow}, T. and {Dailey}, J. and {Eisenhardt}, P. and {Fabinsky}, B. and {Fajardo-Acosta}, S. and {Fowler}, J. and {Gelino}, C. and {Grillmair}, C. and {Heinrichsen}, I. and {Kendall}, M. and {Kirkpatrick}, J. Davy and {Liu}, F. and {Masci}, F. and {McCallon}, H. and {Nugent}, C.~R. and {Papin}, M. and {Rice}, E. and {Royer}, D. and {Ryan}, T. and {Sevilla}, P. and {Sonnett}, S. and {Stevenson}, R. and {Thompson}, D.~B. and {Wheelock}, S. and {Wiemer}, D. and {Wittman}, M. and {Wright}, E. and {Yan}, L.},
        title = "{Initial Performance of the NEOWISE Reactivation Mission}",
      journal = {\apj},
         year = 2014,
        month = sep,
       volume = {792},
       number = {1},
          eid = {30},
        pages = {30},
          doi = {10.1088/0004-637X/792/1/30},
archivePrefix = {arXiv},
       eprint = {1406.6025},
 primaryClass = {astro-ph.EP},
       adsurl = {https://ui.adsabs.harvard.edu/abs/2014ApJ...792...30M}
}

@ARTICLE{Yan2013,
       author = {{Yan}, Lin and {Donoso}, E. and {Tsai}, Chao-Wei and {Stern}, D. and {Assef}, R.~J. and {Eisenhardt}, P. and {Blain}, A.~W. and {Cutri}, R. and {Jarrett}, T. and {Stanford}, S.~A. and {Wright}, E. and {Bridge}, C. and {Riechers}, D.~A.},
        title = "{Characterizing the mid-infrared extragalactic sky with WISE and SDSS}",
      journal = {\aj},
         year = 2013,
        month = mar,
       volume = {145},
       number = {3},
          eid = {55},
        pages = {55},
          doi = {10.1088/0004-6256/145/3/55},
archivePrefix = {arXiv},
       eprint = {1209.2065},
 primaryClass = {astro-ph.CO},
       adsurl = {https://ui.adsabs.harvard.edu/abs/2013AJ....145...55Y}
}

@ARTICLE{Silverman2012,
       author = {{Silverman}, Jeffrey M. and {Foley}, Ryan J. and {Filippenko}, Alexei V. and {Ganeshalingam}, Mohan and {Barth}, Aaron J. and {Chornock}, Ryan and {Griffith}, Christopher V. and {Kong}, Jason J. and {Lee}, Nicholas and {Leonard}, Douglas C. and {Matheson}, Thomas and {Miller}, Emily G. and {Steele}, Thea N. and {Barris}, Brian J. and {Bloom}, Joshua S. and {Cobb}, Bethany E. and {Coil}, Alison L. and {Desroches}, Louis-Benoit and {Gates}, Elinor L. and {Ho}, Luis C. and {Jha}, Saurabh W. and {Kandrashoff}, Michael T. and {Li}, Weidong and {Mandel}, Kaisey S. and {Modjaz}, Maryam and {Moore}, Matthew R. and {Mostardi}, Robin E. and {Papenkova}, Marina S. and {Park}, Sung and {Perley}, Daniel A. and {Poznanski}, Dovi and {Reuter}, Cassie A. and {Scala}, James and {Serduke}, Franklin J.~D. and {Shields}, Joseph C. and {Swift}, Brandon J. and {Tonry}, John L. and {Van Dyk}, Schuyler D. and {Wang}, Xiaofeng and {Wong}, Diane S.},
        title = "{Berkeley Supernova Ia Program - I. Observations, data reduction and spectroscopic sample of 582 low-redshift Type Ia supernovae}",
      journal = {\mnras},
         year = 2012,
        month = sep,
       volume = {425},
       number = {3},
        pages = {1789-1818},
          doi = {10.1111/j.1365-2966.2012.21270.x},
archivePrefix = {arXiv},
       eprint = {1202.2128},
 primaryClass = {astro-ph.CO},
       adsurl = {https://ui.adsabs.harvard.edu/abs/2012MNRAS.425.1789S}
}

@ARTICLE{Stern2012,
       author = {{Stern}, Daniel and {Assef}, Roberto J. and {Benford}, Dominic J. and {Blain}, Andrew and {Cutri}, Roc and {Dey}, Arjun and {Eisenhardt}, Peter and {Griffith}, Roger L. and {Jarrett}, T.~H. and {Lake}, Sean and {Masci}, Frank and {Petty}, Sara and {Stanford}, S.~A. and {Tsai}, Chao-Wei and {Wright}, E.~L. and {Yan}, Lin and {Harrison}, Fiona and {Madsen}, Kristin},
        title = "{Mid-infrared selection of active galactic nuclei with the wide-field infrared survey explorer. I. characterizing WISE-selected active galactic nuclei in COSMOS}",
      journal = {\apj},
         year = 2012,
        month = jul,
       volume = {753},
       number = {1},
          eid = {30},
        pages = {30},
          doi = {10.1088/0004-637X/753/1/30},
archivePrefix = {arXiv},
       eprint = {1205.0811},
 primaryClass = {astro-ph.CO},
       adsurl = {https://ui.adsabs.harvard.edu/abs/2012ApJ...753...30S}
}

@ARTICLE{Schlafly2011,
       author = {{Schlafly}, Edward F. and {Finkbeiner}, Douglas P.},
        title = "{Measuring Reddening with Sloan Digital Sky Survey Stellar Spectra and Recalibrating SFD}",
      journal = {\apj},
         year = 2011,
        month = aug,
       volume = {737},
       number = {2},
          eid = {103},
        pages = {103},
          doi = {10.1088/0004-637X/737/2/103},
archivePrefix = {arXiv},
       eprint = {1012.4804},
 primaryClass = {astro-ph.GA},
       adsurl = {https://ui.adsabs.harvard.edu/abs/2011ApJ...737..103S}
}

@ARTICLE{Wright2010,
       author = {{Wright}, Edward L. and {Eisenhardt}, Peter R.~M. and {Mainzer}, Amy K. and {Ressler}, Michael E. and {Cutri}, Roc M. and {Jarrett}, Thomas and {Kirkpatrick}, J. Davy and {Padgett}, Deborah and {McMillan}, Robert S. and {Skrutskie}, Michael and {Stanford}, S.~A. and {Cohen}, Martin and {Walker}, Russell G. and {Mather}, John C. and {Leisawitz}, David and {Gautier}, Thomas N., III and {McLean}, Ian and {Benford}, Dominic and {Lonsdale}, Carol J. and {Blain}, Andrew and {Mendez}, Bryan and {Irace}, William R. and {Duval}, Valerie and {Liu}, Fengchuan and {Royer}, Don and {Heinrichsen}, Ingolf and {Howard}, Joan and {Shannon}, Mark and {Kendall}, Martha and {Walsh}, Amy L. and {Larsen}, Mark and {Cardon}, Joel G. and {Schick}, Scott and {Schwalm}, Mark and {Abid}, Mohamed and {Fabinsky}, Beth and {Naes}, Larry and {Tsai}, Chao-Wei},
        title = "{The wide-field infrared survey explorer (WISE): mission description and initial on-orbit performance}",
      journal = {\aj},
         year = 2010,
        month = dec,
       volume = {140},
       number = {6},
        pages = {1868-1881},
          doi = {10.1088/0004-6256/140/6/1868},
archivePrefix = {arXiv},
       eprint = {1008.0031},
 primaryClass = {astro-ph.IM},
       adsurl = {https://ui.adsabs.harvard.edu/abs/2010AJ....140.1868W}
}

@ARTICLE{Lu2006,
       author = {{Lu}, Honglin and {Zhou}, Hongyan and {Wang}, Junxian and {Wang}, Tinggui and {Dong}, Xiaobo and {Zhuang}, Zhenquan and {Li}, Cheng},
        title = "{Ensemble Learning for Independent Component Analysis of Normal Galaxy Spectra}",
      journal = {\aj},
         year = 2006,
        month = feb,
       volume = {131},
       number = {2},
        pages = {790-805},
          doi = {10.1086/498711},
archivePrefix = {arXiv},
       eprint = {astro-ph/0510246},
 primaryClass = {astro-ph},
       adsurl = {https://ui.adsabs.harvard.edu/abs/2006AJ....131..790L}
}

@BOOK{Osterbrock2006,
       author = {{Osterbrock}, Donald E. and {Ferland}, Gary J.},
        title = "{Astrophysics of gaseous nebulae and active galactic nuclei}",
         year = 2006,
       adsurl = {https://ui.adsabs.harvard.edu/abs/2006agna.book.....O}
}

@ARTICLE{Burrows2005,
       author = {{Burrows}, David N. and {Hill}, J.~E. and {Nousek}, J.~A. and {Kennea}, J.~A. and {Wells}, A. and {Osborne}, J.~P. and {Abbey}, A.~F. and {Beardmore}, A. and {Mukerjee}, K. and {Short}, A.~D.~T. and {Chincarini}, G. and {Campana}, S. and {Citterio}, O. and {Moretti}, A. and {Pagani}, C. and {Tagliaferri}, G. and {Giommi}, P. and {Capalbi}, M. and {Tamburelli}, F. and {Angelini}, L. and {Cusumano}, G. and {Br{\"a}uninger}, H.~W. and {Burkert}, W. and {Hartner}, G.~D.},
        title = "{The swift x-ray telescope}",
      journal = {\ssr},
         year = 2005,
        month = oct,
       volume = {120},
       number = {3-4},
        pages = {165-195},
          doi = {10.1007/s11214-005-5097-2},
archivePrefix = {arXiv},
       eprint = {astro-ph/0508071},
 primaryClass = {astro-ph},
       adsurl = {https://ui.adsabs.harvard.edu/abs/2005SSRv..120..165B}
}

@ARTICLE{Roming2005,
       author = {{Roming}, Peter W.~A. and {Kennedy}, Thomas E. and {Mason}, Keith O. and {Nousek}, John A. and {Ahr}, Lindy and {Bingham}, Richard E. and {Broos}, Patrick S. and {Carter}, Mary J. and {Hancock}, Barry K. and {Huckle}, Howard E. and {Hunsberger}, S.~D. and {Kawakami}, Hajime and {Killough}, Ronnie and {Koch}, T. Scott and {McLelland}, Michael K. and {Smith}, Kelly and {Smith}, Philip J. and {Soto}, Juan Carlos and {Boyd}, Patricia T. and {Breeveld}, Alice A. and {Holland}, Stephen T. and {Ivanushkina}, Mariya and {Pryzby}, Michael S. and {Still}, Martin D. and {Stock}, Joseph},
        title = "{The swift ultra-violet/optical telescope}",
      journal = {\ssr},
         year = 2005,
        month = oct,
       volume = {120},
       number = {3-4},
        pages = {95-142},
          doi = {10.1007/s11214-005-5095-4},
archivePrefix = {arXiv},
       eprint = {astro-ph/0507413},
 primaryClass = {astro-ph},
       adsurl = {https://ui.adsabs.harvard.edu/abs/2005SSRv..120...95R}
}

@ARTICLE{York2000,
       author = {{York}, Donald G. and {Adelman}, J. and {Anderson}, John E., Jr. and {Anderson}, Scott F. and {Annis}, James and {Bahcall}, Neta A. and {Bakken}, J.~A. and {Barkhouser}, Robert and {Bastian}, Steven and {Berman}, Eileen and {Boroski}, William N. and {Bracker}, Steve and {Briegel}, Charlie and {Briggs}, John W. and {Brinkmann}, J. and {Brunner}, Robert and {Burles}, Scott and {Carey}, Larry and {Carr}, Michael A. and {Castander}, Francisco J. and {Chen}, Bing and {Colestock}, Patrick L. and {Connolly}, A.~J. and {Crocker}, J.~H. and {Csabai}, Istv{\'a}n and {Czarapata}, Paul C. and {Davis}, John Eric and {Doi}, Mamoru and {Dombeck}, Tom and {Eisenstein}, Daniel and {Ellman}, Nancy and {Elms}, Brian R. and {Evans}, Michael L. and {Fan}, Xiaohui and {Federwitz}, Glenn R. and {Fiscelli}, Larry and {Friedman}, Scott and {Frieman}, Joshua A. and {Fukugita}, Masataka and {Gillespie}, Bruce and {Gunn}, James E. and {Gurbani}, Vijay K. and {de Haas}, Ernst and {Haldeman}, Merle and {Harris}, Frederick H. and {Hayes}, J. and {Heckman}, Timothy M. and {Hennessy}, G.~S. and {Hindsley}, Robert B. and {Holm}, Scott and {Holmgren}, Donald J. and {Huang}, Chi-hao and {Hull}, Charles and {Husby}, Don and {Ichikawa}, Shin-Ichi and {Ichikawa}, Takashi and {Ivezi{\'c}}, {\v{Z}}eljko and {Kent}, Stephen and {Kim}, Rita S.~J. and {Kinney}, E. and {Klaene}, Mark and {Kleinman}, A.~N. and {Kleinman}, S. and {Knapp}, G.~R. and {Korienek}, John and {Kron}, Richard G. and {Kunszt}, Peter Z. and {Lamb}, D.~Q. and {Lee}, B. and {Leger}, R. French and {Limmongkol}, Siriluk and {Lindenmeyer}, Carl and {Long}, Daniel C. and {Loomis}, Craig and {Loveday}, Jon and {Lucinio}, Rich and {Lupton}, Robert H. and {MacKinnon}, Bryan and {Mannery}, Edward J. and {Mantsch}, P.~M. and {Margon}, Bruce and {McGehee}, Peregrine and {McKay}, Timothy A. and {Meiksin}, Avery and {Merelli}, Aronne and {Monet}, David G. and {Munn}, Jeffrey A. and {Narayanan}, Vijay K. and {Nash}, Thomas and {Neilsen}, Eric and {Neswold}, Rich and {Newberg}, Heidi Jo and {Nichol}, R.~C. and {Nicinski}, Tom and {Nonino}, Mario and {Okada}, Norio and {Okamura}, Sadanori and {Ostriker}, Jeremiah P. and {Owen}, Russell and {Pauls}, A. George and {Peoples}, John and {Peterson}, R.~L. and {Petravick}, Donald and {Pier}, Jeffrey R. and {Pope}, Adrian and {Pordes}, Ruth and {Prosapio}, Angela and {Rechenmacher}, Ron and {Quinn}, Thomas R. and {Richards}, Gordon T. and {Richmond}, Michael W. and {Rivetta}, Claudio H. and {Rockosi}, Constance M. and {Ruthmansdorfer}, Kurt and {Sandford}, Dale and {Schlegel}, David J. and {Schneider}, Donald P. and {Sekiguchi}, Maki and {Sergey}, Gary and {Shimasaku}, Kazuhiro and {Siegmund}, Walter A. and {Smee}, Stephen and {Smith}, J. Allyn and {Snedden}, S. and {Stone}, R. and {Stoughton}, Chris and {Strauss}, Michael A. and {Stubbs}, Christopher and {SubbaRao}, Mark and {Szalay}, Alexander S. and {Szapudi}, Istvan and {Szokoly}, Gyula P. and {Thakar}, Anirudda R. and {Tremonti}, Christy and {Tucker}, Douglas L. and {Uomoto}, Alan and {Vanden Berk}, Dan and {Vogeley}, Michael S. and {Waddell}, Patrick and {Wang}, Shu-i. and {Watanabe}, Masaru and {Weinberg}, David H. and {Yanny}, Brian and {Yasuda}, Naoki and {SDSS Collaboration}},
        title = "{The Sloan Digital Sky Survey: Technical Summary}",
      journal = {\aj},
         year = 2000,
        month = sep,
       volume = {120},
       number = {3},
        pages = {1579-1587},
          doi = {10.1086/301513},
archivePrefix = {arXiv},
       eprint = {astro-ph/0006396},
 primaryClass = {astro-ph},
       adsurl = {https://ui.adsabs.harvard.edu/abs/2000AJ....120.1579Y}
}

@ARTICLE{Veilleux1987,
       author = {{Veilleux}, Sylvain and {Osterbrock}, Donald E.},
        title = "{Spectral classification of emission-line galaxies}",
      journal = {\apjs},
         year = 1987,
        month = feb,
       volume = {63},
        pages = {295},
          doi = {10.1086/191166},
       adsurl = {https://ui.adsabs.harvard.edu/abs/1987ApJS...63..295V}
}

@ARTICLE{Baldwin1981,
       author = {{Baldwin}, J.~A. and {Phillips}, M.~M. and {Terlevich}, R.},
        title = "{Classification parameters for the emission-line spectra of extragalactic objects.}",
      journal = {\pasp},
         year = 1981,
        month = feb,
       volume = {93},
        pages = {5-19},
          doi = {10.1086/130766},
       adsurl = {https://ui.adsabs.harvard.edu/abs/1981PASP...93....5B}
}

@ARTICLE{Hills1975,
       author = {{Hills}, J.~G.},
        title = "{Possible power source of Seyfert galaxies and QSOs}",
      journal = {\nat},
         year = 1975,
        month = mar,
       volume = {254},
       number = {5498},
        pages = {295-298},
          doi = {10.1038/254295a0},
       adsurl = {https://ui.adsabs.harvard.edu/abs/1975Natur.254..295H}
}

@ARTICLE{Ivanov2001,
       author = {{Ivanov}, Pavel B. and {Novikov}, Igor D.},
        title = "{A new model of a tidally disrupted star}",
      journal = {\apj},
         year = 2001,
        month = mar,
       volume = {549},
       number = {1},
        pages = {467-482},
          doi = {10.1086/319050},
archivePrefix = {arXiv},
       eprint = {astro-ph/0005107},
 primaryClass = {astro-ph},
       adsurl = {https://ui.adsabs.harvard.edu/abs/2001ApJ...549..467I}
}

@ARTICLE{Guillochon2013,
       author = {{Guillochon}, James and {Ramirez-Ruiz}, Enrico},
        title = "{Hydrodynamical simulations to determine the feeding rate of black holes by the tidal disruption of stars: the importance of the impact parameter and stellar structure}",
      journal = {\apj},
         year = 2013,
        month = apr,
       volume = {767},
       number = {1},
          eid = {25},
        pages = {25},
          doi = {10.1088/0004-637X/767/1/25},
archivePrefix = {arXiv},
       eprint = {1206.2350},
 primaryClass = {astro-ph.HE},
       adsurl = {https://ui.adsabs.harvard.edu/abs/2013ApJ...767...25G}
}

@ARTICLE{Diener1997,
       author = {{Diener}, P. and {Frolov}, V.~P. and {Khokhlov}, A.~M. and {Novikov}, I.~D. and {Pethick}, C.~J.},
        title = "{Relativistic tidal interaction of stars with a rotating black hole}",
      journal = {\apj},
         year = 1997,
        month = apr,
       volume = {479},
       number = {1},
        pages = {164-178},
          doi = {10.1086/303875},
       adsurl = {https://ui.adsabs.harvard.edu/abs/1997ApJ...479..164D}
}

@ARTICLE{Coughlin2019,
       author = {{Coughlin}, Eric R. and {Nixon}, C.~J.},
        title = "{Partial stellar disruption by a supermassive black hole: is the light curve really proportional to $t^{-9/4}$?}",
      journal = {\apjl},
         year = 2019,
        month = sep,
       volume = {883},
       number = {1},
          eid = {L17},
        pages = {L17},
          doi = {10.3847/2041-8213/ab412d},
archivePrefix = {arXiv},
       eprint = {1907.03034},
 primaryClass = {astro-ph.GA},
       adsurl = {https://ui.adsabs.harvard.edu/abs/2019ApJ...883L..17C}
}

@ARTICLE{LinZheyu2024,
       author = {{Lin}, Zhe-Yu and {Jiang}, Ning and {Wang}, Ting-Gui and {Kong}, Xu and {Li}, Dongyue and {He}, Han and {Wang}, Yibo and {Zhu}, Jiazheng and {Li}, Wentao and {Jiang}, Ji-an and {Singh}, Avinash and {Teja}, Rishabh Singh and {Sahu}, D.~K. and {Jin}, Chichuan and {Maeda}, Keiichi and {Huang}, Shifeng},
        title = "{The unluckiest star: a spectroscopically confirmed repeated partial tidal disruption event AT 2022dbl}",
      journal = {\apjl},
         year = 2024,
        month = aug,
       volume = {971},
       number = {1},
          eid = {L26},
        pages = {L26},
          doi = {10.3847/2041-8213/ad638e},
archivePrefix = {arXiv},
       eprint = {2405.10895},
 primaryClass = {astro-ph.HE},
       adsurl = {https://ui.adsabs.harvard.edu/abs/2024ApJ...971L..26L}
}

@ARTICLE{LiuChang2025,
       author = {{Liu}, Chang and {Yarza}, Ricardo and {Ramirez-Ruiz}, Enrico},
        title = "{Repeating partial tidal encounters of sun-like stars leading to their complete disruption}",
      journal = {\apj},
         year = 2025,
        month = jan,
       volume = {979},
       number = {1},
          eid = {40},
        pages = {40},
          doi = {10.3847/1538-4357/ad9b0b},
archivePrefix = {arXiv},
       eprint = {2406.01670},
 primaryClass = {astro-ph.HE},
       adsurl = {https://ui.adsabs.harvard.edu/abs/2025ApJ...979...40L}
}

@ARTICLE{Bandopadhyay2024,
       author = {{Bandopadhyay}, Ananya and {Coughlin}, Eric R. and {Nixon}, C.~J. and {Pasham}, Dheeraj R.},
        title = "{Repeating nuclear transients from repeating partial tidal disruption events: reproducing ASASSN-14ko and AT2020vdq}",
      journal = {\apj},
         year = 2024,
        month = oct,
       volume = {974},
       number = {1},
          eid = {80},
        pages = {80},
          doi = {10.3847/1538-4357/ad6a5a},
archivePrefix = {arXiv},
       eprint = {2406.03675},
 primaryClass = {astro-ph.HE},
       adsurl = {https://ui.adsabs.harvard.edu/abs/2024ApJ...974...80B}
}

@ARTICLE{LawSmith2020,
       author = {{Law-Smith}, Jamie A.~P. and {Coulter}, David A. and {Guillochon}, James and {Mockler}, Brenna and {Ramirez-Ruiz}, Enrico},
        title = "{Stellar tidal disruption events with abundances and realistic structures (STARS): library of fallback rates}",
      journal = {\apj},
         year = 2020,
        month = dec,
       volume = {905},
       number = {2},
          eid = {141},
        pages = {141},
          doi = {10.3847/1538-4357/abc489},
archivePrefix = {arXiv},
       eprint = {2007.10996},
 primaryClass = {astro-ph.HE},
       adsurl = {https://ui.adsabs.harvard.edu/abs/2020ApJ...905..141L}
}

@ARTICLE{MacLeod2012GaiantTDE,
       author = {{MacLeod}, Morgan and {Guillochon}, James and {Ramirez-Ruiz}, Enrico},
        title = "{The tidal disruption of giant stars and their contribution to the flaring supermassive black hole population}",
      journal = {\apj},
         year = 2012,
        month = oct,
       volume = {757},
       number = {2},
          eid = {134},
        pages = {134},
          doi = {10.1088/0004-637X/757/2/134},
archivePrefix = {arXiv},
       eprint = {1206.2922},
 primaryClass = {astro-ph.HE},
       adsurl = {https://ui.adsabs.harvard.edu/abs/2012ApJ...757..134M}
}

@ARTICLE{Drake2009,
       author = {{Drake}, A.~J. and {Djorgovski}, S.~G. and {Mahabal}, A. and {Beshore}, E. and {Larson}, S. and {Graham}, M.~J. and {Williams}, R. and {Christensen}, E. and {Catelan}, M. and {Boattini}, A. and {Gibbs}, A. and {Hill}, R. and {Kowalski}, R.},
        title = "{First results from the catalina real-time transient survey}",
      journal = {\apj},
         year = 2009,
        month = may,
       volume = {696},
       number = {1},
        pages = {870-884},
          doi = {10.1088/0004-637X/696/1/870},
archivePrefix = {arXiv},
       eprint = {0809.1394},
 primaryClass = {astro-ph},
       adsurl = {https://ui.adsabs.harvard.edu/abs/2009ApJ...696..870D}
}

@ARTICLE{Bonnerot2021,
       author = {{Bonnerot}, C. and {Stone}, N.~C.},
        title = "{Formation of an accretion flow}",
      journal = {\ssr},
         year = 2021,
        month = feb,
       volume = {217},
       number = {1},
          eid = {16},
        pages = {16},
          doi = {10.1007/s11214-020-00789-1},
archivePrefix = {arXiv},
       eprint = {2008.11731},
 primaryClass = {astro-ph.HE},
       adsurl = {https://ui.adsabs.harvard.edu/abs/2021SSRv..217...16B}
}

@ARTICLE{Frank1976,
       author = {{Frank}, J. and {Rees}, M.~J.},
        title = "{Effects of massive black holes on dense stellar systems.}",
      journal = {\mnras},
         year = 1976,
        month = sep,
       volume = {176},
        pages = {633-647},
          doi = {10.1093/mnras/176.3.633},
       adsurl = {https://ui.adsabs.harvard.edu/abs/1976MNRAS.176..633F}
}

@ARTICLE{Kochanek2016,
       author = {{Kochanek}, C.~S.},
        title = "{Abundance anomalies in tidal disruption events}",
      journal = {\mnras},
         year = 2016,
        month = may,
       volume = {458},
       number = {1},
        pages = {127-134},
          doi = {10.1093/mnras/stw267},
archivePrefix = {arXiv},
       eprint = {1512.03065},
 primaryClass = {astro-ph.HE},
       adsurl = {https://ui.adsabs.harvard.edu/abs/2016MNRAS.458..127K}
}

@ARTICLE{YangChenwei2017,
       author = {{Yang}, Chenwei and {Wang}, Tinggui and {Ferland}, Gary J. and {Dou}, Liming and {Zhou}, Hongyan and {Jiang}, Ning and {Sheng}, Zhenfeng},
        title = "{The Carbon and Nitrogen Abundance Ratio in the Broad Line Region of Tidal Disruption Events}",
      journal = {\apj},
         year = 2017,
        month = sep,
       volume = {846},
       number = {2},
          eid = {150},
        pages = {150},
          doi = {10.3847/1538-4357/aa8598},
archivePrefix = {arXiv},
       eprint = {1708.03548},
 primaryClass = {astro-ph.GA},
       adsurl = {https://ui.adsabs.harvard.edu/abs/2017ApJ...846..150Y}
}

@ARTICLE{Ramsden2023,
       author = {{Ramsden}, P. and {Moore}, T. and {Aamer}, A. and {Fulton}, M. and {Smith}, K.~W. and {Yaron}, O.},
        title = "{ePESSTO+ transient classification report for 2023-11-05}",
      journal = {Transient Name Server Classification Report},
         year = 2023,
        month = nov,
       volume = {2023-2854},
        pages = {1},
       adsurl = {https://ui.adsabs.harvard.edu/abs/2023TNSCR2854....1R}
}

@ARTICLE{Gehrels2004,
       author = {{Gehrels}, N. and {Chincarini}, G. and {Giommi}, P. and {Mason}, K.~O. and {Nousek}, J.~A. and {Wells}, A.~A. and {White}, N.~E. and {Barthelmy}, S.~D. and {Burrows}, D.~N. and {Cominsky}, L.~R. and {Hurley}, K.~C. and {Marshall}, F.~E. and {M{\'e}sz{\'a}ros}, P. and {Roming}, P.~W.~A. and {Angelini}, L. and {Barbier}, L.~M. and {Belloni}, T. and {Campana}, S. and {Caraveo}, P.~A. and {Chester}, M.~M. and {Citterio}, O. and {Cline}, T.~L. and {Cropper}, M.~S. and {Cummings}, J.~R. and {Dean}, A.~J. and {Feigelson}, E.~D. and {Fenimore}, E.~E. and {Frail}, D.~A. and {Fruchter}, A.~S. and {Garmire}, G.~P. and {Gendreau}, K. and {Ghisellini}, G. and {Greiner}, J. and {Hill}, J.~E. and {Hunsberger}, S.~D. and {Krimm}, H.~A. and {Kulkarni}, S.~R. and {Kumar}, P. and {Lebrun}, F. and {Lloyd-Ronning}, N.~M. and {Markwardt}, C.~B. and {Mattson}, B.~J. and {Mushotzky}, R.~F. and {Norris}, J.~P. and {Osborne}, J. and {Paczynski}, B. and {Palmer}, D.~M. and {Park}, H. -S. and {Parsons}, A.~M. and {Paul}, J. and {Rees}, M.~J. and {Reynolds}, C.~S. and {Rhoads}, J.~E. and {Sasseen}, T.~P. and {Schaefer}, B.~E. and {Short}, A.~T. and {Smale}, A.~P. and {Smith}, I.~A. and {Stella}, L. and {Tagliaferri}, G. and {Takahashi}, T. and {Tashiro}, M. and {Townsley}, L.~K. and {Tueller}, J. and {Turner}, M.~J.~L. and {Vietri}, M. and {Voges}, W. and {Ward}, M.~J. and {Willingale}, R. and {Zerbi}, F.~M. and {Zhang}, W.~W.},
        title = "{The swift gamma-ray burst mission}",
      journal = {\apj},
         year = 2004,
        month = aug,
       volume = {611},
       number = {2},
        pages = {1005-1020},
          doi = {10.1086/422091},
archivePrefix = {arXiv},
       eprint = {astro-ph/0405233},
 primaryClass = {astro-ph},
       adsurl = {https://ui.adsabs.harvard.edu/abs/2004ApJ...611.1005G}
}

@inproceedings{Jiang2018NGPS,
author = {{Jiang}, Hai-Jiao and {Hu}, Zhong-Wen and {Xu}, Ming-Ming and {Dai}, Songxin and {Zhang}, Huatao and {Wang}, Lei and {Chen}, Yi},
title = {{The preliminary design of the next generation Palomar spectrograph for 200-inch Hale telescope}},
volume = {10702},
booktitle = {Ground-based and Airborne Instrumentation for Astronomy VII},
editor = {Christopher J. Evans and Luc Simard and Hideki Takami},
organization = {International Society for Optics and Photonics},
publisher = {SPIE},
pages = {107022L},
year = {2018},
doi = {10.1117/12.2312550},
URL = {https://doi.org/10.1117/12.2312550}
}

@ARTICLE{Ricci2017,
       author = {{Ricci}, C. and {Trakhtenbrot}, B. and {Koss}, M.~J. and {Ueda}, Y. and {Del Vecchio}, I. and {Treister}, E. and {Schawinski}, K. and {Paltani}, S. and {Oh}, K. and {Lamperti}, I. and {Berney}, S. and {Gandhi}, P. and {Ichikawa}, K. and {Bauer}, F.~E. and {Ho}, L.~C. and {Asmus}, D. and {Beckmann}, V. and {Soldi}, S. and {Balokovi{\'c}}, M. and {Gehrels}, N. and {Markwardt}, C.~B.},
        title = "{BAT AGN Spectroscopic Survey. V. X-Ray Properties of the Swift/BAT 70-month AGN Catalog}",
      journal = {\apjs},
         year = 2017,
        month = dec,
       volume = {233},
       number = {2},
          eid = {17},
        pages = {17},
          doi = {10.3847/1538-4365/aa96ad},
archivePrefix = {arXiv},
       eprint = {1709.03989},
 primaryClass = {astro-ph.HE},
       adsurl = {https://ui.adsabs.harvard.edu/abs/2017ApJS..233...17R}
}

@ARTICLE{ZFPS,
       author = {{Masci}, Frank J. and {Laher}, Russ R. and {Rusholme}, Benjamin and {Shupe}, David and {Paladini}, Roberta and {Groom}, Steve and {Wold}, Avery and {Miller}, Adam A. and {Drake}, Andrew},
        title = "{A New Forced Photometry Service for the Zwicky Transient Facility}",
      journal = {arXiv e-prints},
         year = 2023,
        month = may,
          eid = {arXiv:2305.16279},
        pages = {arXiv:2305.16279},
          doi = {10.48550/arXiv.2305.16279},
archivePrefix = {arXiv},
       eprint = {2305.16279},
 primaryClass = {astro-ph.IM},
       adsurl = {https://ui.adsabs.harvard.edu/abs/2023arXiv230516279M}
}

@ARTICLE{Payne2021,
       author = {{Payne}, Anna V. and {Shappee}, Benjamin J. and {Hinkle}, Jason T. and {Vallely}, Patrick J. and {Kochanek}, Christopher S. and {Holoien}, Thomas W. -S. and {Auchettl}, Katie and {Stanek}, K.~Z. and {Thompson}, Todd A. and {Neustadt}, Jack M.~M. and {Tucker}, Michael A. and {Armstrong}, James D. and {Brimacombe}, Joseph and {Cacella}, Paulo and {Cornect}, Robert and {Denneau}, Larry and {Fausnaugh}, Michael M. and {Flewelling}, Heather and {Grupe}, Dirk and {Heinze}, A.~N. and {Lopez}, Laura A. and {Monard}, Berto and {Prieto}, Jose L. and {Schneider}, Adam C. and {Sheppard}, Scott S. and {Tonry}, John L. and {Weiland}, Henry},
        title = "{ASASSN-14ko is a periodic nuclear transient in ESO 253-G003}",
      journal = {\apj},
         year = 2021,
        month = apr,
       volume = {910},
       number = {2},
          eid = {125},
        pages = {125},
          doi = {10.3847/1538-4357/abe38d},
archivePrefix = {arXiv},
       eprint = {2009.03321},
 primaryClass = {astro-ph.HE},
       adsurl = {https://ui.adsabs.harvard.edu/abs/2021ApJ...910..125P}
}

@ARTICLE{Demircan1991,
       author = {{Demircan}, Osman and {Kahraman}, Goksel},
        title = "{Stellar Mass / Luminosity and Mass / Radius Relations}",
      journal = {\apss},
         year = 1991,
        month = jul,
       volume = {181},
       number = {2},
        pages = {313-322},
          doi = {10.1007/BF00639097},
       adsurl = {https://ui.adsabs.harvard.edu/abs/1991Ap&SS.181..313D}
}

@MISC{Becker2015,
       author = {{Becker}, Andrew},
        title = "{HOTPANTS: High Order Transform of PSF ANd Template Subtraction}",
 howpublished = {Astrophysics Source Code Library, record ascl:1504.004},
         year = 2015,
        month = apr,
          eid = {ascl:1504.004},
        pages = {ascl:1504.004},
archivePrefix = {ascl},
       eprint = {1504.004},
       adsurl = {https://ui.adsabs.harvard.edu/abs/2015ascl.soft04004B}
}

@ARTICLE{Astropy,
       author = {{Astropy Collaboration} and {Price-Whelan}, Adrian M. and {Lim}, Pey Lian and {Earl}, Nicholas and {Starkman}, Nathaniel and {Bradley}, Larry and {Shupe}, David L. and {Patil}, Aarya A. and {Corrales}, Lia and {Brasseur}, C.~E. and {N{\"o}the}, Maximilian and {Donath}, Axel and {Tollerud}, Erik and {Morris}, Brett M. and {Ginsburg}, Adam and {Vaher}, Eero and {Weaver}, Benjamin A. and {Tocknell}, James and {Jamieson}, William and {van Kerkwijk}, Marten H. and {Robitaille}, Thomas P. and {Merry}, Bruce and {Bachetti}, Matteo and {G{\"u}nther}, H. Moritz and {Aldcroft}, Thomas L. and {Alvarado-Montes}, Jaime A. and {Archibald}, Anne M. and {B{\'o}di}, Attila and {Bapat}, Shreyas and {Barentsen}, Geert and {Baz{\'a}n}, Juanjo and {Biswas}, Manish and {Boquien}, M{\'e}d{\'e}ric and {Burke}, D.~J. and {Cara}, Daria and {Cara}, Mihai and {Conroy}, Kyle E. and {Conseil}, Simon and {Craig}, Matthew W. and {Cross}, Robert M. and {Cruz}, Kelle L. and {D'Eugenio}, Francesco and {Dencheva}, Nadia and {Devillepoix}, Hadrien A.~R. and {Dietrich}, J{\"o}rg P. and {Eigenbrot}, Arthur Davis and {Erben}, Thomas and {Ferreira}, Leonardo and {Foreman-Mackey}, Daniel and {Fox}, Ryan and {Freij}, Nabil and {Garg}, Suyog and {Geda}, Robel and {Glattly}, Lauren and {Gondhalekar}, Yash and {Gordon}, Karl D. and {Grant}, David and {Greenfield}, Perry and {Groener}, Austen M. and {Guest}, Steve and {Gurovich}, Sebastian and {Handberg}, Rasmus and {Hart}, Akeem and {Hatfield-Dodds}, Zac and {Homeier}, Derek and {Hosseinzadeh}, Griffin and {Jenness}, Tim and {Jones}, Craig K. and {Joseph}, Prajwel and {Kalmbach}, J. Bryce and {Karamehmetoglu}, Emir and {Ka{\l}uszy{\'n}ski}, Miko{\l}aj and {Kelley}, Michael S.~P. and {Kern}, Nicholas and {Kerzendorf}, Wolfgang E. and {Koch}, Eric W. and {Kulumani}, Shankar and {Lee}, Antony and {Ly}, Chun and {Ma}, Zhiyuan and {MacBride}, Conor and {Maljaars}, Jakob M. and {Muna}, Demitri and {Murphy}, N.~A. and {Norman}, Henrik and {O'Steen}, Richard and {Oman}, Kyle A. and {Pacifici}, Camilla and {Pascual}, Sergio and {Pascual-Granado}, J. and {Patil}, Rohit R. and {Perren}, Gabriel I. and {Pickering}, Timothy E. and {Rastogi}, Tanuj and {Roulston}, Benjamin R. and {Ryan}, Daniel F. and {Rykoff}, Eli S. and {Sabater}, Jose and {Sakurikar}, Parikshit and {Salgado}, Jes{\'u}s and {Sanghi}, Aniket and {Saunders}, Nicholas and {Savchenko}, Volodymyr and {Schwardt}, Ludwig and {Seifert-Eckert}, Michael and {Shih}, Albert Y. and {Jain}, Anany Shrey and {Shukla}, Gyanendra and {Sick}, Jonathan and {Simpson}, Chris and {Singanamalla}, Sudheesh and {Singer}, Leo P. and {Singhal}, Jaladh and {Sinha}, Manodeep and {Sip{\H{o}}cz}, Brigitta M. and {Spitler}, Lee R. and {Stansby}, David and {Streicher}, Ole and {{\v{S}}umak}, Jani and {Swinbank}, John D. and {Taranu}, Dan S. and {Tewary}, Nikita and {Tremblay}, Grant R. and {de Val-Borro}, Miguel and {Van Kooten}, Samuel J. and {Vasovi{\'c}}, Zlatan and {Verma}, Shresth and {de Miranda Cardoso}, Jos{\'e} Vin{\'\i}cius and {Williams}, Peter K.~G. and {Wilson}, Tom J. and {Winkel}, Benjamin and {Wood-Vasey}, W.~M. and {Xue}, Rui and {Yoachim}, Peter and {Zhang}, Chen and {Zonca}, Andrea and {Astropy Project Contributors}},
        title = "{The Astropy Project: Sustaining and Growing a Community-oriented Open-source Project and the Latest Major Release (v5.0) of the Core Package}",
      journal = {\apj},
         year = 2022,
        month = aug,
       volume = {935},
       number = {2},
          eid = {167},
        pages = {167},
          doi = {10.3847/1538-4357/ac7c74},
archivePrefix = {arXiv},
       eprint = {2206.14220},
 primaryClass = {astro-ph.IM},
       adsurl = {https://ui.adsabs.harvard.edu/abs/2022ApJ...935..167A}
}

@ARTICLE{CIGALE,
       author = {{Boquien}, M. and {Burgarella}, D. and {Roehlly}, Y. and {Buat}, V. and {Ciesla}, L. and {Corre}, D. and {Inoue}, A.~K. and {Salas}, H.},
        title = "{CIGALE: a python Code Investigating GALaxy Emission}",
      journal = {\aap},
         year = 2019,
        month = feb,
       volume = {622},
          eid = {A103},
        pages = {A103},
          doi = {10.1051/0004-6361/201834156},
archivePrefix = {arXiv},
       eprint = {1811.03094},
 primaryClass = {astro-ph.GA},
       adsurl = {https://ui.adsabs.harvard.edu/abs/2019A&A...622A.103B}
}

@ARTICLE{Stalevski2012,
       author = {{Stalevski}, Marko and {Fritz}, Jacopo and {Baes}, Maarten and {Nakos}, Theodoros and {Popovi{\'c}}, Luka {\v{C}}.},
        title = "{3D radiative transfer modelling of the dusty tori around active galactic nuclei as a clumpy two-phase medium}",
      journal = {\mnras},
         year = 2012,
        month = mar,
       volume = {420},
       number = {4},
        pages = {2756-2772},
          doi = {10.1111/j.1365-2966.2011.19775.x},
archivePrefix = {arXiv},
       eprint = {1109.1286},
 primaryClass = {astro-ph.CO},
       adsurl = {https://ui.adsabs.harvard.edu/abs/2012MNRAS.420.2756S}
}

@ARTICLE{Stalevski2016,
       author = {{Stalevski}, Marko and {Ricci}, Claudio and {Ueda}, Yoshihiro and {Lira}, Paulina and {Fritz}, Jacopo and {Baes}, Maarten},
        title = "{The dust covering factor in active galactic nuclei}",
      journal = {\mnras},
         year = 2016,
        month = may,
       volume = {458},
       number = {3},
        pages = {2288-2302},
          doi = {10.1093/mnras/stw444},
archivePrefix = {arXiv},
       eprint = {1602.06954},
 primaryClass = {astro-ph.GA},
       adsurl = {https://ui.adsabs.harvard.edu/abs/2016MNRAS.458.2288S}
}

@ARTICLE{1995PASP..107..375O,
       author = {{Oke}, J.~B. and {Cohen}, J.~G. and {Carr}, M. and {Cromer}, J. and {Dingizian}, A. and {Harris}, F.~H. and {Labrecque}, S. and {Lucinio}, R. and {Schaal}, W. and {Epps}, H. and {Miller}, J.},
        title = "{The Keck low-resolution imaging spectrometer}",
      journal = {\pasp},
         year = 1995,
        month = apr,
       volume = {107},
        pages = {375},
          doi = {10.1086/133562},
       adsurl = {https://ui.adsabs.harvard.edu/abs/1995PASP..107..375O}
}

@INPROCEEDINGS{Tody1986,
       author = {{Tody}, Doug},
        title = "{The IRAF Data Reduction and Analysis System}",
    booktitle = {Instrumentation in astronomy VI},
         year = 1986,
       editor = {{Crawford}, David L.},
       series = {Society of Photo-Optical Instrumentation Engineers (SPIE) Conference Series},
       volume = {627},
        month = jan,
        pages = {733},
          doi = {10.1117/12.968154},
       adsurl = {https://ui.adsabs.harvard.edu/abs/1986SPIE..627..733T}
}

@INPROCEEDINGS{1993ASPC...52..173T,
       author = {{Tody}, Doug},
        title = "{IRAF in the Nineties}",
    booktitle = {Astronomical Data Analysis Software and Systems II},
         year = 1993,
       editor = {{Hanisch}, R.~J. and {Brissenden}, R.~J.~V. and {Barnes}, J.},
       series = {Astronomical Society of the Pacific Conference Series},
       volume = {52},
        month = jan,
        pages = {173},
       adsurl = {https://ui.adsabs.harvard.edu/abs/1993ASPC...52..173T}
}

@ARTICLE{2019PASP..131h4503P,
       author = {{Perley}, Daniel A.},
        title = "{Fully Automated Reduction of Longslit Spectroscopy with the Low Resolution Imaging Spectrometer at the Keck Observatory}",
      journal = {\pasp},
         year = 2019,
        month = aug,
       volume = {131},
       number = {1002},
        pages = {084503},
          doi = {10.1088/1538-3873/ab215d},
archivePrefix = {arXiv},
       eprint = {1903.07629},
 primaryClass = {astro-ph.IM},
       adsurl = {https://ui.adsabs.harvard.edu/abs/2019PASP..131h4503P}
}

@ARTICLE{ASASSN18ap,
       author = {{Wang}, Yibo and {Wang}, Tinggui and {Jiang}, Ning and {Zhang}, Xiaer and {Zhu}, Jiazheng and {Shu}, Xinwen and {Huang}, Shifeng and {Zhang}, FaBao and {Sheng}, Zhenfeng and {Lin}, Zheyu},
        title = "{ASASSN-18ap: A Dusty Tidal Disruption Event Candidate with an Early Bump in the Light Curve}",
      journal = {\apj},
         year = 2024,
        month = may,
       volume = {966},
       number = {1},
          eid = {136},
        pages = {136},
          doi = {10.3847/1538-4357/ad2ae4},
archivePrefix = {arXiv},
       eprint = {2312.12015},
 primaryClass = {astro-ph.HE},
       adsurl = {https://ui.adsabs.harvard.edu/abs/2024ApJ...966..136W}
}

@ARTICLE{Ilic2020,
       author = {{Ili{\'c}}, D. and {Oknyansky}, V. and {Popovi{\'c}}, L. {\v{C}}. and {Tsygankov}, S.~S. and {Belinski}, A.~A. and {Tatarnikov}, A.~M. and {Dodin}, A.~V. and {Shatsky}, N.~I. and {Ikonnikova}, N.~P. and {Raki{\'c}}, N. and {Kova{\v{c}}evi{\'c}}, A. and {Mar{\v{c}}eta-Mandi{\'c}}, S. and {Burlak}, M.~A. and {Mishin}, E.~O. and {Metlova}, N.~V. and {Potanin}, S.~A. and {Zheltoukhov}, S.~G.},
        title = "{A flare in the optical spotted in the changing-look Seyfert NGC 3516}",
      journal = {\aap},
         year = 2020,
        month = jun,
       volume = {638},
          eid = {A13},
        pages = {A13},
          doi = {10.1051/0004-6361/202037532},
archivePrefix = {arXiv},
       eprint = {2004.01308},
 primaryClass = {astro-ph.GA},
       adsurl = {https://ui.adsabs.harvard.edu/abs/2020A&A...638A..13I}
}

@ARTICLE{Raki2022,
       author = {{Raki{\'c}}, N.},
        title = "{Kinematics of the H {\ensuremath{\alpha}} and H {\ensuremath{\beta}} broad-line region in an SDSS sample of type-1 AGNs}",
      journal = {\mnras},
         year = 2022,
        month = oct,
       volume = {516},
       number = {2},
        pages = {1624-1634},
          doi = {10.1093/mnras/stac2259},
archivePrefix = {arXiv},
       eprint = {2208.04359},
 primaryClass = {astro-ph.GA},
       adsurl = {https://ui.adsabs.harvard.edu/abs/2022MNRAS.516.1624R}
}

@ARTICLE{Ilic2023,
       author = {{Ili{\'c}}, Dragana and {Raki{\'c}}, Nemanja and {Popovi{\'c}}, Luka {\v{C}}.},
        title = "{Fantastic Fits with fantasy of Active Galactic Nuclei Spectra: Exploring the Fe II Emission near the H{\ensuremath{\alpha}} Line}",
      journal = {\apjs},
         year = 2023,
        month = jul,
       volume = {267},
       number = {1},
          eid = {19},
        pages = {19},
          doi = {10.3847/1538-4365/acd783},
archivePrefix = {arXiv},
       eprint = {2306.12782},
 primaryClass = {astro-ph.GA},
       adsurl = {https://ui.adsabs.harvard.edu/abs/2023ApJS..267...19I}
}

@ARTICLE{Ryu2020pTDE,
       author = {{Ryu}, Taeho and {Krolik}, Julian and {Piran}, Tsvi and {Noble}, Scott C.},
        title = "{Tidal disruptions of main-sequence stars. III. stellar mass dependence of the character of partial disruptions}",
      journal = {\apj},
         year = 2020,
        month = dec,
       volume = {904},
       number = {2},
          eid = {100},
        pages = {100},
          doi = {10.3847/1538-4357/abb3ce},
archivePrefix = {arXiv},
       eprint = {2001.03503},
 primaryClass = {astro-ph.HE},
       adsurl = {https://ui.adsabs.harvard.edu/abs/2020ApJ...904..100R}
}

@ARTICLE{MacLeod2013,
       author = {{MacLeod}, Morgan and {Ramirez-Ruiz}, Enrico and {Grady}, Sean and {Guillochon}, James},
        title = "{Spoon-feeding giant stars to supermassive black holes: episodic mass transfer from evolving stars and their contribution to the quiescent activity of galactic nuclei}",
      journal = {\apj},
         year = 2013,
        month = nov,
       volume = {777},
       number = {2},
          eid = {133},
        pages = {133},
          doi = {10.1088/0004-637X/777/2/133},
archivePrefix = {arXiv},
       eprint = {1307.2900},
 primaryClass = {astro-ph.HE},
       adsurl = {https://ui.adsabs.harvard.edu/abs/2013ApJ...777..133M}
}

@ARTICLE{Somalwar2025,
       author = {{Somalwar}, Jean J. and {Ravi}, Vikram and {Yao}, Yu-Han and {Guolo}, Muryel and {Graham}, Matthew and {Hammerstein}, Erica and {Lu}, Wenbin and {Nicholl}, Matt and {Sharma}, Yashvi and {Stein}, Robert and {van Velzen}, Sjoert and {Bellm}, Eric C. and {Coughlin}, Michael W. and {Groom}, Steven L. and {Masci}, Frank J. and {Riddle}, Reed},
        title = "{The first systematically identified repeating partial tidal disruption event}",
      journal = {\apj},
         year = 2025,
        month = jun,
       volume = {985},
       number = {2},
          eid = {175},
        pages = {175},
          doi = {10.3847/1538-4357/adcc19},
archivePrefix = {arXiv},
       eprint = {2310.03782},
 primaryClass = {astro-ph.HE},
       adsurl = {https://ui.adsabs.harvard.edu/abs/2025ApJ...985..175S}
}

@ARTICLE{Wevers2023,
       author = {{Wevers}, T. and {Coughlin}, E.~R. and {Pasham}, D.~R. and {Guolo}, M. and {Sun}, Y. and {Wen}, S. and {Jonker}, P.~G. and {Zabludoff}, A. and {Malyali}, A. and {Arcodia}, R. and {Liu}, Z. and {Merloni}, A. and {Rau}, A. and {Grotova}, I. and {Short}, P. and {Cao}, Z.},
        title = "{Live to die another day: the rebrightening of AT 2018fyk as a repeating partial tidal disruption event}",
      journal = {\apjl},
         year = 2023,
        month = jan,
       volume = {942},
       number = {2},
          eid = {L33},
        pages = {L33},
          doi = {10.3847/2041-8213/ac9f36},
archivePrefix = {arXiv},
       eprint = {2209.07538},
 primaryClass = {astro-ph.HE},
       adsurl = {https://ui.adsabs.harvard.edu/abs/2023ApJ...942L..33W}
}

@ARTICLE{SunLuming2024,
       author = {{Sun}, Lu-Ming and {Jiang}, Ning and {Dou}, Li-Ming and {Shu}, Xinwen and {Zhu}, Jiazheng and {Dong}, Subo and {Buckley}, David and {Bradley Cenko}, S. and {Fan}, Xiaohui and {Gromadzki}, Mariusz and {Liu}, Zhu and {Wang}, Jianguo and {Wang}, Tinggui and {Wang}, Yibo and {Wu}, Tao and {Yang}, Lei and {Zhang}, Fabao and {Zhang}, Wenjie and {Zhang}, Xiaer},
        title = "{Recurring tidal disruption events a decade apart in IRAS F01004-2237}",
      journal = {\aap},
         year = 2024,
        month = dec,
       volume = {692},
          eid = {A262},
        pages = {A262},
          doi = {10.1051/0004-6361/202452380},
archivePrefix = {arXiv},
       eprint = {2410.09720},
 primaryClass = {astro-ph.HE},
       adsurl = {https://ui.adsabs.harvard.edu/abs/2024A&A...692A.262S}
}

@ARTICLE{LiuZhu2024,
       author = {{Liu}, Zhu and {Ryu}, Taeho and {Goodwin}, A.~J. and {Rau}, A. and {Homan}, D. and {Krumpe}, M. and {Merloni}, A. and {Grotova}, I. and {Anderson}, G.~E. and {Malyali}, A. and {Miller-Jones}, J.~C.~A.},
        title = "{Rapid evolution of the recurrence time in the repeating partial tidal disruption event eRASSt J045650.3{\ensuremath{-}}203750}",
      journal = {\aap},
         year = 2024,
        month = mar,
       volume = {683},
          eid = {L13},
        pages = {L13},
          doi = {10.1051/0004-6361/202348682},
archivePrefix = {arXiv},
       eprint = {2401.14091},
 primaryClass = {astro-ph.HE},
       adsurl = {https://ui.adsabs.harvard.edu/abs/2024A&A...683L..13L}
}

@ARTICLE{HuangShifeng2023ASASN14ko,
       author = {{Huang}, Shi-Feng and {Jiang}, Ning and {Shen}, Rong-Feng and {Wang}, Ting-Gui and {Sheng}, Zhen-Feng},
        title = "{Dissonance in harmony: the UV/optical periodic outbursts of ASASSN-14ko exhibit repeated bumps and rebrightenings}",
      journal = {\apjl},
         year = 2023,
        month = oct,
       volume = {956},
       number = {2},
          eid = {L46},
        pages = {L46},
          doi = {10.3847/2041-8213/acffc5},
archivePrefix = {arXiv},
       eprint = {2310.03236},
 primaryClass = {astro-ph.HE},
       adsurl = {https://ui.adsabs.harvard.edu/abs/2023ApJ...956L..46H}
}

@ARTICLE{HuangShifeng2025ASASN14ko,
       author = {{Huang}, Shi-Feng and {Wang}, Ting-Gui and {Jiang}, Ning and {Shen}, Rong-Feng and {Chen}, Zhaohao and {Wang}, Yuanming and {Zhu}, Jiazheng and {Wang}, Yibo and {Jiang}, Yunguo and {Shu}, Xinwen and {Ding}, Hucheng and {Fang}, Xiongjun and {Wang}, Yifan and {Lin}, Jie and {Xu}, Jingran and {Chen}, Xu and {Lin}, Zheyu and {Sheng}, Zhenfeng},
        title = "{Unveiling the cosmic dance of repeated nuclear transient ASASSN-14ko: insights from multiwavelength observations}",
      journal = {\apj},
         year = 2025,
        month = aug,
       volume = {988},
       number = {2},
          eid = {237},
        pages = {237},
          doi = {10.3847/1538-4357/ade795},
archivePrefix = {arXiv},
       eprint = {2506.15039},
 primaryClass = {astro-ph.HE},
       adsurl = {https://ui.adsabs.harvard.edu/abs/2025ApJ...988..237H}
}

@ARTICLE{Payne2023ASASSN14ko,
       author = {{Payne}, Anna V. and {Auchettl}, Katie and {Shappee}, Benjamin J. and {Kochanek}, Christopher S. and {Boyd}, Patricia T. and {Holoien}, Thomas W. -S. and {Fausnaugh}, Michael M. and {Ashall}, Chris and {Hinkle}, Jason T. and {Vallely}, Patrick J. and {Stanek}, K.~Z. and {Thompson}, Todd A.},
        title = "{Chandra, HST/STIS, NICER, Swift, and TESS detail the flare evolution of the repeating nuclear transient ASASSN-14ko}",
      journal = {\apj},
         year = 2023,
        month = jul,
       volume = {951},
       number = {2},
          eid = {134},
        pages = {134},
          doi = {10.3847/1538-4357/acd455},
archivePrefix = {arXiv},
       eprint = {2206.11278},
 primaryClass = {astro-ph.HE},
       adsurl = {https://ui.adsabs.harvard.edu/abs/2023ApJ...951..134P}
}

@ARTICLE{Payne2022ASASSN14ko,
       author = {{Payne}, Anna V. and {Shappee}, Benjamin J. and {Hinkle}, Jason T. and {Holoien}, Thomas W. -S. and {Auchettl}, Katie and {Kochanek}, Christopher S. and {Stanek}, K.~Z. and {Thompson}, Todd A. and {Tucker}, Michael A. and {Armstrong}, James D. and {Boyd}, Patricia T. and {Brimacombe}, Joseph and {Cornect}, Robert and {Huber}, Mark E. and {Jha}, Saurabh W. and {Lin}, Chien-Cheng},
        title = "{The rapid x-ray and UV evolution of ASASSN-14ko}",
      journal = {\apj},
         year = 2022,
        month = feb,
       volume = {926},
       number = {2},
          eid = {142},
        pages = {142},
          doi = {10.3847/1538-4357/ac480c},
archivePrefix = {arXiv},
       eprint = {2104.06414},
 primaryClass = {astro-ph.HE},
       adsurl = {https://ui.adsabs.harvard.edu/abs/2022ApJ...926..142P}
}

@ARTICLE{YaoPhilippe2025,
       author = {{Yao}, Philippe Z. and {Quataert}, Eliot and {Jiang}, Yan-Fei and {Lu}, Wenbin and {White}, Christopher J.},
        title = "{Star‑disk collisions: implications for quasi-periodic eruptions and other transients near supermassive black holes}",
      journal = {\apj},
         year = 2025,
        month = jan,
       volume = {978},
       number = {1},
          eid = {91},
        pages = {91},
          doi = {10.3847/1538-4357/ad8911},
archivePrefix = {arXiv},
       eprint = {2407.14578},
 primaryClass = {astro-ph.HE},
       adsurl = {https://ui.adsabs.harvard.edu/abs/2025ApJ...978...91Y}
}

@ARTICLE{Li2013,
       author = {{Li}, Gong-Jie and {Loeb}, Abraham},
        title = "{Accumulated tidal heating of stars over multiple pericentre passages near SgrA*}",
      journal = {\mnras},
         year = 2013,
        month = mar,
       volume = {429},
       number = {4},
        pages = {3040-3046},
          doi = {10.1093/mnras/sts567},
archivePrefix = {arXiv},
       eprint = {1209.1104},
 primaryClass = {astro-ph.GA},
       adsurl = {https://ui.adsabs.harvard.edu/abs/2013MNRAS.429.3040L}
}

@ARTICLE{ChenJinHong2024,
       author = {{Chen}, Jin-Hong and {Dai}, Lixin and {Liu}, Shang-Fei and {Ou}, Jian-Wen},
        title = "{Fate of the Remnant in Tidal Stripping Event: Repeating and Nonrepeating}",
      journal = {\apj},
         year = 2024,
        month = dec,
       volume = {977},
       number = {1},
          eid = {80},
        pages = {80},
          doi = {10.3847/1538-4357/ad8b24},
archivePrefix = {arXiv},
       eprint = {2408.10925},
 primaryClass = {astro-ph.HE},
       adsurl = {https://ui.adsabs.harvard.edu/abs/2024ApJ...977...80C}
}

@ARTICLE{Cufari2023,
       author = {{Cufari}, M. and {Nixon}, C.~J. and {Coughlin}, Eric R.},
        title = "{Tidal capture of stars by supermassive black holes: implications for periodic nuclear transients and quasi-periodic eruptions}",
      journal = {\mnras},
         year = 2023,
        month = mar,
       volume = {520},
       number = {1},
        pages = {L38-L41},
          doi = {10.1093/mnrasl/slad001},
archivePrefix = {arXiv},
       eprint = {2301.01300},
 primaryClass = {astro-ph.HE},
       adsurl = {https://ui.adsabs.harvard.edu/abs/2023MNRAS.520L..38C}
}

@ARTICLE{Manukian2013,
       author = {{Manukian}, Haik and {Guillochon}, James and {Ramirez-Ruiz}, Enrico and {O'Leary}, Ryan M.},
        title = "{Turbovelocity Stars: Kicks Resulting from the Tidal Disruption of Solitary Stars}",
      journal = {\apjl},
         year = 2013,
        month = jul,
       volume = {771},
       number = {2},
          eid = {L28},
        pages = {L28},
          doi = {10.1088/2041-8205/771/2/L28},
archivePrefix = {arXiv},
       eprint = {1305.4634},
 primaryClass = {astro-ph.HE},
       adsurl = {https://ui.adsabs.harvard.edu/abs/2013ApJ...771L..28M}
}

@ARTICLE{Linial2024,
       author = {{Linial}, Itai and {Quataert}, Eliot},
        title = "{Period evolution of repeating transients in galactic nuclei}",
      journal = {\mnras},
         year = 2024,
        month = jan,
       volume = {527},
       number = {2},
        pages = {4317-4329},
          doi = {10.1093/mnras/stad3470},
archivePrefix = {arXiv},
       eprint = {2309.15849},
 primaryClass = {astro-ph.HE},
       adsurl = {https://ui.adsabs.harvard.edu/abs/2024MNRAS.527.4317L}
}

@ARTICLE{Linial2023,
       author = {{Linial}, Itai and {Metzger}, Brian D.},
        title = "{EMRI + TDE = QPE: Periodic X-Ray Flares from Star-Disk Collisions in Galactic Nuclei}",
      journal = {\apj},
         year = 2023,
        month = nov,
       volume = {957},
       number = {1},
          eid = {34},
        pages = {34},
          doi = {10.3847/1538-4357/acf65b},
archivePrefix = {arXiv},
       eprint = {2303.16231},
 primaryClass = {astro-ph.HE},
       adsurl = {https://ui.adsabs.harvard.edu/abs/2023ApJ...957...34L}
}

@ARTICLE{Golightly2019,
       author = {{Golightly}, Elen C.~A. and {Coughlin}, Eric R. and {Nixon}, C.~J.},
        title = "{Tidal Disruption Events: The Role of Stellar Spin}",
      journal = {\apj},
         year = 2019,
        month = feb,
       volume = {872},
       number = {2},
          eid = {163},
        pages = {163},
          doi = {10.3847/1538-4357/aafd2f},
archivePrefix = {arXiv},
       eprint = {1901.03717},
 primaryClass = {astro-ph.HE},
       adsurl = {https://ui.adsabs.harvard.edu/abs/2019ApJ...872..163G}
}

@ARTICLE{McQuillan2014,
       author = {{McQuillan}, A. and {Mazeh}, T. and {Aigrain}, S.},
        title = "{Rotation Periods of 34,030 Kepler Main-sequence Stars: The Full Autocorrelation Sample}",
      journal = {\apjs},
         year = 2014,
        month = apr,
       volume = {211},
       number = {2},
          eid = {24},
        pages = {24},
          doi = {10.1088/0067-0049/211/2/24},
archivePrefix = {arXiv},
       eprint = {1402.5694},
 primaryClass = {astro-ph.SR},
       adsurl = {https://ui.adsabs.harvard.edu/abs/2014ApJS..211...24M}
}

@ARTICLE{Bandopadhyay2025,
       author = {{Bandopadhyay}, Ananya and {Coughlin}, Eric R. and {Nixon}, C.~J.},
        title = "{Repeated tidal interactions between stars and supermassive black holes: mass transfer, stability, and implications for repeating partial tidal disruption events}",
      journal = {\apj},
         year = 2025,
        month = jul,
       volume = {987},
       number = {1},
          eid = {16},
        pages = {16},
          doi = {10.3847/1538-4357/add9a5},
archivePrefix = {arXiv},
       eprint = {2504.18614},
 primaryClass = {astro-ph.HE},
       adsurl = {https://ui.adsabs.harvard.edu/abs/2025ApJ...987...16B}
}

@ARTICLE{Cufari2022,
       author = {{Cufari}, M. and {Coughlin}, Eric R. and {Nixon}, C.~J.},
        title = "{Using the Hills mechanism to generate repeating partial tidal disruption events and ASASSN-14ko}",
      journal = {\apjl},
         year = 2022,
        month = apr,
       volume = {929},
       number = {2},
          eid = {L20},
        pages = {L20},
          doi = {10.3847/2041-8213/ac6021},
archivePrefix = {arXiv},
       eprint = {2203.08162},
 primaryClass = {astro-ph.HE},
       adsurl = {https://ui.adsabs.harvard.edu/abs/2022ApJ...929L..20C}
}

@ARTICLE{JiangNing2025PS16dtm,
       author = {{Jiang}, Ning and {Luo}, Di and {Zhu}, Jiazheng and {Cutri}, Roc M.},
        title = "{The Extraordinary Long-lasting Infrared Echo of PS16dtm Reveals an Extremely Energetic Nuclear Outburst}",
      journal = {\apjl},
         year = 2025,
        month = feb,
       volume = {980},
       number = {1},
          eid = {L17},
        pages = {L17},
          doi = {10.3847/2041-8213/adaeb9},
archivePrefix = {arXiv},
       eprint = {2501.15966},
 primaryClass = {astro-ph.HE},
       adsurl = {https://ui.adsabs.harvard.edu/abs/2025ApJ...980L..17J}
}

@ARTICLE{PhyilippeYao2025TidallyHeated,
       author = {{Yao}, Philippe Z. and {Quataert}, Eliot},
        title = "{Mass Transfer in Tidally Heated Stars Orbiting Massive Black Holes and Implications for Repeating Nuclear Transients}",
      journal = {arXiv e-prints},
         year = 2025,
        month = may,
          eid = {arXiv:2505.10611},
        pages = {arXiv:2505.10611},
          doi = {10.48550/arXiv.2505.10611},
archivePrefix = {arXiv},
       eprint = {2505.10611},
 primaryClass = {astro-ph.HE},
       adsurl = {https://ui.adsabs.harvard.edu/abs/2025arXiv250510611Y}
}

@ARTICLE{ZhongShiyan2023,
       author = {{Zhong}, Shiyan and {Hayasaki}, Kimitake and {Li}, Shuo and {Berczik}, Peter and {Spurzem}, Rainer},
        title = "{Exploring the Origin of Stars on Bound and Unbound Orbits Causing Tidal Disruption Events}",
      journal = {\apj},
         year = 2023,
        month = dec,
       volume = {959},
       number = {1},
          eid = {19},
        pages = {19},
          doi = {10.3847/1538-4357/ad0122},
archivePrefix = {arXiv},
       eprint = {2011.09400},
 primaryClass = {astro-ph.HE},
       adsurl = {https://ui.adsabs.harvard.edu/abs/2023ApJ...959...19Z}
}

@ARTICLE{Linial2025,
       author = {{Linial}, Itai and {Metzger}, Brian D. and {Quataert}, Eliot},
        title = "{QPEs from EMRI debris streams impacting accretion disks in galactic nuclei}",
      journal = {\apj},
         year = 2025,
        month = oct,
       volume = {991},
       number = {2},
          eid = {147},
        pages = {147},
          doi = {10.3847/1538-4357/adfa0e},
archivePrefix = {arXiv},
       eprint = {2506.10096},
 primaryClass = {astro-ph.HE},
       adsurl = {https://ui.adsabs.harvard.edu/abs/2025ApJ...991..147L}
}

@ARTICLE{Shapovalova2012,
       author = {{Shapovalova}, A.~I. and {Popovi{\'c}}, L. {\v{C}}. and {Burenkov}, A.~N. and {Chavushyan}, V.~H. and {Ili{\'c}}, D. and {Kova{\v{c}}evi{\'c}}, A. and {Kollatschny}, W. and {Kova{\v{c}}evi{\'c}}, J. and {Bochkarev}, N.~G. and {Valdes}, J.~R. and {Torrealba}, J. and {Le{\'o}n-Tavares}, J. and {Mercado}, A. and {Ben{\'\i}tez}, E. and {Carrasco}, L. and {Dultzin}, D. and {de la Fuente}, E.},
        title = "{Spectral Optical Monitoring of the Narrow-line Seyfert 1 Galaxy Ark 564}",
      journal = {\apjs},
         year = 2012,
        month = sep,
       volume = {202},
       number = {1},
          eid = {10},
        pages = {10},
          doi = {10.1088/0067-0049/202/1/10},
archivePrefix = {arXiv},
       eprint = {1207.1782},
 primaryClass = {astro-ph.CO},
       adsurl = {https://ui.adsabs.harvard.edu/abs/2012ApJS..202...10S}
}

@ARTICLE{Kovavcevic2010,
       author = {{Kova{\v{c}}evi{\'c}}, Jelena and {Popovi{\'c}}, Luka {\v{C}}. and {Dimitrijevi{\'c}}, Milan S.},
        title = "{Analysis of Optical Fe II Emission in a Sample of Active Galactic Nucleus Spectra}",
      journal = {\apjs},
         year = 2010,
        month = jul,
       volume = {189},
       number = {1},
        pages = {15-36},
          doi = {10.1088/0067-0049/189/1/15},
archivePrefix = {arXiv},
       eprint = {1004.2212},
 primaryClass = {astro-ph.CO},
       adsurl = {https://ui.adsabs.harvard.edu/abs/2010ApJS..189...15K}
}

@ARTICLE{Yu2024,
       author = {{Yu}, Fang-Yuan and {Lai}, Dong},
        title = "{Binary stars approaching supermassive black holes: tidal breakup, double stellar disruptions, and stellar collision}",
      journal = {\apj},
         year = 2024,
        month = dec,
       volume = {977},
       number = {2},
          eid = {268},
        pages = {268},
          doi = {10.3847/1538-4357/ad93a6},
archivePrefix = {arXiv},
       eprint = {2409.09597},
 primaryClass = {astro-ph.HE},
       adsurl = {https://ui.adsabs.harvard.edu/abs/2024ApJ...977..268Y}
}

@ARTICLE{LiuChang2023,
       author = {{Liu}, Chang and {Mockler}, Brenna and {Ramirez-Ruiz}, Enrico and {Yarza}, Ricardo and {Law-Smith}, Jamie A.~P. and {Naoz}, Smadar and {Melchor}, Denyz and {Rose}, Sanaea},
        title = "{Tidal Disruption Events from Eccentric Orbits and Lessons Learned from the Noteworthy ASASSN-14ko}",
      journal = {\apj},
         year = 2023,
        month = feb,
       volume = {944},
       number = {2},
          eid = {184},
        pages = {184},
          doi = {10.3847/1538-4357/acafe1},
archivePrefix = {arXiv},
       eprint = {2206.13494},
 primaryClass = {astro-ph.HE},
       adsurl = {https://ui.adsabs.harvard.edu/abs/2023ApJ...944..184L}
}

@ARTICLE{Bonnerot2022,
       author = {{Bonnerot}, Cl{\'e}ment and {Pessah}, Martin E. and {Lu}, Wen-Bin},
        title = "{From pericenter and back: full debris stream evolution in tidal disruption events}",
      journal = {\apjl},
         year = 2022,
        month = may,
       volume = {931},
       number = {1},
          eid = {L6},
        pages = {L6},
          doi = {10.3847/2041-8213/ac6950},
archivePrefix = {arXiv},
       eprint = {2112.08384},
 primaryClass = {astro-ph.HE},
       adsurl = {https://ui.adsabs.harvard.edu/abs/2022ApJ...931L...6B}
}

@ARTICLE{Fabricant2019,
       author = {{Fabricant}, Daniel and {Fata}, Robert and {Epps}, Harland and {Gauron}, Thomas and {Mueller}, Mark and {Zajac}, Joseph and {Amato}, Stephen and {Barberis}, Jack and {Bergner}, Henry and {Brennan}, Patricia and {Brown}, Warren and {Chilingarian}, Igor and {Geary}, John and {Kradinov}, Vladimir and {McLeod}, Brian and {Smith}, Matthew and {Woods}, Deborah},
        title = "{Binospec: a wide-field imaging spectrograph for the MMT}",
      journal = {\pasp},
         year = 2019,
        month = jul,
       volume = {131},
       number = {1001},
        pages = {075004},
          doi = {10.1088/1538-3873/ab1d78},
archivePrefix = {arXiv},
       eprint = {1905.03320},
 primaryClass = {astro-ph.IM},
       adsurl = {https://ui.adsabs.harvard.edu/abs/2019PASP..131g5004F}
}

@ARTICLE{Wu2025,
       author = {{Wu}, Mingxin and {Jiang}, Ning and {Zhu}, Jiazheng and {Luo}, Di and {Dou}, Liming and {Wang}, Tinggui},
        title = "{A Torus Remnant Revealed by the Infrared Echo of Tidal Disruption Event AT 2019qiz: Implications for the Missing Energy and Quasiperiodic Eruption Formation}",
      journal = {\apjl},
         year = 2025,
        month = aug,
       volume = {988},
       number = {2},
          eid = {L77},
        pages = {L77},
          doi = {10.3847/2041-8213/adf229},
archivePrefix = {arXiv},
       eprint = {2507.13251},
 primaryClass = {astro-ph.HE},
       adsurl = {https://ui.adsabs.harvard.edu/abs/2025ApJ...988L..77W}
}

@ARTICLE{LuQuataert2023,
       author = {{Lu}, Wen-Bin and {Quataert}, Eliot},
        title = "{Quasi-periodic eruptions from mildly eccentric unstable mass transfer in galactic nuclei}",
      journal = {\mnras},
         year = 2023,
        month = oct,
       volume = {524},
       number = {4},
        pages = {6247-6266},
          doi = {10.1093/mnras/stad2203},
archivePrefix = {arXiv},
       eprint = {2210.08023},
 primaryClass = {astro-ph.HE},
       adsurl = {https://ui.adsabs.harvard.edu/abs/2023MNRAS.524.6247L}
}

@ARTICLE{LuWenbin2021,
       author = {{Lu}, Wen-Bin and {Fuller}, Jim and {Raveh}, Yael and {Perets}, Hagai B. and {Li}, Ting S. and {Hosek}, Jr., Matthew W. and {Do}, Tuan},
        title = "{The former companion of hyper-velocity star S5-HVS1}",
      journal = {\mnras},
         year = 2021,
        month = may,
       volume = {503},
       number = {1},
        pages = {603-613},
          doi = {10.1093/mnras/stab459},
archivePrefix = {arXiv},
       eprint = {2005.12300},
 primaryClass = {astro-ph.HE},
       adsurl = {https://ui.adsabs.harvard.edu/abs/2021MNRAS.503..603L}
}

@ARTICLE{GALEX,
       author = {{Martin}, D. Christopher and {Fanson}, James and {Schiminovich}, David and {Morrissey}, Patrick and {Friedman}, Peter G. and {Barlow}, Tom A. and {Conrow}, Tim and {Grange}, Robert and {Jelinsky}, Patrick N. and {Milliard}, Bruno and {Siegmund}, Oswald H.~W. and {Bianchi}, Luciana and {Byun}, Yong-Ik and {Donas}, Jose and {Forster}, Karl and {Heckman}, Timothy M. and {Lee}, Young-Wook and {Madore}, Barry F. and {Malina}, Roger F. and {Neff}, Susan G. and {Rich}, R. Michael and {Small}, Todd and {Surber}, Frank and {Szalay}, Alex S. and {Welsh}, Barry and {Wyder}, Ted K.},
        title = "{The Galaxy Evolution Explorer: A Space Ultraviolet Survey Mission}",
      journal = {\apjl},
         year = 2005,
        month = jan,
       volume = {619},
       number = {1},
        pages = {L1-L6},
          doi = {10.1086/426387},
archivePrefix = {arXiv},
       eprint = {astro-ph/0411302},
 primaryClass = {astro-ph},
       adsurl = {https://ui.adsabs.harvard.edu/abs/2005ApJ...619L...1M}
}

@ARTICLE{2MASS,
       author = {{Skrutskie}, M.~F. and {Cutri}, R.~M. and {Stiening}, R. and {Weinberg}, M.~D. and {Schneider}, S. and {Carpenter}, J.~M. and {Beichman}, C. and {Capps}, R. and {Chester}, T. and {Elias}, J. and {Huchra}, J. and {Liebert}, J. and {Lonsdale}, C. and {Monet}, D.~G. and {Price}, S. and {Seitzer}, P. and {Jarrett}, T. and {Kirkpatrick}, J.~D. and {Gizis}, J.~E. and {Howard}, E. and {Evans}, T. and {Fowler}, J. and {Fullmer}, L. and {Hurt}, R. and {Light}, R. and {Kopan}, E.~L. and {Marsh}, K.~A. and {McCallon}, H.~L. and {Tam}, R. and {Van Dyk}, S. and {Wheelock}, S.},
        title = "{The Two Micron All Sky Survey (2MASS)}",
      journal = {\aj},
         year = 2006,
        month = feb,
       volume = {131},
       number = {2},
        pages = {1163-1183},
          doi = {10.1086/498708},
       adsurl = {https://ui.adsabs.harvard.edu/abs/2006AJ....131.1163S}
}

@ARTICLE{HeZhicheng2021,
       author = {{He}, Zhi-Cheng and {Jiang}, Ning and {Wang}, Ting-Gui and {Liu}, Guilin and {Sun}, Mouyuan and {Guo}, Hengxiao and {Shen}, Lu and {Cai}, Zhenyi and {Shu}, Xinwen and {Sheng}, Zhenfeng and {Liang}, Zhixiong and {Xu}, Youhua},
        title = "{An extraordinary response of iron emission to the central outburst in a tidal disruption event candidate}",
      journal = {\apjl},
         year = 2021,
        month = feb,
       volume = {907},
       number = {2},
          eid = {L29},
        pages = {L29},
          doi = {10.3847/2041-8213/abd7fd},
archivePrefix = {arXiv},
       eprint = {2009.05243},
 primaryClass = {astro-ph.GA},
       adsurl = {https://ui.adsabs.harvard.edu/abs/2021ApJ...907L..29H}
}

@ARTICLE{Ansky2024,
       author = {{S{\'a}nchez-S{\'a}ez}, P. and {Hern{\'a}ndez-Garc{\'\i}a}, L. and {Bernal}, S. and {Bayo}, A. and {Calistro Rivera}, G. and {Bauer}, F.~E. and {Ricci}, C. and {Merloni}, A. and {Graham}, M.~J. and {Cartier}, R. and {Ar{\'e}valo}, P. and {Assef}, R.~J. and {Concas}, A. and {Homan}, D. and {Krumpe}, M. and {Lira}, P. and {Malyali}, A. and {Mart{\'\i}nez-Aldama}, M.~L. and {Mu{\~n}oz Arancibia}, A.~M. and {Rau}, A. and {Bruni}, G. and {F{\"o}rster}, F. and {Pavez-Herrera}, M. and {Tub{\'\i}n-Arenas}, D. and {Brightman}, M.},
        title = "{SDSS1335+0728: The awakening of a {\ensuremath{\sim}}{}10$^{6}$ M$_{{\ensuremath{\odot}}}$ black hole}",
      journal = {\aap},
         year = 2024,
        month = aug,
       volume = {688},
          eid = {A157},
        pages = {A157},
          doi = {10.1051/0004-6361/202347957},
archivePrefix = {arXiv},
       eprint = {2406.11983},
 primaryClass = {astro-ph.GA},
       adsurl = {https://ui.adsabs.harvard.edu/abs/2024A&A...688A.157S}
}

@software{eMCP,
       author = {{Moldon}, Javier},
        title = "{eMCP: e-MERLIN CASA pipeline}",
 howpublished = {Astrophysics Source Code Library, record ascl:2109.006},
         year = 2021,
        month = sep,
          eid = {ascl:2109.006},
archivePrefix = {ascl},
       eprint = {2109.006},
       adsurl = {https://ui.adsabs.harvard.edu/abs/2021ascl.soft09006M}
}

@ARTICLE{Munoz2025,
       author = {{Munoz-Arancibia}, A.},
        title = "{ZTF discovery of a re-brightening or new flaring event towards AT 2023uqm / ZTF23abikzfr ($z$ = 0.238)}",
      journal = {Transient Name Server AstroNote},
         year = 2025,
        month = apr,
       volume = {120},
        pages = {1},
       adsurl = {https://ui.adsabs.harvard.edu/abs/2025TNSAN.120....1M}
}

@ARTICLE{JiangNing2025QPE,
       author = {{Jiang}, Ning and {Pan}, Zhen},
        title = "{Embers of active galactic nuclei: tidal disruption events and quasiperiodic eruptions}",
      journal = {\apjl},
         year = 2025,
        month = apr,
       volume = {983},
       number = {1},
          eid = {L18},
        pages = {L18},
          doi = {10.3847/2041-8213/adc456},
archivePrefix = {arXiv},
       eprint = {2503.17609},
 primaryClass = {astro-ph.HE},
       adsurl = {https://ui.adsabs.harvard.edu/abs/2025ApJ...983L..18J}
}

@ARTICLE{Metzger2017EMRI,
       author = {{Metzger}, Brian D. and {Stone}, Nicholas C.},
        title = "{Periodic Accretion-powered Flares from Colliding EMRIs as TDE Imposters}",
      journal = {\apj},
         year = 2017,
        month = jul,
       volume = {844},
       number = {1},
          eid = {75},
        pages = {75},
          doi = {10.3847/1538-4357/aa7a16},
archivePrefix = {arXiv},
       eprint = {1705.00643},
 primaryClass = {astro-ph.HE},
       adsurl = {https://ui.adsabs.harvard.edu/abs/2017ApJ...844...75M}
}

@ARTICLE{Klimek2004NLSy1,
       author = {{Klimek}, Elizabeth S. and {Gaskell}, C. Martin and {Hedrick}, Cecelia H.},
        title = "{Optical Variability of Narrow-Line Seyfert 1 Galaxies}",
      journal = {\apj},
         year = 2004,
        month = jul,
       volume = {609},
       number = {1},
        pages = {69-79},
          doi = {10.1086/420809},
archivePrefix = {arXiv},
       eprint = {astro-ph/0403334},
 primaryClass = {astro-ph},
       adsurl = {https://ui.adsabs.harvard.edu/abs/2004ApJ...609...69K}
}

@ARTICLE{AiYanLi2010,
       author = {{Ai}, Y.~L. and {Yuan}, W. and {Zhou}, H.~Y. and {Wang}, T.~G. and {Dong}, X.-B. and {Wang}, J.~G. and {Lu}, H.~L.},
        title = "{Dependence of the Optical/Ultraviolet Variability on the Emission-line Properties and Eddington Ratio in Active Galactic Nuclei}",
      journal = {\apjl},
         year = 2010,
        month = jun,
       volume = {716},
       number = {1},
        pages = {L31-L35},
          doi = {10.1088/2041-8205/716/1/L31},
archivePrefix = {arXiv},
       eprint = {1005.0901},
 primaryClass = {astro-ph.CO},
       adsurl = {https://ui.adsabs.harvard.edu/abs/2010ApJ...716L..31A}
}

@ARTICLE{AiYanLi2013,
       author = {{Ai}, Y.~L. and {Yuan}, W. and {Zhou}, H. and {Wang}, T.~G. and {Dong}, X.-B. and {Wang}, J.~G. and {Lu}, H.~L.},
        title = "{A Comparative Study of Optical/Ultraviolet Variability of Narrow-line Seyfert 1 and Broad-line Seyfert 1 Active Galactic Nuclei}",
      journal = {\aj},
         year = 2013,
        month = apr,
       volume = {145},
       number = {4},
          eid = {90},
        pages = {90},
          doi = {10.1088/0004-6256/145/4/90},
archivePrefix = {arXiv},
       eprint = {1301.4739},
 primaryClass = {astro-ph.CO},
       adsurl = {https://ui.adsabs.harvard.edu/abs/2013AJ....145...90A}
}

@ARTICLE{OJ2871988,
       author = {{Sillanpaa}, A. and {Haarala}, S. and {Valtonen}, M.~J. and {Sundelius}, B. and {Byrd}, G.~G.},
        title = "{OJ 287: Binary Pair of Supermassive Black Holes}",
      journal = {\apj},
         year = 1988,
        month = feb,
       volume = {325},
        pages = {628},
          doi = {10.1086/166033},
       adsurl = {https://ui.adsabs.harvard.edu/abs/1988ApJ...325..628S}
}

@ARTICLE{Chen2024QPOblazar,
       author = {{Chen}, Zhao-Hao and {Jiang}, Yunguo},
        title = "{Quasi-periodic oscillation analysis for a sample of blazars at the optical band}",
      journal = {\aap},
         year = 2024,
        month = sep,
       volume = {689},
          eid = {A35},
        pages = {A35},
          doi = {10.1051/0004-6361/202347593},
       adsurl = {https://ui.adsabs.harvard.edu/abs/2024A&A...689A..35C}
}

@ARTICLE{Panic2024,
       author = {{Ward}, Charlotte and {Gezari}, Suvi and {Nugent}, Peter and {Kerr}, Matthew and {Eracleous}, Michael and {Frederick}, Sara and {Hammerstein}, Erica and {Graham}, Matthew J. and {van Velzen}, Sjoert and {Kasliwal}, Mansi M. and {Laher}, Russ R. and {Masci}, Frank J. and {Purdum}, Josiah and {Racine}, Benjamin and {Smith}, Roger},
        title = "{Panic at the ISCO: Time-varying Double-peaked Broad Lines from Evolving Accretion Disks Are Common among Optically Variable AGNs}",
      journal = {\apj},
         year = 2024,
        month = feb,
       volume = {961},
       number = {2},
          eid = {172},
        pages = {172},
          doi = {10.3847/1538-4357/ad147d},
archivePrefix = {arXiv},
       eprint = {2309.02516},
 primaryClass = {astro-ph.GA},
       adsurl = {https://ui.adsabs.harvard.edu/abs/2024ApJ...961..172W}
}

@ARTICLE{Chen2024QPOquasar,
       author = {{Chen}, Yong-Jie and {Zhai}, Shuo and {Liu}, Jun-Rong and {Guo}, Wei-Jian and {Peng}, Yue-Chang and {Li}, Yan-Rong and {Songsheng}, Yu-Yang and {Du}, Pu and {Hu}, Chen and {Wang}, Jian-Min},
        title = "{Searching for quasar candidates with periodic variations from the Zwicky Transient Facility: results and implications}",
      journal = {\mnras},
         year = 2024,
        month = feb,
       volume = {527},
       number = {4},
        pages = {12154-12177},
          doi = {10.1093/mnras/stad3981},
archivePrefix = {arXiv},
       eprint = {2206.11497},
 primaryClass = {astro-ph.GA},
       adsurl = {https://ui.adsabs.harvard.edu/abs/2024MNRAS.52712154C}
}

@ARTICLE{Graham2015opticalQPO,
       author = {{Graham}, Matthew J. and {Djorgovski}, S.~G. and {Stern}, Daniel and {Drake}, Andrew J. and {Mahabal}, Ashish A. and {Donalek}, Ciro and {Glikman}, Eilat and {Larson}, Steve and {Christensen}, Eric},
        title = "{A systematic search for close supermassive black hole binaries in the Catalina Real-time Transient Survey}",
      journal = {\mnras},
         year = 2015,
        month = oct,
       volume = {453},
       number = {2},
        pages = {1562-1576},
          doi = {10.1093/mnras/stv1726},
archivePrefix = {arXiv},
       eprint = {1507.07603},
 primaryClass = {astro-ph.GA},
       adsurl = {https://ui.adsabs.harvard.edu/abs/2015MNRAS.453.1562G}
}

@ARTICLE{JiangNing2022TickTock,
       author = {{Jiang}, Ning and {Yang}, Huan and {Wang}, Tinggui and {Zhu}, Jiazheng and {Lyu}, Zhenwei and {Dou}, Liming and {Wang}, Yibo and {Wang}, Jianguo and {Pan}, Zhen and {Liu}, Hui and {Shu}, Xinwen and {Zheng}, Zhenya},
        title = "{Tick-Tock: The Imminent Merger of a Supermassive Black Hole Binary}",
      journal = {arXiv e-prints},
         year = 2022,
        month = jan,
          eid = {arXiv:2201.11633},
        pages = {arXiv:2201.11633},
          doi = {10.48550/arXiv.2201.11633},
archivePrefix = {arXiv},
       eprint = {2201.11633},
 primaryClass = {astro-ph.HE},
       adsurl = {https://ui.adsabs.harvard.edu/abs/2022arXiv220111633J}
}

@ARTICLE{Kaur2023diskInstability,
       author = {{Kaur}, Karamveer and {Stone}, Nicholas C. and {Gilbaum}, Shmuel},
        title = "{Magnetically dominated discs in tidal disruption events and quasi-periodic eruptions}",
      journal = {\mnras},
         year = 2023,
        month = sep,
       volume = {524},
       number = {1},
        pages = {1269-1290},
          doi = {10.1093/mnras/stad1894},
archivePrefix = {arXiv},
       eprint = {2211.00704},
 primaryClass = {astro-ph.HE},
       adsurl = {https://ui.adsabs.harvard.edu/abs/2023MNRAS.524.1269K}
}

@ARTICLE{PanXin2022,
       author = {{Pan}, Xin and {Li}, Shuang-Liang and {Cao}, Xinwu and {Miniutti}, Giovanni and {Gu}, Minfeng},
        title = "{A Disk Instability Model for the Quasi-periodic Eruptions of GSN 069}",
      journal = {\apjl},
         year = 2022,
        month = apr,
       volume = {928},
       number = {2},
          eid = {L18},
        pages = {L18},
          doi = {10.3847/2041-8213/ac5faf},
archivePrefix = {arXiv},
       eprint = {2203.12137},
 primaryClass = {astro-ph.GA},
       adsurl = {https://ui.adsabs.harvard.edu/abs/2022ApJ...928L..18P}
}

@ARTICLE{PanXin2021,
       author = {{Pan}, Xin and {Li}, Shuang-Liang and {Cao}, Xinwu},
        title = "{The Effects of Large-scale Magnetic Fields on the Model for Repeating Changing-look AGNs}",
      journal = {\apj},
         year = 2021,
        month = apr,
       volume = {910},
       number = {2},
          eid = {97},
        pages = {97},
          doi = {10.3847/1538-4357/abe766},
archivePrefix = {arXiv},
       eprint = {2103.00828},
 primaryClass = {astro-ph.GA},
       adsurl = {https://ui.adsabs.harvard.edu/abs/2021ApJ...910...97P}
}

@ARTICLE{Raj2021,
       author = {{Raj}, A. and {Nixon}, C.~J.},
        title = "{Disk Tearing: Implications for Black Hole Accretion and AGN Variability}",
      journal = {\apj},
         year = 2021,
        month = mar,
       volume = {909},
       number = {1},
          eid = {82},
        pages = {82},
          doi = {10.3847/1538-4357/abdc25},
archivePrefix = {arXiv},
       eprint = {2101.05825},
 primaryClass = {astro-ph.HE},
       adsurl = {https://ui.adsabs.harvard.edu/abs/2021ApJ...909...82R}
}

@ARTICLE{Dodd2025,
       author = {{Dodd}, Sierra A. and {Huang}, Xiaoshan and {Davis}, Shane W. and {Ramirez-Ruiz}, Enrico},
        title = "{Perturbing AGN Accretion Disks with Stars and Moderately Massive Black Holes: Implications for Changing-Look AGN and Quasi-Periodic Eruptions}",
      journal = {arXiv e-prints},
         year = 2025,
        month = jun,
          eid = {arXiv:2506.19900},
        pages = {arXiv:2506.19900},
          doi = {10.48550/arXiv.2506.19900},
archivePrefix = {arXiv},
       eprint = {2506.19900},
 primaryClass = {astro-ph.HE},
       adsurl = {https://ui.adsabs.harvard.edu/abs/2025arXiv250619900D}
}

@ARTICLE{HuangXiaoshan2025,
       author = {{Huang}, Xiao-Shan and {Linial}, Itai and {Jiang}, Yan-Fei},
        title = "{Multiband Emission from Star-Disk Collisions and Implications for Quasiperiodic Eruptions}",
      journal = {\apj},
         year = 2025,
        month = nov,
       volume = {993},
       number = {2},
          eid = {186},
        pages = {186},
          doi = {10.3847/1538-4357/ae07ca},
archivePrefix = {arXiv},
       eprint = {2506.11231},
 primaryClass = {astro-ph.HE},
       adsurl = {https://ui.adsabs.harvard.edu/abs/2025ApJ...993..186H}
}

@ARTICLE{Bardeen1972,
       author = {{Bardeen}, James M. and {Press}, William H. and {Teukolsky}, Saul A.},
        title = "{Rotating black holes: locally nonrotating frames, energy extraction, and scalar synchrotron radiation}",
      journal = {\apj},
         year = 1972,
        month = dec,
       volume = {178},
        pages = {347-370},
          doi = {10.1086/151796},
       adsurl = {https://ui.adsabs.harvard.edu/abs/1972ApJ...178..347B}
}

@ARTICLE{Hayasaki2013,
       author = {{Hayasaki}, Kimitake and {Stone}, Nicholas and {Loeb}, Abraham},
        title = "{Finite, intense accretion bursts from tidal disruption of stars on bound orbits}",
      journal = {\mnras},
         year = 2013,
        month = sep,
       volume = {434},
       number = {2},
        pages = {909-924},
          doi = {10.1093/mnras/stt871},
archivePrefix = {arXiv},
       eprint = {1210.1333},
 primaryClass = {astro-ph.HE},
       adsurl = {https://ui.adsabs.harvard.edu/abs/2013MNRAS.434..909H}
}

@ARTICLE{Hinkle2020Declinerate,
       author = {{Hinkle}, Jason T. and {Holoien}, Thomas W.-S. and {Shappee}, Benjamin. J. and {Auchettl}, Katie and {Kochanek}, Christopher S. and {Stanek}, K.~Z. and {Payne}, Anna V. and {Thompson}, Todd A.},
        title = "{Examining a peak-luminosity/decline-rate relationship for tidal disruption events}",
      journal = {\apjl},
         year = 2020,
        month = may,
       volume = {894},
       number = {1},
          eid = {L10},
        pages = {L10},
          doi = {10.3847/2041-8213/ab89a2},
archivePrefix = {arXiv},
       eprint = {2001.08215},
 primaryClass = {astro-ph.HE},
       adsurl = {https://ui.adsabs.harvard.edu/abs/2020ApJ...894L..10H}
}

@ARTICLE{Hoogendam2024,
       author = {{Hoogendam}, W.~B. and {Hinkle}, J.~T. and {Shappee}, B.~J. and {Auchettl}, K. and {Kochanek}, C.~S. and {Stanek}, K.~Z. and {Maksym}, W.~P. and {Tucker}, M.~A. and {Huber}, M.~E. and {Morrell}, N. and {Burns}, C.~R. and {Hey}, D. and {Holoien}, T.~W.-S. and {Prieto}, J.~L. and {Stritzinger}, M. and {Do}, A. and {Polin}, A. and {Ashall}, C. and {Brown}, P.~J. and {DerKacy}, J.~M. and {Ferrari}, L. and {Galbany}, L. and {Hsiao}, E.~Y. and {Kumar}, S. and {Lu}, J. and {Stevens}, C.~P.},
        title = "{Discovery and follow-up of ASASSN-23bd (AT 2023clx): the lowest redshift and luminosity optically selected tidal disruption event}",
      journal = {\mnras},
         year = 2024,
        month = jun,
       volume = {530},
       number = {4},
        pages = {4501-4518},
          doi = {10.1093/mnras/stae1121},
archivePrefix = {arXiv},
       eprint = {2401.05490},
 primaryClass = {astro-ph.HE},
       adsurl = {https://ui.adsabs.harvard.edu/abs/2024MNRAS.530.4501H}
}

@ARTICLE{Park2020,
       author = {{Park}, Gwanwoo and {Hayasaki}, Kimitake},
        title = "{Tidal disruption flares from stars on marginally bound and unbound orbits}",
      journal = {\apj},
         year = 2020,
        month = sep,
       volume = {900},
       number = {1},
          eid = {3},
        pages = {3},
          doi = {10.3847/1538-4357/ab9ebb},
archivePrefix = {arXiv},
       eprint = {2001.04548},
 primaryClass = {astro-ph.HE},
       adsurl = {https://ui.adsabs.harvard.edu/abs/2020ApJ...900....3P}

      
}

@ARTICLE{ZhongShiyan2025,
       author = {{Zhong}, Shi-Yan},
        title = "{Modeling the UV/optical light curve of re-brightening tidal disruption events}",
      journal = {\apj},
         year = 2025,
        month = apr,
       volume = {983},
       number = {2},
          eid = {131},
        pages = {131},
          doi = {10.3847/1538-4357/adc005},
archivePrefix = {arXiv},
       eprint = {2412.12549},
 primaryClass = {astro-ph.HE},
       adsurl = {https://ui.adsabs.harvard.edu/abs/2025ApJ...983..131Z}
}

@techreport{LickShane,
    author = {{Miller}, J. S. and {Stone}, R. P. S.},
    title = "{The Kast double spectrograph}",
    institution ={Santa Cruz: Lick Obs.} ,
    year = 1994,
    number= {66}
}

@ARTICLE{Filippenko1982,
       author = {{Filippenko}, A.~V.},
        title = "{The importance of atmospheric differential refraction in spectrophotometry.}",
      journal = {\pasp},
         year = 1982,
        month = aug,
       volume = {94},
        pages = {715-721},
          doi = {10.1086/131052},
       adsurl = {https://ui.adsabs.harvard.edu/abs/1982PASP...94..715F}
}

@Inbook{Yuan2022,
author="Yuan, Wei-Min
and Zhang, Chen
and Chen, Yong
and Ling, Zhi-Xing",
editor="Bambi, Cosimo
and Santangelo, Andrea",
title="The Einstein Probe mission",
bookTitle="Handbook of X-ray and Gamma-ray Astrophysics",
year="2022",
publisher="Springer Nature ",
address="Singapore",
pages="1--30",
isbn="978-981-16-4544-0",
doi="10.1007/978-981-16-4544-0_151-1",
url="https://doi.org/10.1007/978-981-16-4544-0_151-1"
}

@ARTICLE{Jasen2001,
       author = {{Jansen}, F. and {Lumb}, D. and {Altieri}, B. and {Clavel}, J. and {Ehle}, M. and {Erd}, C. and {Gabriel}, C. and {Guainazzi}, M. and {Gondoin}, P. and {Much}, R. and {Munoz}, R. and {Santos}, M. and {Schartel}, N. and {Texier}, D. and {Vacanti}, G.},
        title = "{XMM-Newton observatory. I. the spacecraft and operations}",
      journal = {\aap},
         year = 2001,
        month = jan,
       volume = {365},
        pages = {L1-L6},
          doi = {10.1051/0004-6361:20000036},
       adsurl = {https://ui.adsabs.harvard.edu/abs/2001A&A...365L...1J}
}

@ARTICLE{Kron1980,
       author = {{Kron}, R.~G.},
        title = "{Photometry of a complete sample of faint galaxies.}",
      journal = {\apjs},
         year = 1980,
        month = jun,
       volume = {43},
        pages = {305-325},
          doi = {10.1086/190669},
       adsurl = {https://ui.adsabs.harvard.edu/abs/1980ApJS...43..305K}
}

@ARTICLE{Buzzoni1984,
       author = {{Buzzoni}, B. and {Delabre}, B. and {Dekker}, H. and {Dodorico}, S. and {Enard}, D. and {Focardi}, P. and {Gustafsson}, B. and {Nees}, W. and {Paureau}, J. and {Reiss}, R.},
        title = "{The ESO Faint Object Spectrograph and Camera / EFOSC}",
      journal = {The Messenger},
         year = 1984,
        month = dec,
       volume = {38},
        pages = {9},
       adsurl = {https://ui.adsabs.harvard.edu/abs/1984Msngr..38....9B}
}


\end{document}


\begin{frontmatter}



\title{Supplementary materials} 


\end{frontmatter}
\section{Observations and reductions} 
\subsection{Optical/UV photometry} 
Multiband optical/UV photometric data were obtained from public surveys, including ZTF and ATLAS, as well as from our own follow-up campaign using Swift/UVOT, and the $2.5\,\rm m$ Wide Field Survey Telescope (WFST~\cite{WangTinggui2023}). The detailed data-reduction procedure is described below. 

\emph{ATLAS photometry}.
We first obtained point-spread-function (PSF) profile-fitting photometry from the ATLAS Forced Photometry Website\footnote{https://fallingstar-data.com/forcedphot/}. This forced-photometry method fits the PSF profile at user-specified coordinates on the differential images. For the retrieved ATLAS $o$- and $c$-band light curves, we filtered out exposures with sky backgrounds brighter than 18 mag$\, {\rm arcsec}^{-2}$, and manually eliminated any remaining outliers. Finally, we binned the light curves in 0.5-day intervals to improve the signal-to-noise ratio (S/N). The ATLAS survey offers the longest temporal coverage of the periodic outbursts, spanning from MJD 57300 to the present.

\emph{ZTF photometry}. 
We acquired the ZTF light curves of AT2023uqm through the ZTF forced-photometry service\footnote{https://ztfweb.ipac.caltech.edu/cgi-bin/requestForcedPhotometry.cgi} \citep{Masci2019}. The photometric data were filtered and the 
light curves were constructed following the criteria outlined by Masci et al.~\citep{ZFPS}. Also, the acquired light curves were binned each day to improve the S/N. 

\emph{WFST}. We also initiated a follow-up campaign with the WFST in the $u$, $g$, and $r$ bands, beginning on 2025 May 9 UTC, using an exposure time of $90\,\rm s$ per frame. Follow-up observations with WFST are particularly valuable when AT2023uqm fades below $\sim$20 mag, making detection with ZTF increasingly difficult. For image subtraction, we used Pan-STARRS~\citep{Chambers2016} $g$- and $r$-band stack images as reference templates, and employed {\tt HOTPANTS}~\citep{Becker2015} to perform the subtraction. PSF photometry was then conducted on the resulting difference images using the Photutils package from the Astropy~\citep{Astropy} for the $g$ and $r$ bands. For the $u$-band data, we performed aperture photometry using a $5''$-radius aperture, and subtracted the host-galaxy flux ($u = 20.31 \pm 0.22$ mag from SDSS DR17) to isolate the transient component.

\emph{Swift/UVOT}. 
We proposed several target-of-opportunity (ToO) observations (ToO IDs 21580, 21876, 22417, 22593, 22712, and 22835) to monitor the multiwavelength evolution using the UVOT onboard the Swift Observatory \citep{Roming2005}. All UVOT data were processed with HEASoft 6.35.1. Images were first summed via the \texttt{uvotimsum} task, and then source and background counts were extracted --- using a circular aperture of radius $10^{\prime\prime}$ for the source and a $30^{\prime\prime}$-radius aperture placed in a nearby source-free region for the background --- by running \texttt{uvotsource} to produce the final light curves.  The host-galaxy contribution was subtracted by removing the flux from the quiescent state (MJD 60640--60725), prior to the recent outburst.

\subsection{WISE mid-infrared photometry} 
Multiple mid-infrared (MIR) flares corresponding to the optical ones were also identified\footnote{https://www.wis-tns.org/astronotes/astronote/2025-121} in the light curves from the Near-Earth Object WISE Reactivation mission (NEOWISE-R~\cite{Mainzer2014}), the successor to the Wide-field Infrared Survey Explorer (WISE~\cite{Wright2010}). 
We obtained single-exposure photometry following the procedure described in Section~2.3 of Jiang et al.~\citep{MIRONGI}. The data were then binned every six-month intervals, and the average flux before $\rm MJD\,58500$ was used as a baseline to subtract the host-galaxy contribution. Distinct delayed infrared flares corresponding to the third and fourth optical outbursts were clearly detected. The earlier outbursts were too faint for a confident detection, and the most recent outburst occurred after the final available WISE data point.

\subsection{X-ray observations} 
\label{sec:Xray-obs}
We monitored the long-term X-ray evolution using the XRT \citep{Burrows2005}. The tasks \texttt{xrtpipeline} and \texttt{xrtproducts} were run to generate light curves and spectra. Source counts were extracted from a circular region of radius $47^{\prime\prime}$ centered on the target, while background counts were measured from an annular region with an inner radius of $100^{\prime\prime}$ and an outer radius of $200^{\prime\prime}$. For data in which the source was not significantly detected in a single observation, we derived a 3$\sigma$ upper limit on the X-ray flux using the WebPIMMS tool\footnote{https://heasarc.gsfc.nasa.gov/cgi-bin/Tools/w3pimms/w3pimms.pl} and assuming a power-law X-ray spectrum with a photon index of 1.75 \citep{Ricci2017}. To improve the S/N, we stacked the event files for the quiescent state with MJD 60651--60721 (ObsIDs 00018962001--00018962011) and UV/optical flare state with MJD 607936--60890 (ObsIDs 0018962011--0018962039). In these two stacked files, the source was detected above the 3$\sigma$ level using the \texttt{ximage} task. Stacked spectra for UV/optical quiescent and flare were extracted by \texttt{xselect}.

We also triggered a ToO observation with XMM-Newton, executed on MJD 60833 (ObsID 0971190301) for a total time of 28.5 ks. The XMM-Newton data were reduced with the Science Analysis System (SAS) version 22.1. We first ran \texttt{cifbuild} and \texttt{odfingest} to generate the calibration index file and ingest the observation data files, respectively. Finally, light curves and spectra were extracted with the \texttt{xmmextractor} task. 

We obtained follow-up observations with Einstein Probe (EP~\citep{Yuan2022}) and processed the resulting data for AT2023uqm using the Follow-up X-ray Telescope Data Analysis Software (FXTDAS) v1.20. Data reduction was performed with the \texttt{fxtchain} pipeline to produce calibrated light curves and spectra.

\subsection{Radio observations }
We observed AT2023uqm using e-MERLIN on 17 May and 20 September 2025, under the DDT program with code RR19007. The observations were performed at the C-band, centered at $\sim$5.1 GHz. For the first epoch, to solve for the time-dependent complex gains, we used the nearby phase calibration source J0743+1714, while the standard calibrator J1407+2827 and J1331+3030 was used as bandpass calibrator and to set the flux-density scale, respectively. The data were reduced and calibrated following standard e-MERLIN pipeline. The final cleaned map reaches a root-mean-square (RMS) noise of $\sim$20 $\mu$Jy\,beam$^{-1}$. We do not detect any radio emission component at the optical position of AT2023uqm, with a 5$\sigma$ upper limit of $\sim$100 $\mu$Jy. The second epoch also resulted in a nondetection against a residual RMS noise of 26$\,\rm \mu Jy$.

\subsection{Optical spectroscopy} 
Eight spectra were obtained of the target; the basic information is provided in Table~\ref{tab:specinfo}. One of the spectra, acquired during the fourth flare by Ramsden et al.~\citep{Ramsden2023} using the ESO Faint Object Spectrograph and Camera 2 (EFOSC2~\citep{Buzzoni1984}) mounted on the New Technology Telescope (NTT) on 2023-11-05 UTC, was retrieved from the Transient Name Server\footnote{https://www.wis-tns.org/object/2023uqm}. This spectrum is characterized primarily by Balmer emission lines with intermediate widths.

Seven spectra were obtained of AT2023uqm during its fifth flare. Three of these were taken with the Next Generation Palomar Spectrograph (NGPS~\cite{Jiang2018NGPS}) mounted on the Palomar 5\,m Hale Telescope. NGPS was designed to replace the older Double Spectrograph~\citep{Oke1982}, offering higher resolution and efficiency with a wavelength coverage of 3200--10,300~\AA. However, owing to the unavailability of the blue camera, the current coverage is limited to 5500--10,300 \AA. Since NGPS has only recently begun its scientific observations, no dedicated pipeline is yet available. As a result, in this paper we use the extracted one-dimensional (1D) spectra from the quick-look GUI of the observing system; this does not affect the results of the spectroscopic analysis. All spectra exhibit features similar to that taken during the previous outburst. 

One spectrum was obtained with the Low Resolution Imaging Spectrometer (LRIS; \citealp{1995PASP..107..375O}) on the Keck~I 10\,m telescope at the W. M. Keck Observatory on May 25, 2025 (MJD 60820.600); it has an atmospheric dispersion corrector. 
After bias and flat-field corrections, the flux beams in each red- and blue-arm spectral image were extracted following standard routines within IRAF\footnote{{IRAF} is distributed by the National Optical Astronomy Observatories, which are operated by the Association of Universities for Research in Astronomy, Inc., under cooperative agreement with the National Science Foundation (NSF).}~\citep{Tody1986, 1993ASPC...52..173T}. 
A typical RMS accuracy of $\sim$0.3\,\AA\ was achieved in the wavelength calibration. Spectroscopic data reduction using the \texttt{LPipe} \citep{2019PASP..131h4503P} pipeline provides a consistent result. 
Flux calibration was carried out using observations of spectrophotometric standard stars, namely BD+284211 for the blue arm and BD+174708 for the red arm, observed on the same night, at similar airmasses, and with an identical instrument configuration.

Two spectra were acquired with the Kast double spectrograph~\citep{LickShane} mounted on the 3~m Shane telescope at Lick Observatory, California.  These observations utilized the $2^{\prime\prime}$-wide slit, the D57 dichroic, the 600/4310 grism, and the 300/7500 grating.  This instrument configuration has a combined wavelength range of $\sim$ 3600--10,700$\,\AA$, and a spectral resolving power of $R \approx 800$.  To minimize slit losses caused by atmospheric dispersion~\citep{Filippenko1982}, the slit was oriented at or near the parallactic angle. The data were reduced following standard techniques for CCD processing and spectrum extraction~\citep{Silverman2012} utilizing standard IRAF routines and custom Python and IDL codes\footnote{https://github.com/ishivvers/TheKastShiv}.  Low-order polynomial fits to comparison-lamp spectra were used to calibrate the wavelength scale, and small adjustments derived from night-sky lines in the target frames were applied. The spectra were flux calibrated using observations of appropriate spectrophotometric standard stars observed on the same night, at similar airmasses, and with an identical instrument configuration.

We obtained one spectrum using the BINOSPEC spectrograph~\citep{Fabricant2019} mounted on the 6.5\,m MMT on 2025 August 30, in which a 270 ($R\approx 1400$) grating at a central wavelength of 6500~\AA\ and a 1$^{\prime\prime}$-wide slit were used. The data were reduced by the SAO staff using the standard Binospec IDL pipeline. We note that the spectral shape at the blue end may be slightly unreliable based on our experience, though this does not affect our scientific conclusions.

\section{Data analysis}
\subsection{Host SED fitting} 
We collected multiband photometry for the host galaxy of AT2023uqm from several archival sources, including GALEX~\citep{GALEX}, SDSS~\citep{York2000}, 2MASS~\citep{2MASS}, and WISE, as listed in Table~\ref{tab:host}. All photometry was measured within the Kron aperture~\citep{Kron1980} to ensure consistency in the spectral energy distribution (SED) construction. To model the host-galaxy SED, we utilized the Python package Code Investigating GALaxy Emission (CIGALE~\cite{CIGALE}), following the procedure outlined in Section 3.2 of Wang et al.~\citep{ASASSN18ap}, with the exception that we adopted the AGN model from Stalevski et al.~\citep{Stalevski2012,Stalevski2016}. The W3 and W4 photometry is contaminated by nearby galaxies and  therefore excluded from the analysis. The resulting fit, shown in Figure~\ref{fig:cigale}, is well described by a combination of stellar components and an AGN contribution. However, the necessity of including the AGN component, and the degree to which it contributes to the host galaxy’s SED, remains uncertain. This is because the W1 and W2 bands are also affected by contamination from the nearby source, although to a lesser extent than W3 and W4. In fact, the SED fitting favors no AGN contribution if these two bands are further excluded.

\subsection{Infrared echo} 
Delayed IR flares were detected in AT2023uqm, believed to be nuclear dust echoes in response to the intrinsic emission. Similar delayed dust echoes have been observed in other TDEs~\citep{JiangNing2016, vV2016, ATLAS17jrp, Short2023, Wu2025}, and can be used to estimate the nuclear dust scale and the intrinsic bolometric luminosity~\citep{JiangNing2025PS16dtm}. Adapting the dust-echo model from Jiang et al.~\citep{JiangNing2025PS16dtm}, we further simplified the model by assuming that the dust scale remains constant, unaffected by sublimation. Using this simplified model, we performed Markov chain Monte Carlo (MCMC) fitting with \emph{emcee} to model the infrared light curve, assuming the bolometric luminosity can be converted from monochromatic optical flux via an uncertain factor (i.e., $L_{\rm bol}=\lambda L_{\rm ZTF-r}$)\footnote{The parameters $\lambda$, $\epsilon_\Omega$ (solid angle covering factor of the dust distribution), and $\epsilon_{\rm dust}$ (the fraction of incident radiation reprocessed into IR radiation by the illuminated dust) are combined into a single fitting parameter, as they are degenerate with each other.} and applying a uniform prior. To obtain a continuous $L_{\text{ZTF-r}}$ for the IR light-curve modeling, we linearly interpolate between the observed $L_{\text{ZTF-r}}$ data points. The distribution of the posterior samples is shown in Figure~\ref{fig:IR-para}, with the best-fit IR light curve shown in Figure~\ref{fig:bestir}. The IR light curves are well modeled by a spherical dust shell with a radius of $\sim 0.025$ pc.

\subsection{X-ray data analysis}
We fitted the X-ray spectra from both Swift/XRT and XMM-Newton using the \texttt{xspec v12.15.0}. The Galactic column density was fixed at $6.22\times 10^{20}\,\text{cm}^{-2}$ \citep{HI4PI2016}, and we modeled the data with \texttt{tbabs*zashift*powerlaw}. For the stacked quiescent spectrum (see Section~\ref{sec:Xray-obs}), we obtained a photon index of $\Gamma=2.66^{+1.01}_{-0.90}$ (cstat/d.o.f. = 6.64/15, effective exposure time 22.49 ks) with an unabsorbed flux of $1.08\pm{0.28}\times 10^{-13}~\rm erg~s^{-1}~cm^{-2}$, corresponding to a luminosity of $(1.84\pm0.48) \times 10^{43}\,\rm erg\,s^{-1}$. For the flare state,  we derived $\Gamma=2.48^{+2.02}_{-1.79}$ (cstat/d.o.f. = 9.48/16, effective exposure time 48.39 ks) and the unabsorbed flux of $(2.83\pm{1.05})\times 10^{-14}~\rm erg~s^{-1}~cm^{-2}$. The XMM-Newton spectrum yielded  $\Gamma=1.49^{+0.96}_{-0.83}$ (cstat/d.o.f. = 13.55/12, effective exposure time 14.74 ks) and the unabsorbed flux of $(1.87\pm{0.54})\times 10^{-14}~\rm erg~s^{-1}~cm^{-2}$. 

\subsection{Spectroscopic analysis}
\subsubsection{Spectral decomposition and emission-line fitting }
For the P200/NGPS and low-S/N NTT/EFOSC spectra, we directly use a linear function to represent the underlying continuum when fitting the emission lines of interest. For the Shane/Kast spectra, as well as the high-quality Keck and MMT spectra, we model both the continuum and emission lines following the procedure outlined in Section 3.1.1 of Wang et al.~\citep{ASASSN18ap}, which is briefly summarized below. First, Galactic extinction was corrected using the dust map of Schlafly \& Finkbeiner~\citep{Schlafly2011} and the extinction curve from Fitzpatrick \citep{Fitzpatrick1999}. Next, the continuum was modeled as a combination of a blackbody component, host stellar templates from Lu et al.~\citep{Lu2006}, and \feii\ templates, using the TWFIT procedure\footnote{https://github.com/wybustc/twfit}. Prominent emission lines and telluric regions were masked during the fitting process. As shown in Figure \ref{fig:specfit}, this approach fits the overall continuum well but could not accurately match the strengths of each \feii\ line. The best-fit temperature for the blackbody is\footnote{This result is from the Keck spectrum, while the one given by the MMT spectrum is not useful owing to poor flux calibration at the blue end.}$ \sim 16,000\,\rm K$, consistent with results obtained from fitting the multiband photometry. 

After subtracting the modeled continuum from the spectra, each emission line was fitted using multiple Gaussian components\footnote{The line structure may reflect the geometry of the emitting region. However, detailed investigation of this possibility is beyond the scope of this paper.
 }, including one narrow component ($\rm FWHM<800\,km\ s^{-1}$) and up to three broad components ($\rm FWHM\,>\,1000\,km\ s^{-1}$). The number of broad components used for each line was determined via an F-test.  For lines outside the wavelength range of modeled continuum (e.g., \Bowenoiii), we locally approximated the continuum with a low-order polynomial during line fitting.

Strong \niii\ Bowen fluorescence lines (notably $\rm N\,\textsc{iii}\,\lambda4640$) are typically detected in the H+He spectral class of TDEs (e.g., \cite{Leloudas2019,vV2021ApJ}) and in the Bowen fluorescence flare (BFF; \cite{Trakhtenbrot2019NewClassFlare}), a phenomenon also potentially linked to TDEs. These lines are thought to be key discriminators between TDEs and imposters, as strong \niii\ emission may indicate nitrogen abundance enhancement from the disruption of a main-sequence star~\citep{Kochanek2016}. However, while strong Bowen fluorescence lines (e.g., \Bowenoiii) were detected in the Keck spectrum of AT2023uqm, the \niii$\,\lambda4640$ feature was blended with \feii\ and not prominently detected. Furthermore, our simple continuum fitting inadequately reproduced the \feii\ lines, preventing conclusive assessment of the \niii\ emission.


To improve the \feii\ line fitting, we used the Fully Automated pythoN tool for AGN Spectra (FANTASY; \cite{Ilic2020,Raki2022,Ilic2023}), which simultaneously fits the underlying broken power-law continuum and sets of predefined emission lines. The \feii\ model in FANTASY accounts for complex emission by grouping lines sharing the same lower energy level, with ratios constrained by transition oscillator strengths, building on the models of Kovavcevic et al.~\citep{Kovavcevic2010} and Shapovalova et al.~\citep{Shapovalova2012}. During the fitting, we included one narrow and one broad component for the \feii\ models, as well as for the coronal lines. An additional broad component was included for the Balmer lines, helium lines, and \niii\ line. Lines within the same set were constrained to have the same  width and velocity offset. Before fitting, we subtracted the starlight contribution derived from \texttt{TWFIT} and used prior emission-line fits to constrain initial parameters,  facilitating convergence to an optimized solution. As shown in Figure~\ref{fig:specfit} (middle panel), the model excellently reproduces the starlight-subtracted Keck spectrum, revealing a significant \niii\ emission component. Using FANTASY’s improved continuum and \feii\ models, we refitted the emission lines that are not strongly contaminated by the \feii\ component, allowing for more flexibility. For others, we directly used the results from FANTASY.

\subsubsection{Spectroscopic features, comparison, and evolution}
Intermediate-width Balmer emission lines were detected in all eight spectra acquired. As shown in Figure \ref{fig:specCompare}, these lines (FWHM $\approx 1300\,\rm km\,s^{-1}$ )\footnote{The emission-line fitting results typically reveal two kinematic components: a narrower one at $\sim 1100\,\rm km\,s^{-1}$ and a broader one at $\sim 2500\,\rm km\,s^{-1}$.} are narrower than those typically detected in TDEs, which usually exhibit FWHM values around 10,000$\,\rm km\,s^{-1}$ near the optical peak (e.g., \citep{vV2020SSR, Charalampopoulos2022}). Nevertheless, BFFs typically show similar intermediate-width components, as does the well-known repeating nucleus transient ASASSN-14ko. Both BFFs and the individual ASASSN-14ko occurred in galaxies with pre-outburst AGN activity, suggesting that the intermediate-width lines may originate from the original broad-line region. For AT2023uqm, weak AGN activity was plausible prior to the outburst, and the strong IR echoes also indicate a rich dust and gas environment around the nucleus. Hence, the intermediate-width component in this case is likely associated with the nuclear gas responding to the outburst. Considering that the spectra were taken during the fourth and fifth flares\footnote{As discussed in Section 3.2.5, there could be many more weaker outbursts.}, part of the emitting gas may also originate from material expelled by the previous outbursts, possibly in the form of an outflow
\footnote{Assuming recombination is the dominant process for \ha\ emission, the observed \ha\ luminosity can be written as $L_{\mathrm{H}\alpha}=hv_{\mathrm{H}\alpha}\alpha_{32} n_e n_p V= hv_{\mathrm{H\alpha}} \alpha_{32} n_e M_H/m_p$, where $\alpha_{32} $ is the effective recombination coefficient, $n_e$ and $n_p$ are respectively the electron and proton number densities, and $V$ is the emitting volume. The required hydrogen mass is then $M_{\rm H} = (m_p L_{\mathrm{H}\alpha}) / (hv_{H\alpha} \alpha_{32} n_e)\approx 0.63\,(10^{10}\, {\rm cm}^{-3}/n_e)\ {\rm M_\odot} $, where we adopt $L_{\mathrm{H}\alpha } = 2.6\times 10^{42}\, \rm erg\,s^{-1}$ (measured from the Keck spectrum) and $\alpha_{32}= 1.14\times 10^{-13} \, \rm cm^{-3}\,s^{-1}$(assuming Case B recombination at $T=10^4\,\rm K$; Osterbrock and Ferland~\citep{Osterbrock2006}).  Assuming a typical electron number density for the broad-line region, given that the gas expelled by the TDE is expected to reside at a similar location, we obtain $M_\mathrm{H}\approx 0.63\, \rm M_\odot $. However, this estimate carries substantial uncertainty owing to the difficulty in constraining $n_e$.}.

Strong \feii\ emission was also detected, particularly in the Keck spectrum,  where it is dominated by a broad component with FWHM $\rm \approx 2000 \, km \, s^{-1}$. \feii\ emission is commonly observed in the optical spectra of narrow-line Seyfert 1 galaxies (NLSY1s) and also in transients discovered in the same family~\citep{Frederick2021}. Among previously discovered TDEs, only PS1-10adi~\citep{Kankare2017,JiangNing2019,HeZhicheng2021} and PS16dtm~\citep{Blanchard2017,Petrushevska2023}, the most probable TDE candidates in AGN, show strong \feii\ emission. Additionally, weak \feii\ emission is present in the spectra of ASASSN-14ko (see Figure~\ref{fig:specCompare}). \feii\ emission is typically thought to originate from the inner region of the dusty torus, where iron in the dust grains is sublimated into the gas in response to the central outburst~\citep{HeZhicheng2021}. Based on the FWHM of \feii\ emission, the emitting region, under virial assumaption, is $\sim 0.01$ pc ($M_{\rm BH} = 10^7\,\rm M_\odot$). 
This scale is comparable to, but smaller than, the torus scale derived from IR echo modeling (see Section S2.2), consistent with the proposed scenario.

The Keck spectrum and NGPS spectrum obtained on June 19 also revealed high-ionization coronal lines, such as $\rm [Fe\, \textsc{x}]$, $\rm [Fe\, \textsc{xi}]$, and $\rm [Fe\, \textsc{vii}]$. Coronal lines indicate the presence of strong soft X-ray emission, and prominent coronal lines are typically considered indicators of TDEs~\cite{Komossa2008, WangTinggui2012}. Based on the emission-line fitting results, the luminosity of $\rm [Fe\, \textsc{x}]\, \lambda 6374$ was derived to be  $\sim \rm 10^{41} \, erg \, s^{-1}$, consistent with values observed in TDEs but too luminous for supernovae (see Figure 12 of Wang et al.~\citep{ASASSN18ap}).

Another interesting feature in the acquired spectra is the blueshifted \oiii\ emission with an offset of $\sim \rm 430 \, km\,s^{-1}$ relative to the narrow \oiii. This blueshifted component suggests the presence of a gas outflow, which helps resolve the potential discrepancy between the strong coronal lines and the nondetection of X-rays, as the outflow along the line of sight would absorb the central X-ray radiation\footnote{The absorption by the outflow expelled during the event depends sensitively on its geometry. Under a simple and conservative assumption of a quasispherical distribution, the column density along the line of sight can be estimated as $N_H \approx 0.1\, {\rm M_\odot} / (4\pi m_p R_h^2)\approx 10^{25}\,\rm cm^{-2}$, where $R_h$ is the blackbody radius inferred from the recent outburst. Given a typical X-ray luminosity of $10^{43-44}$ erg\,s$^{-1}$ for TDEs, the material expelled by the event itself is sufficient to significantly obscure the X-ray emission. }. 

Finally, we examine the spectral evolution. We first focus on the behavior of $\mathrm{H}\alpha$ and $\mathrm{H}\beta$, both of which are detected in all spectra. During the decay phase of the fifth flare, the normalized $\mathrm{H}\alpha$ profile shows no significant change across the five spectra, though it is narrower than in the NTT/EFOSC spectrum of the fourth flare, a difference potentially attributable to variations in instrumental resolution. 
To track the strength evolution, we used the narrow \oiii\ line as a constant flux reference. This calibration reveals a clear decline in the broad components of $\mathrm{H}\alpha$ and $\mathrm{H}\beta$ that correlates with the fading optical outburst (see Table~\ref{tab:hahbevolution}). While recent studies have suggested that the \oiii\ line in some targets may respond quickly to the central outburst in a year-long timescale (e.g., ~\citep{YangChenwei2017,Onori2022,ASASSN18ap,Ansky2024}), the decline (or even disappearance ) of high-order Balmer lines in the late-time MMT spectra supports the above result. Additionally, Bowen lines, coronal lines, and helium lines also nearly disappeared in the MMT spectra. This coordinated spectral evolution firmly establishes that these characteristic emission features are physically linked to the optical outburst.


\section{Discussion} 
This section provides supplementary details on the orbital evolution of the star, the requirement for giant stars in certain scenarios, and the application of the rpTDE model to these stars. Alternative scenarios for AT2023uqm, aside from the rpTDE interpretation, are also discussed and ultimately disfavored.


\subsection{The evolution of period and pericenter} 
As derived in Section 3.2.2, no significant period evolution is detected within current observational limits. In the rpTDE scenario, the orbital energy and, consequently, the period can be perturbed by several effects. First, tidal interaction between star and SMBH can inject energy into the star at the expense of orbital energy \citep{Li2013}. Second, the stellar remnant can gain orbital energy due to asymmetric mass loss \citep{Manukian2013}, particularly at large impact parameters, where mass loss becomes significant.


Numerical simulations indicate that the combined effect of these two processes typically results in a maximum loss of orbital energy amounting to only a few percent of the star’s binding energy \citep{Cufari2023, ChenJinHong2024}, on the order of  $0.01\,(GM_*/R_*)$. The energy gain due to mass loss scales as  $\sim (G \Delta M)/R_* $~\citep{ChenJinHong2024}, becoming significant in severe disruptions. Based on the mass loss estimated in Section 3.2.4, the total change in orbital energy from these two effects is approximately
\begin{align}
|\Delta E_{orb}|\lesssim 0.1 \frac{GM_*}{R_*}\approx 1.9\times10^{14} \frac{M_*}{M_\odot} \frac{R_\odot}{R_*}\,\rm erg\,g^{-1}\, .
\end{align}
In combination with the orbital energy derived in Section 4.1, the corresponding rate of period evolution is $|\dot{P}| \approx (3/2)(|\Delta E_{\rm orb}|/E_{\rm orb}) \approx 3 \times 10^{-4}$, which is below the observational uncertainty derived in Section 3.2.2 (see Table~\ref{tab:period} ) and thus consistent with the data.

In addition to these effects, the orbital period can evolve due to cumulative encounters with other stars in the stellar cusp around the SMBH or to interactions with an accretion disk. The former effect is negligible owing to the small orbital semimajor axis. The latter has been explored in the context of QPEs and ASASSN-14ko \citep{Linial2023, Linial2024}, with a predicted period change rate given by
\begin{align} 
\dot{P}&\approx -\frac{3\pi \Sigma_{\rm disk}(r_p) R_*^2a}{M_*r_p} \\
& \approx 3\times10^{-5}\frac{\Sigma_{\rm disk}}{10^{4}\, \rm g\,cm^{-2}}*\frac{\beta}{0.6}\left(\frac{R_*}{\rm R_\odot}\right)\left(\frac{M_*}{1\,\rm M_\odot}\right)^{-2/3} 
\end{align}

If the involved disk originated from previous pTDE processes, the disk surface density ($\Sigma_{\rm disk} \approx \Delta M/(\pi r_p^2) \lesssim 0.1\rm\, M_\odot/(\pi r_p^2) \approx 2\times10^{4}\,\rm g\,cm^{-2}$) would be low enough to produce a negligible period evolution rate. Alternatively, considering the possibility of weak AGN activity prior to the outburst, we adopt a standard $\alpha$-disk with $\alpha = 0.1$, $H/r \approx 0.1$, and an accretion rate of $0.01 \dot{M}_{\rm Edd}$. 
In this case, for a black hole mass of $10^{7}\,\rm M_\odot$ (the fiducial value used) and a solar-like star,  the disk surface density at $r_p$ is about $\Sigma_{\rm disk}(r_p) \approx 1 \times 10^2 \, \rm g\,cm^{-2}$, resulting in a negligible period evolution rate.



Now consider the possible evolution of the pericenter distance. Orbital angular momentum can be transferred into the star via tidal interaction, but the maximum angular momentum a star can retain is far below the orbital angular momentum, typically several orders of magnitude lower \citep{MacLeod2013}. Star-disk collisions could also, in principle, reduce orbital angular momentum, but the expected effect is on the order of $r_p/(3a)$ times the period evolution rate and can thus be safely neglected. Therefore, the pericenter distance can be treated as approximately constant.

Finally, we consider the case of a giant star, which is required in some scenarios (Sections 5.1 and S3.2). The results indicate that, similarly, orbital evolution and changes in pericenter distance are negligible across the observed outbursts.

\subsection{Light-curve morphology and the star's radius} 
\label{sec:StarRadius}
As mentioned in the main text, if the light curve closely tracks the fallback rate at pericenter, the extreme observed ratio of peak separation to the period ($\Delta t_{\rm peak}/P$) would require a star with a radius tens of times larger than that of a main-sequence star --- thus, a giant star. We illustrate this in detail here and further investigate the evolution of the star’s radius under the same assumption.

First, based on the discussion in Section 5.1, the ratio of peak separation to the period can be expressed as
\begin{align}
    \frac{\Delta t_{\rm peak} } {P} & = [t_{fb}(E_{\rm orb}+dE)- t_{\rm fb}(E_{\rm orb}-dE)]/P\\
     & \approx 2.39 \left(\frac{f_E}{1}\right) \left(\frac{10^7 {\rm M}_\odot}{M_{\rm BH}}\right)^{1/3} \left(\frac{M_*}{{\rm M}_\odot}\right)^{2/3}\left(\frac{{\rm R}_\odot}{R_*}\right)\left(\frac{\beta}{0.6}\right)^2  \label{eq:Pratio} \\
     & \approx 2.39 \left(\frac{f_E}{1}\right) \left(\frac{10^7 {\rm M}_\odot}{M_{\rm BH}}\right)^{1/3}   \left(\frac{M_*}{{\rm M}_\odot}\right)^{\frac{2}{3}-q} \left(\frac{\beta}{0.6}\right)^2\, . \label{eq:Pratio-mr}
\end{align}
Equations \ref{eq:Pratio} and \ref{eq:Pratio-mr} assume that $dE \ll E_{\rm orb}$, since the duration between the two peaks of each flare is much shorter than the orbital period. Equation \eqref{eq:Pratio-mr} further adopts a mass-radius relation for the star of the form $R_*/{\rm R}_\odot = (M_*/{\rm M}_\odot)^q$. In the case of a solar-like star disrupted by a $10^7\,\rm M_\odot$ BH with an impact parameter of 0.6 and $f_E \approx 1$, the ratio is\footnote{Here, the $dE \ll E_{\rm orb}$ condition is violated, leading to an actual value exceeding the estimated one, thus worsening the discrepancy noted in the text.} about 2.39, much higher than the observed ratio of 0.038. For typical main-sequence stars, using the empirical mass-radius relation \cite{Demircan1991}, $q$ is 0.945 for $M_* < 1.66\,{\rm M}_\odot$ and 0.555 for larger mass. Hence, the minimum value of $\Delta t_{\rm peak}/P$ occurs for a main-sequence star with a mass of 1.66 $\rm M_\odot$, yielding a value of about 2, exceeding the observed ratio. Therefore, a main-sequence star is disfavored under the assumption that the light curve closely tracks the fallback rate, except in the unlikely case of an extremely low SMBH mass and $f_E$. According to Equation \eqref{eq:Pratio}, the most promising scenario is to expand the star’s radius by tens to hundreds of times that of a main-sequence star (i.e., an evolved giant star). Such a star is considered more appropriate for producing repeated pTDEs and has been proposed as the star that was disrupted in ASASSN-14ko~\citep{LiuChang2023}. 

Before reaching a conclusion for the above results, an additional factor, the stellar spin, must be considered, as it can also influence the energy spread of the debris \citep{Golightly2019}. In particular, a retrograde spin relative to the orbital angular momentum would narrow the energy spread, although it would also make the star more resistant to tidal disruption. However, observations show that most stars rotate at less than 10\% of their breakup velocity \citep{McQuillan2014},  corresponding to a maximum modification of the energy spread of $\sim 10$\%. Additionally, tidal interactions with the SMBH can spin up the star, but in the prograde direction, which would increase the energy spread rather than decrease it. Therefore, even when stellar spin is taken into account, a main-sequence star remains disfavored. 

To account for all relevant parameters simultaneously, we perform an MCMC 
fitting to the observed peak positions using the Python package \emph{emcee}. During the fitting procedure, we excluded the poorly sampled third and fifth flares. Table S4 lists the parameters considered along with their prior and posterior distributions.  Specifically, we set the star’s original radius before the first observed flare as $R_* = 10^{r_{\rm star}}R_{\rm MS}$, where $R_{\rm MS}$ is the radius of a main-sequence star with a given mass based on the empirical relation from Demircan \& Kahraman~\citep{Demircan1991}. We assume that the stellar radius changes by a factor of $f_{\rm star}$ during each passage, with $f_{\rm star}$ held constant for all passages. The factor relating half of the energy spread between the fallback peaks, $f_E$, is also assumed to be the same across multiple passages. Additionally, we require the orbital eccentricity to be $\geq 0.5$ to exclude nearly circular orbits and the unexpected case where $r_p > a_{\rm orb}$. Owing to the uncertainty in the black hole mass (see Section 3.1), we adopted a conservatively wide prior range of $10^6$--$10^9\,\rm M_\odot$. Based on the simulation results of Guillochon et al.~\citep{Guillochon2013}, we set the prior range for $\beta$ as [0.5, 2], below which the stripped mass would be negligible, and above which most stars would be completely disrupted. A loose prior range of [0.5, 2] was also adopted for $f_E$, following results of simulations (e.g., \cite{Ryu2020pTDE}).

The fitting results are shown in the left panel of Figure~\ref{fig:mcmcfitting}, the corresponding corner plots are shown in Figure~\ref{fig:corner-peak}, and the posterior distributions are summarized in Table S4. The fit is insensitive to the stellar mass while favoring a radius much larger than that of a typical main-sequence star. Within the  90\% confidence interval, we find that $r_{\rm star}=1.41^{+0.38}_{-0.43}$, corresponding to $R_*= 26^{+36}_{-16}\, R_{\rm MS}$, consistent with the evolved giant-star scenario described above. 
We find that the parameters $\beta$, $M_{\rm BH}$, and $f_E$ did not converge, instead tending toward one-sided values that produce low $\Delta t_{\rm peak}/P$, despite the conservatively wide priors adopted. This lack of convergence likely stems from an oversimplified model. For example, very low $\beta$ values are disfavored as they would not produce sufficient mass loss to power the observed high luminosity, but our fitting focuses solely on the peak timing. Furthermore, the uniform priors may be unsuitable for several parameters. These issues should be addressed in future dedicated work. Given the wide parameter ranges explored, our core conclusion that the progenitor was a giant star (assuming light curves closely track fallback rate) remains robust. 

Furthermore, we attempt to assess the radius evolution of the star across multiple pTDE passages. The above MCMC fitting to the timing of the peaks yields a best-fit stellar expansion factor of $1.00^{+0.02}_{-0.02}$. Nevertheless, assuming a constant $f_E$ for different passages is inaccurate. In reality, according to its definition, $f_E$ should decrease as mass is increasingly being stripped. Consequently, we refine our method to utilize flare boundaries instead of peaks. In this case, $f_E$, representing half of the total debris energy spread relative to the typical value $(GM_{\rm BH}/r_p^2)R_*$, tends to stay consistent across various flares, regardless of the radiation timescale from fallback material. The optimal fit outcomes with this refined approach are detailed in Table S4 and depicted in Figures~\ref{fig:mcmcfitting} and \ref{fig:corner-boundary}. The revised fit yields an updated factor 
$ 1.03^{+0.016}_{-0.017}$, which is consistent with the runaway manner of the mass-loss evolution.

\subsection{The giant-star case} 

\subsubsection{The origin of the giant star?} 
As we discussed in Section 5 and carefully illustrated in Section S3.2, if the light curve closely tracks the fallback rate at pericenter, the extreme observed ratio of peak time separation to period would demand the star disrupted as a giant star with a radius at least ten times that of the main-sequence star. However, for a less massive black hole ($<10^8\,\rm M_\odot$), giants only can contribute up to 10\% of the observable tidal disruption flares~\citep{MacLeod2012GaiantTDE}. 
In this section, we investigate whether a main-sequence star could achieve the large radius via alternative mechanisms, such as star-disk collisions~\citep{YaoPhilippe2025} or tidal heating~\citep{Li2013}, rather than through standard evolutionary processes. 

As explored by some works~\citep{YaoPhilippe2025}, repeated star-disk collisions can inflate a small portion of a main-sequence star’s outer layers, resulting in a large envelope. However, such an inflated envelope typically lacks sufficient mass to account for the luminous flares observed in AT2023uqm\footnote{For instance, in a heated solar-type star, only about $10^{-4}\, \rm M_\odot$ lies at $R > 1.1\,{\rm R}_\odot$~\citep{YaoPhilippe2025}, which is far too little to power the observed emission.}. In the tidal heating scenario, a star scattered into orbits with a pericenter greater than its tidal disruption radius can expand to a larger size and eventually be disrupted due to tidal interactions with the SMBH after many orbits~\citep{Li2013}. The tidal heating effect is highly sensitive to $\beta$, with stars having small values of $\beta$ experiencing very weak tidal interactions. Typically, an initial $\beta_{ini} \gtrsim 0.2$ is required for a star to expand to several times its original radius over its lifetime~\citep{Li2013,PhyilippeYao2025TidallyHeated}. If the large radius required is primarily due to tidal heating,  Equation~\eqref{eq:Pratio} requires $\beta_{\rm ini} < 0.06$,  inconsistent with the limits imparted by the star's lifetime. Therefore, an evolved giant star remains the most plausible scenario. 

\subsubsection{The mass-loss evolution for the giant-star case} 
An evolved giant star typically has a dense core. However, this has a limited impact on the early-time evolution of mass loss since the extended envelope is much larger than the core. MacLeod et al.~\citep{MacLeod2013} investigated mass-loss evolution  during multiple pericenter passages for a 1.4 $\rm M_\odot$ giant star\footnote{The study did not explicitly simulate multiple passages. Instead, it models the stellar-structure response to mass loss through winds and estimates mass-loss evolution using empirical relations.} and found that mass loss initially increases in a runaway manner, followed by a decline as the stellar core’s gravity becomes dominant. Consequently, a giant progenitor, inferred from the extreme ratio of flare duration to recurrence period, also explains the observed trend of increasing mass loss in AT2023uqm. Moreover, this implies that both the mass loss and flare luminosity should eventually peak and then decline. Finding this transition would strongly indicate the disruption of a giant star. However, whether a declining phase occurs under various initial conditions remains uncertain and requires further numerical simulations. 


Using the empirical relations provided in Equations (3) and (4) of MacLeod et al.~\citep{MacLeod2013}, along with the stellar radius expansion factor derived in Section S3.2, we plot the temporal evolution of mass loss for different initial impact parameters\footnote{The initial impact parameter is associated with the first flare observed by ZTF.}  in Figure~S15. The results indicate that a grazing encounter with an initial $\beta \approx 0.51$ best matches the observed temporal behavior of AT2023uqm. Additionally, the absolute mass loss (assuming $\eta=0.01$) is well reproduced by a progenitor with a stellar mass of $3 \, \rm M_\odot$ and a core-to-envelope mass ratio of 0.2. This is a reasonable value, falling within the expected mass range. However, large uncertainties still exist in the empirical relation.
\subsubsection{The low orbital eccentricity}

Here, we attempt to constrain the orbital eccentricity of the giant star, which can be expressed as
\begin{align}
    e & = 1- \frac{r_p}{a_{\rm orb}} \\ 
      & \approx 1- 0.44\frac{f_E}{1}\frac{\beta}{0.6}\left(\frac{M_*/M_{\rm BH}}{10^{-7}}\right)^{1/3}\frac{P/\Delta t_{\rm peak}}{26.3} \, ,
\end{align}
where we have substituted Equation~\eqref{eq:Pratio}. This yields a relatively low eccentricity. We also compute the posterior distribution of eccentricity using the results from the two MCMC fits discussed above. Similar low eccentricities around 0.7 (median value; see Figure S10) are obtained, suggesting a moderately circularized orbit. TDEs on such eccentric orbits would keep all the debris bound~\citep{Hayasaki2013}.

Traditionally, stars are thought to be scattered into the vicinity of the SMBH via two-body relaxation in the nuclear stellar cluster~\citep{Frank1976,ZhongShiyan2023}, which typically produces nearly parabolic orbits. An alternative formation channel is the Hills mechanism, in which the tidal breakup of stellar binaries also generally produces highly eccentric orbits ($\gtrsim 0.99$;~\citep{Cufari2022}). However, lower-eccentricity orbits may arise from the breakup of tight stellar binaries~\citep{LuQuataert2023}. Pre-existing AGN activity may also facilitate the production of stars on low-eccentricity orbits~\citep{JiangNing2025QPE}. Additionally, repeated star-disk collisions can dissipate orbital energy and circularize stellar orbits. A detailed exploration of whether such an orbit can be formed through different channels is beyond the scope of this work and will be presented in a dedicated paper. Given the short orbital period and low eccentricity, gravitational-wave (GW) radiation may play an important role in the late stages of orbital evolution. Because the GW decay timescale is much longer than the orbital period, a series of much weaker flares may have occurred prior to the first flare detected by ZTF. We indeed find tentative evidence for such activity in the historical light curve (see Section 3.2.5).

It is worth noting that the acquired eccentricities are based on the assumption that the light curve closely tracks the fallback rate at pericenter, a scenario that necessitates a giant star. As discussed in Section 5, the presence of an accretion disk would relax the constraints on the star's radius requirements. Thus, higher orbital eccentricities can be achieved for giant stars with smaller radius.

\subsection{Alternative scenarios}
Given the strong periodicity observed in AT2023uqm, we primarily explore scenarios involving an SMBH with a companion, for which the orbital period could naturally account for the observed periodicity. Alternatively, in light of its spectroscopic similarity to NLSy1 galaxies, we also investigate whether intrinsic AGN variability could produce a similar event, although the AGN activity prior to the event appears to be weak (see Section~3.1).

\subsubsection{SMBH with a companion}
First, we consider a star orbiting the SMBH, a system known as a stellar extreme mass-ratio inspiral (EMRI). Besides the transient accretion powered by the violent tidal disruption of the star, steady accretion can also be established through Roche-lobe overflow if the star is on a nearly circular orbit. Periodic collisions could occur between consecutive EMRIs, specifically between a star spiraling in and another migrating outward, which would enhance mass loss and produce periodic accretion flares. However, the period of such flares would be at least decades, which is too long to align with the observations of AT2023uqm~\citep{Metzger2017EMRI}.

On the other hand, if the stellar companion exists in an AGN with a pre-existing accretion disk\footnote{Alternatively, the disk may be fed by tidal disruption events or by Roche-lobe overflow from a star in a circular orbit, rather than being associated with an AGN.}, a misaligned orbit would naturally lead to periodic star-disk interactions. These interactions could produce periodic flares through three main channels: direct star-disk interaction (e.g., ~\citep{Linial2023}), collisions between stripped stellar debris and the accretion disk (e.g., ~\citep{LuQuataert2023, YaoPhilippe2025, Linial2025}), and enhanced accretion triggered by the star's perturbation of the disk (e.g., ~\cite{Dodd2025}). However, the large flare energy and dominant optical-band emission observed in the AT2023uqm outburst are generally incompatible with direct star-disk interaction~\citep{Linial2023,HuangXiaoshan2025}. Even when considering the more luminous process powered by collisions between stripped stellar debris and the disk, such a channel typically also cannot produce optical outbursts as luminous as AT2023uqm~\citep{Linial2025}, though it remains uncertain whether certain conditions could give rise to the observed properties. Furthermore, in this scenario, the mass loss stripped from the star would quickly saturate to a constant value over multiple collisions~\citep{YaoPhilippe2025}, in contrast to the observed runaway increase in flare strength in AT2023uqm. As for the enhanced accretion scenario driven by spiral density waves excited by star-disk interactions, reproducing an amplitude as large as that of AT2023uqm would require the star's pericenter to lie near the tidal disruption radius~\citep{Dodd2025}. Whether such a rapid and powerful outburst could occur before the star is tidally disrupted requires further investigation through dedicated simulations. Moreover, this scenario does not naturally explain the exponentially increasing flare energy seen in AT2023uqm, given the nearly unchanged stellar, orbital, and disk properties. We do note that the star would be heated and expand during the interaction, thereby increasing its perturbation cross section for subsequent encounters; however, this effect would quickly converge to a constant value within less than four collisions~\citep{YaoPhilippe2025}. 

Beyond a stellar companion to the central SMBH, a massive black hole companion could more naturally account for the required high luminosity by perturbing a pre-existing accretion disk, owing to its larger interaction cross-section. However, as noted above, this scenario also fails to reproduce the specific flare strength evolution behavior observed in AT2023uqm. To match the large amplitude of the event, the companion black hole would need to have a mass at least that of an IMBH. Furthermore, a black hole binary with such a mass range and the short orbital separation required would be short-lived, existing only briefly before merging due to GW emission (see detailed discussion in \cite{Payne2021,Dodd2025}). This makes it an unlikely origin for AT2023uqm. Finally, if the companion were of similar mass to the central SMBH, an additional problem arises: such a binary would be expected to produce a significant peculiar velocity shift in the spectra, which is not observed.

\subsubsection{AGN variability}
Given the spectroscopic similarity of AT2023uqm to NLSy1s, we consider here whether AGN-related variability could be its origin. Compared to AGNs with broader emission lines, NLSy1s tend to exhibit smaller variability amplitudes~\citep{Klimek2004NLSy1, AiYanLi2010, AiYanLi2013}. Leaving aside the periodicity for the moment, the large amplitude (about 1 mag for the recent flare), short timescale of months, and smooth light curve observed in AT2023uqm are difficult to explain with normal AGN variability. Recently, ambiguous nuclear transients with large amplitudes and smooth, flare-like light curves have also been found in the NLSy1 family~\citep{Frederick2019}. However, these were more likely attributed to tidal disruption events, although AGN variability cannot be completely ruled out.

Regarding periodicity, periodic variability in AGNs has attracted significant attention owing to its potential link to SMBH binary candidates. Indeed, quasiperiodic oscillations (QPOs) on timescales of months to years have been observed in the optical emission of blazars (e.g., ~\citep{OJ2871988, Chen2024QPOblazar}) and normal quasars~\citep{Graham2015opticalQPO, JiangNing2022TickTock, Panic2024, Chen2024QPOquasar}). However, unlike AT2023uqm, these QPOs typically appear as sinusoidal functions, which might manifest as outbursts if limited by instrument detection thresholds. More importantly, such QPOs generally do not exhibit the same degree of perfect periodicity as AT2023uqm. They are often fractional and less smooth. Therefore, from an observational perspective, it is unlikely that AT2023uqm is caused by AGN activity.

On the theoretical side, although disk instability models (e.g., ~\cite{Sniegowska2020, PanXin2021, Raj2021, PanXin2022, Kaur2023diskInstability}) have been proposed to explain quasiperiodic behavior including repeating changing-look AGNs and X-ray QPEs, these mechanisms do not naturally account for the runaway manner observed in AT2023uqm. In summary, AGN variability is an unlikely origin for AT2023uqm, although it cannot be entirely excluded;  a more detailed investigation should be undertaken in future work.

\clearpage
\section{Supplementary Tables and Figures}
\label{sec1}

\begin{table}[htbp]
  \centering
  \scriptsize
  \begin{threeparttable}
    \caption{Gaussian fitting parameters}
    \label{tab:gaussianFitting}
    \begin{tabular}{ccccccccccccc}
      \toprule
Flare & $t_{\rm peak1}$ &  $t_{\rm peak2}$ &   $ \Delta t_{\rm peak}$ & $ t_{\rm dip}$ & $t1$ & $t2$ & $\sigma_{1}$ & $\sigma_{2}$ & $\rm  A1_{ZTFg}$ & $\rm A2_{ZTFg}$ &  $\rm A1_{ZTFr}$ & $\rm A2_{ZTFr}$ \\ 
\midrule
Flare1   & -8.2$\pm$1.1    &     9.4$\pm$0.6    &     17.6$\pm$1.3    &     -3.4$\pm$1.3    &     -8.4$\pm$1.2    &     9.4$\pm$0.6    &     2.0$\pm$0.8    &     13.3$\pm$0.4    &     1.6$\pm$0.5    &     1.9$\pm$0.1    &     1.0$\pm$0.3    &     1.2$\pm$0.0\\
Flare2   & -10.8$\pm$0.3    &     12.0$\pm$0.7    &     22.7$\pm$0.8    &     -1.6$\pm$0.6    &     -10.9$\pm$0.3    &     12.0$\pm$0.7    &     4.2$\pm$0.4    &     9.2$\pm$0.8    &     3.7$\pm$0.3    &     3.0$\pm$0.2    &     2.2$\pm$0.1    &     1.8$\pm$0.1\\
Flare3   & -14.2$\pm$0.4    &     7.2$\pm$2.0    &     21.4$\pm$2.0    &     1.4$\pm$0.3    &     -14.3$\pm$0.4    &     7.2$\pm$2.0    &     3.0$\pm$0.1    &     12.0$\pm$0.7    &     8.5$\pm$0.0    &     3.2$\pm$0.6    &     5.2$\pm$0.8    &     1.9$\pm$0.3\\
Flare4   & -15.9$\pm$0.2    &     0.7$\pm$0.7    &     16.5$\pm$0.7    &     -5.0$\pm$0.7    &     -16.9$\pm$0.2    &     1.0$\pm$0.7    &     5.0$\pm$0.2    &     13.3$\pm$0.6    &     7.2$\pm$0.3    &     6.8$\pm$0.3    &     4.4$\pm$0.2    &     4.2$\pm$0.2\\
Flare5   & ---    &     16.7$\pm$1.2    &    ---    &     -0.9$\pm$1.0 \tnote{*}   &     9.9$\pm$0.3    &     16.9$\pm$1.3    &     37.4$\pm$2.2    &     6.1$\pm$1.0    &     ---    &     ---    &     4.7$\pm$0.3    &     3.7$\pm$0.4\\
      \bottomrule
    \end{tabular}
    \begin{tablenotes}
      \item The first four columns ($t_{\rm peak1}$, $t_{\rm peak2}$, $\Delta t_{\rm peak}$, and $t_{\rm dip}$) list the times of the first peak, second peak, peak separation, and dip between the two peaks, respectively, as determined from the best-fitting light curve modeled with two Gaussian components: $A_1 e^{-(t-t_1)^2/(2\sigma_1^2)} + A_2 e^{-(t-t_2)^2/(2\sigma_2^2)}$. The corresponding best-fit Gaussian parameters are provided in the remaining columns. For the third flare, the sparse sampling and poor photometric quality result in highly unreliable fitting parameters; therefore, these results are not used in any subsequent analysis. For the fifth flare, although the first peak was not directly observed, a two-Gaussian model was still adopted to obtain the best-fitting light curve. In this case, the first Gaussian component is extremely broad and primarily accounts for the declining tail of the outburst, rather than representing the true contribution of the missed first peak. This does not affect our analysis, as only the timing properties derived from the best-fitting curve are used. All times listed in the table are relative to $\rm MJD=58666 + 527 \times (n_{\rm peak}-1)$.
      
      \item[*] For the fifth flare, the dip time was not obtained from the two-Gaussian fit to the full flare profile owing to the absence of the first peak. Instead, it was estimated using a second-order polynomial fit to the subset of data points bracketing the dip (see Section 3.2.1).
      
  \end{tablenotes}
  \end{threeparttable}
\end{table}


\begin{table}[htbp]
  \centering
  \scriptsize
  \begin{threeparttable}
    \caption{Period fitting results } 
    \label{tab:period}
    \begin{tabular}{cccccccc}
      \toprule
feature & method & $t_0$ & $P_0$  & $\dot{P}$ \\ 
      \midrule
$\rm t_{peak1}$ & func1&$58657.80\pm0.38$&$524.45\pm0.15$& --\\
$\rm t_{dip}$   & func1&$58663.76\pm2.10$&$526.75\pm0.87$& --\\
$\rm t_{peak2}$ & func1&$58675.93\pm4.95$&$525.76\pm2.39$& --\\
\hline  
$t_{peak1}$ & func2&$58657.81\pm1.15$&$524.42\pm1.58$&$0.00003\pm0.00155$\\
$t_{dip} $  & func2&$58664.90\pm4.06$&$524.86\pm5.01$&$0.0017\pm0.0044$\\
$t_{peak2}$ & func2&$58677.25\pm7.22$&$520.83\pm12.32$&$0.0053\pm0.0129$\\

      \bottomrule
    \end{tabular}
  \end{threeparttable}
\end{table}

\begin{table}[htbp]
  \centering
  \fontsize{6.8pt}{8.5pt}\selectfont
  \begin{threeparttable}
    \caption{Basic information for spectroscopic observations}
    \label{tab:specinfo}
    \begin{tabular}{cccccccc}
      \toprule
DATE & Instrument & Grating\tnote{1}  & Slit width & Exposure time & S/N\tnote{2} & Wavelength range & resolution\tnote{3}   \\  
 &   &    &  (arcsec)    &   (s)   & $\rm pixel^{-1}$  & (\AA)     &     (\AA)  \\ 
      \midrule
2023-11-05& NTT/EFOSC &    Gr\#13                          &  1.5   & 1200          &  4.7       & 3670--9270   &  23.01   \\ 
2025-05-21& P200/NGPS &    ---                             &  1.5   & 1200          &  ---       & 5630--10,235  &  1.71(R), 2.08(I) \\ 
2025-05-25& Keck/LRIS &   600/4000(blue), 400/8500(red)    &  1.0   & 1200  & 35         & 3150--10,245  &  3.8-4.1(blue), 6.9 (red) \\
2025-05-29& Shane/Kast&   600/4310(blue), 300/7500(red)    &  2.0   & 3660(blue),3600(red) & 5.8   &3624--10,754   &  4.7(blue),11.9(red) \\
2025-06-04& Shane/Kast&   600/4310(blue), 300/7500(red)    &  2.0   & 3660(blue),3600(red) & 12    &3620--10,750   &  4.7(blue),11.9(red) \\ 
2025-06-19& P200/NGPS &    ---                             &  1.5   & 1800  &  ---       & 5630--10,235  & 1.71(R), 2.08(I)  \\ 
2025-08-30& MMT/Binospec&  270                             &  1.0   & 3600          & 26.7       & 3800--8700   &  4.85 \\ 
2025-10-05& P200/NGPS &    ---                             &  1.5   & 2400          &  ---       & 5630--10,235  & 1.71(R), 2.08(I)  \\
      \bottomrule
    \end{tabular}
    \begin{tablenotes}
      \item[1]The P200/NGPS uses fixed volume phase holographic (VPH) gratings. 
      \item[2]The S/N was calculated around 6500 \AA\ in the observed frame. For the three P200/NGPS observations, we used quick-look results that did not produce the flux error, so no S/N values are provided here.
      \item[3]All resolutions are given for a slit width of 1$^{\prime\prime}$, except for P200/NGPS, which is given for a slit width of 0.56$^{\prime\prime}$.
    \end{tablenotes}
  \end{threeparttable}
\end{table}

\clearpage

\begin{table}[htbp]
  \centering
  \scriptsize
  \begin{threeparttable}
    \caption{Prior and posterior distributions of parameters for MCMC fitting} 
    \label{tab:MCMCfit}
    \begin{tabular}{cccccccc}
      \toprule
Parameter &      Prior &    Posterior    & Posterior-$3\sigma$   & Description \\
      \midrule
    \multicolumn{5}{c}{MCMC fitting for flare peaks } \\ 
      \midrule
$M_{\rm BH} $  &  [6, 9]    &   $8.5^{+0.4}_{-0.6}$  &     $ 8.5^{+0.5}_{-1.0}$                & Supermassive black hole mass \\ 
$m_{\rm star}$ &  [0.3, 15] &   $5.9^{+5.4}_{-3.5}$  &     $ 5.9^{+7.7}_{-4.3}$                 & $M_* = m_{\rm star}\,\rm  M_\odot$ \\ 
$r_{\rm star}$ &  [0  , 2]  &   $1.4^{+0.2}_{-0.3}$  &     $ 1.4^{+0.4}_{-0.4}$                 & $R_* = 10^{r_{\rm star}}R_{\rm MS}$ \\ 
$\beta$    &  [0.5, 2]  &   $0.8^{+0.4}_{-0.2}$  &     $ 0.8^{+0.6}_{-0.3}$                & The impact parameter for the first passage\\
$P_{\rm orb}$  &  [500,540] &  $524.0^{+0.2}_{-0.2}$ &     $ 524.0^{+0.3}_{-0.3}$               & The observed period  \\ 
$t_0$      &  [58640, 58690] &  $58667.4^{+0.3}_{-0.3}$  &  $ 58667.4^{+0.5}_{-0.5}$         & The MJD of the first flare \\ 
$f_{E}$    &  [0.5, 2]  &    $0.8^{+0.4}_{-0.2}$  &   $ 0.8^{+0.7}_{-0.3}$                 & The factor for energy spread (see text) \\ 
$f_{\rm star}$ &  [0.9,1.1] &  $1.00^{+0.02}_{-0.02}$   &  $ 1.00^{+0.03}_{-0.03}$                    & The expansion factor for the star's radius \\
      \midrule 
        \multicolumn{5}{c}{MCMC fitting for flare boundaries } \\ 
      \midrule 
$M_{BH} $  &  [6, 9]    &   $8.2^{+0.6}_{-0.8}$  &     $8.2^{+0.8}_{-1.3}$                & Supermassive black hole mass \\ 
$m_{star}$ &  [0.3, 15] &   $7.0^{+5.0}_{-4.3}$  &     $7.1^{+7.0}_{-5.8}$                 & $M_* = m_{\rm star}\,\rm  M_\odot$ \\ 
$r_{\rm star}$ &  [0  , 2]  &   $1.5^{+0.3}_{-0.3}$  &     $1.5^{+0.4}_{-0.6}$                 & $R_* = 10^{r_{\rm star}}R_{\rm MS}$ \\ 
$\beta$    &  [0.5, 2]  &   $1.3^{+0.4}_{-0.4}$  &     $1.3^{+0.6}_{-0.6}$               & The impact parameter for the first passage\\
$P_{orb}$  &  [500,540] &  $523.1^{+0.5}_{-0.6}$ &     $523.1^{+0.9}_{-0.9}$                & The observed period  \\ 
$t_0$      &  [58640, 58690] & $58679.5^{+1.1}_{-1.1}$  &   $58679.5^{+1.7}_{-1.8}$          & The MJD of the first flare \\ 
$f_{E}$    &  [0.5, 2]  &   $1.1^{+0.5}_{-0.4}$  &    $ 1.1^{+0.7}_{-0.5} $                 & The factor for energy spread (see text) \\ 
$f_{star}$ &  [0.9,1.1] &  $ 1.03^{+0.02}_{-0.02}$   &  $1.03^{+0.03}_{-0.03}$                     & The expansion factor for the star's radius \\
      \bottomrule
    \end{tabular}
    \begin{tablenotes}
    \item The values in this table have been rounded to reflect the actual uncertainties and are consistent with those displayed in Figures~\ref{fig:corner-peak} and \ref{fig:corner-boundary}.
    \end{tablenotes}
  \end{threeparttable}
\end{table}




\begin{table}[htbp]
  \centering
  \scriptsize
  \begin{threeparttable}
    \caption{Host SED photometry } 
    \label{tab:host}
    \begin{tabular}{cccccccc}
      \toprule
       Survey & Band & Flux \\ 
              &      & (mJy)  \\ 
      \midrule
GALEX & NUV & 0.017 $\pm$ 0.013 \\
SDSS & u & 0.029 $\pm$ 0.011 \\
SDSS & g & 0.075 $\pm$ 0.004 \\
SDSS & r & 0.178 $\pm$ 0.004 \\
SDSS & i & 0.274 $\pm$ 0.006 \\
SDSS & z & 0.344 $\pm$ 0.025 \\
2MASS & J & 0.334 $\pm$ 0.188 \\
2MASS & H & 0.347 $\pm$ 0.224 \\
WISE & W1 & 0.607 $\pm$ 0.042 \\
WISE & W2 & 0.586 $\pm$ 0.051 \\
WISE & W3 & 2.547 $\pm$ 0.285 \\
WISE & W4 & 5.873 $\pm$ 2.442 \\
      \bottomrule
    \end{tabular}
  \begin{tablenotes}
      \item Photometric measurements listed in the table have been corrected for Galactic extinction. W3- and W4-band data were excluded from the SED fitting. 
  \end{tablenotes}
  \end{threeparttable}
\end{table}

\begin{table}[htbp]
  \scriptsize
  \begin{threeparttable}
    \caption{Flux evolution of broad $\mathrm{H}\alpha$, broad $\mathrm{H}\beta$, and coronal line $\rm [Fe\,\textsc{x}]$} 
    \label{tab:hahbevolution}
    \begin{tabular}{cccccccc}
      \toprule
       spectrum & $f_{\mathrm{H}\alpha}$ 
                & $f_{\mathrm{H}\beta}$  
                & $f_{\mathrm{[Fe\,\textsc{x}]}}$
                & $f_{[O\,\textsc{iii}]\,\lambda5007}$   
                & $f_{\mathrm{H}\alpha \rm , calib}$ 
                & $f_{\mathrm{H}\beta \rm , calib}$
                & $f_{\mathrm{[Fe\,\textsc{x}]}\rm , calib}$ \\ 
    (1) & (2) & (3) & (4) & (5) & (6) & (7) & (8)   \\
      \midrule
NTT-231105  &196 $\pm$ 101 &86 $\pm$ 29  &--- &30.9 $\pm$ 9.2&221 $\pm$ 133&97 $\pm$ 44&--- \\ 
P200-250521 &367 $\pm$ 39  &75 $\pm$ 10  &--- &8.3 $\pm$ 3.3&1551 $\pm$ 663&319 $\pm$ 139&--- \\
Keck-250525 &1587 $\pm$ 20 &476 $\pm$ 15 &44.9 $\pm$ 3.2&34.9 $\pm$ 2.9&1587 $\pm$ 20&476 $\pm$ 15&44.9 $\pm$ 3.2 \\ 
Shane-250529&672 $\pm$ 163 &280 $\pm$ 37 &---&9.2 $\pm$ 8.2&2543 $\pm$ 2344&1058 $\pm$ 951&---  \\ 
Shane-250604&1090 $\pm$ 37 &360 $\pm$ 324&---&31.4 $\pm$ 10.0&1210 $\pm$ 401&400 $\pm$ 383&---  \\ 
P200-250619\tnote{*} &476 $\pm$ 52  &158 $\pm$ 9  &8.9 $\pm$ 2.9&16.5 $\pm$ 2.5&1009 $\pm$ 206&336 $\pm$ 61&18.9 $\pm$ 7.0 \\ 
MMT-250830  &1988 $\pm$ 671&613 $\pm$ 26 &<22.5&88.7 $\pm$ 4.9&781 $\pm$ 275&241 $\pm$ 26&<8.8  \\ 
P200-251005 &701 $\pm$ 135 &168 $\pm$ 10 &---&60.5 $\pm$ 4.1&404 $\pm$ 89&96 $\pm$ 12&---  \\
      \bottomrule
    \end{tabular}
  \begin{tablenotes}
      \item All fluxes in the table are in units of $\rm 10^{-17}\, erg\,s^{-1}\,cm^{-2}$. The columns are as follows: (1) the spectrum identification, formatted as ``Telescope-YYMMDD''; (2) broad H$\alpha$ flux; (3) broad H$\beta$ flux; (4) $\rm [Fe\,\textsc{x}]$ coronal line flux, only provided for the two spectra in which the line is robustly detected, along with upper limits for MMT-250830, which has good S/N despite the absence of the line; (5) narrow [O$\,${\textsc{iii}}]\, $\lambda$5007 flux. Columns (6-8) list the fluxes of corresponding lines,  calibrated against the narrow [O$\,${\textsc{iii}}]\, $\lambda$5007 flux. In this calibration, the narrow-line flux is scaled to match that of the high-S/N Keck spectrum.  The three calibrated emission-line fluxes exhibit a clear declining trend as the outburst fades.
      \item[*]The spectral uncertainty was assumed to be 10\% of the flux.
      
  \end{tablenotes}
  \end{threeparttable}
\end{table}

\clearpage 
\begin{figure}[htb]
\centering
\begin{minipage}{0.9\textwidth}
\centering{\includegraphics[angle=0,width=1.0\textwidth]{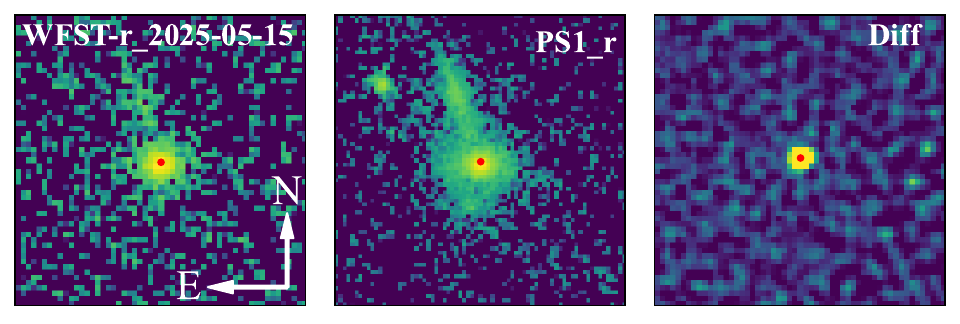}}
\end{minipage}
\caption{\label{fig:Targetposition} Position of the transient event AT2023uqm. From left to right, the three panels show the WFST $r$-band image on 2025-05-10, the pre-outburst PS1 $r$-band image, and the difference image. The red circle (with a radius of $0.1^{\prime\prime}$) marks the transient position, as determined from the difference image, with an uncertainty of $\lesssim 0.1^{\prime\prime}$.    } 
\end{figure} 

\begin{figure}[htb]
\centering
\begin{minipage}{0.9\textwidth}
\centering{\includegraphics[angle=0,width=1.0\textwidth]{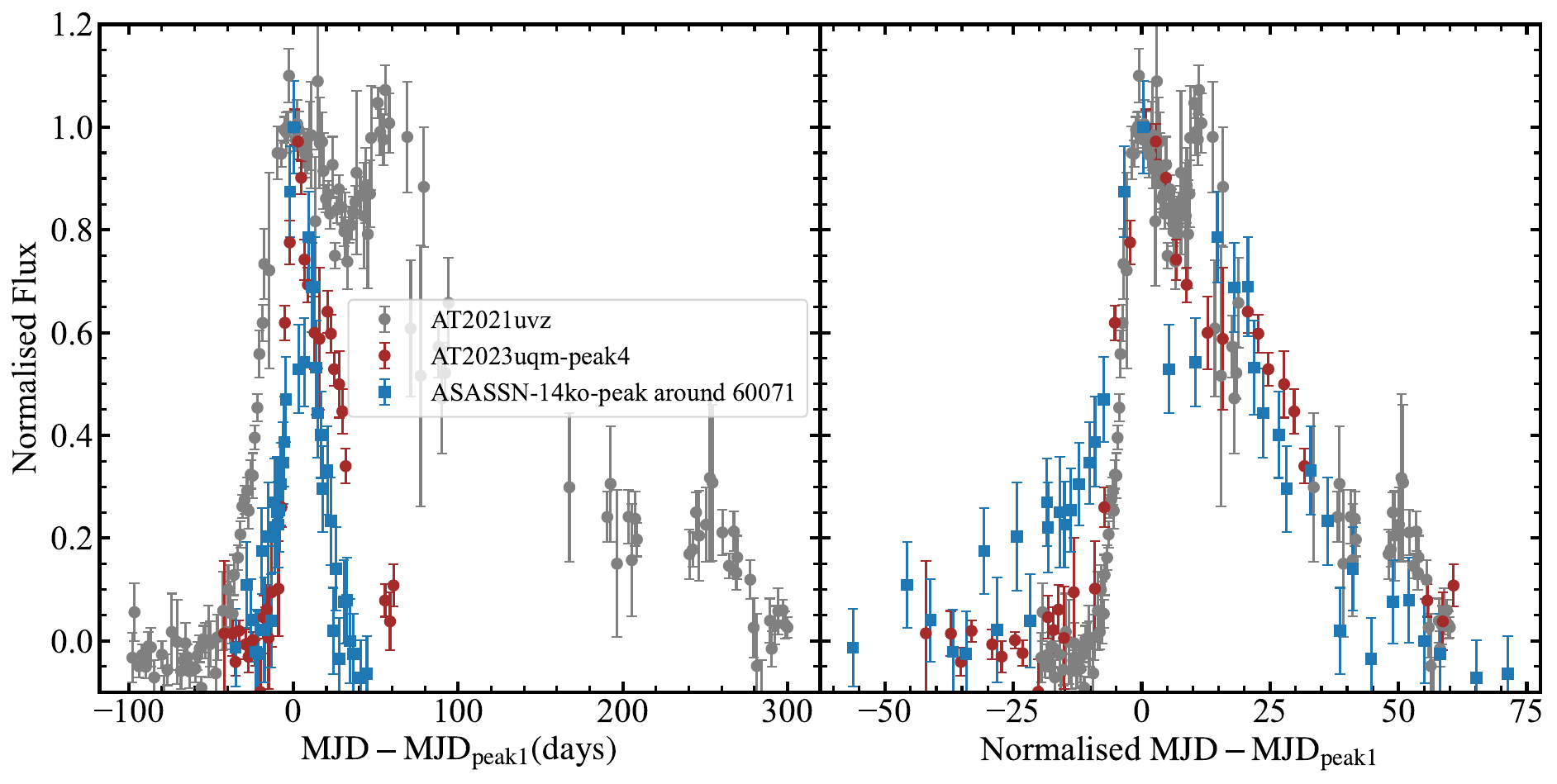}}
\end{minipage}
\caption{\label{fig:profileCompare} Comparison of light-curve profiles for AT2023uqm (red points), ASASSN-14ko (blue squares), and the double-peaked TDE AT2021uvz (gray points). For AT2023uqm, we select the fourth flare, which offers good sampling and S/N. For ASASSN-14ko, we choose the flare around MJD 60071, as its profile closely resembles a double-peaked shape. In the left panel, the maximum flux is normalized, while in the right panel, the duration is also normalized. } 
\end{figure}

\begin{figure*}[htb]
\centering
\begin{minipage}{1.0\textwidth}
\centering{\includegraphics[angle=0,width=1.0\textwidth]{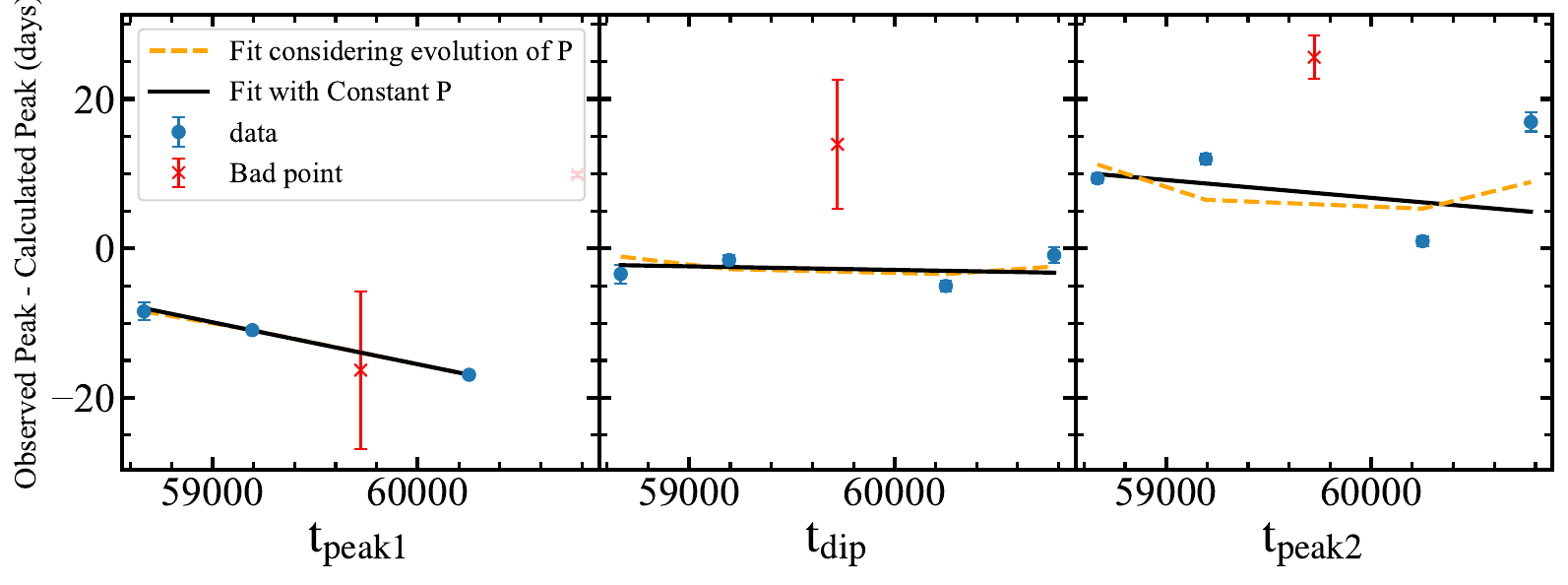}}
\end{minipage}
\caption{\label{fig:O-Cv2} Observed-Calculated plots for three timing features: $t_{\rm peak1}$ (left panel), $t_{\rm dip}$ (middle panel), and $t_{\rm peak2}$ (right panel). The reference calculated times are derived assuming a constant period of 527 days and the first flare occurs at MJD 58666. The blue data points are included in the fitting, while the red crosses indicate the bad points that were excluded because of the reasons discussed in the main text. The black lines show the fitting results assuming a constant period, whereas the orange dashed lines represent the fitting results that allow for period evolution. } 
\end{figure*}

\begin{figure}[htb]
\centering
\begin{minipage}{0.9\textwidth}
\centering{\includegraphics[angle=0,width=1.0\textwidth]{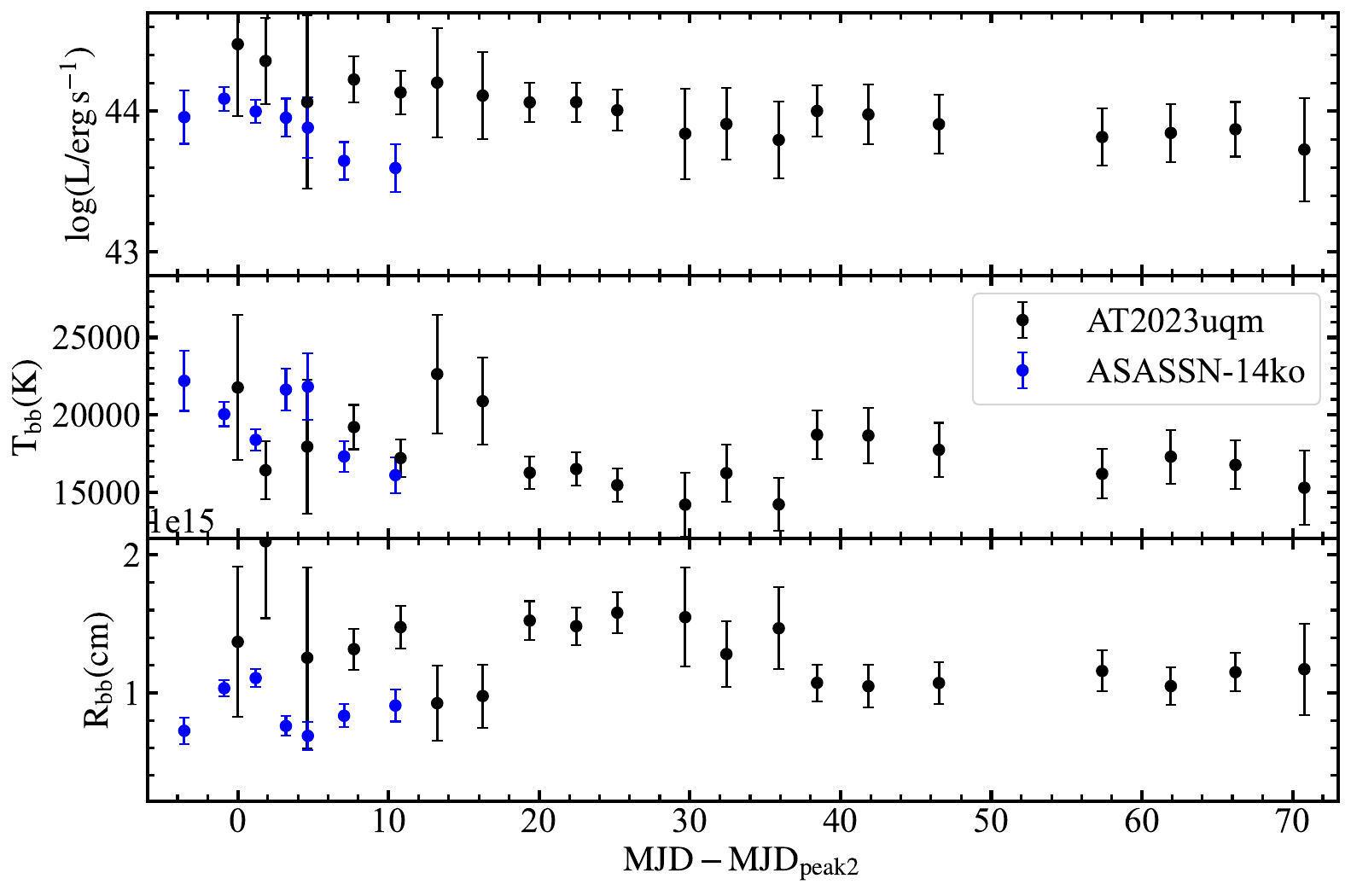}}
\end{minipage}
\caption{\label{fig:BB-fit} Blackbody fitting results for the multiband photometry of the fifth flare of AT2023uqm (black points), compared with ASASSN-14ko (blue points). From top to bottom, the panels display the blackbody luminosity, temperature, and radius, respectively. For ASASSN-14ko, we show results for the outburst around MJD~60071, where the optical light curve displays a roughly double-peaked profile. The reference times ($\rm MJD_{peak2}$) are 60793.94 for AT2023uqm and 60082 for ASASSN-14ko. } 
\end{figure}

\begin{figure}[htb]
\centering
\begin{minipage}{0.88\textwidth}
\centering{\includegraphics[angle=0,width=1.0\textwidth]{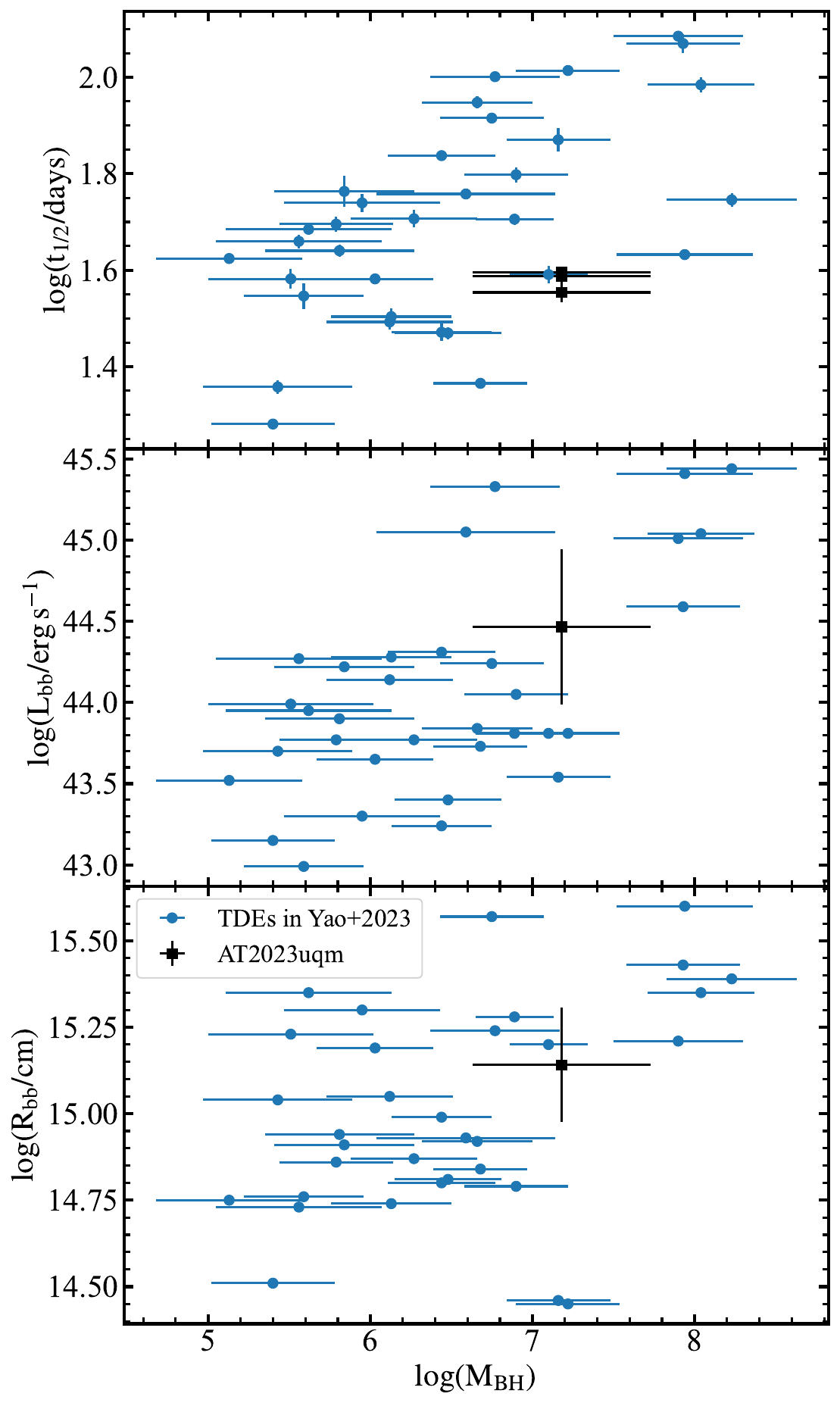}}
\end{minipage}
\caption{\label{fig:BBcompare} Comparison of AT2023uqm's properties with TDEs from Yao et al.~\citep{Yao2023}, including $t_{1/2}$ (rest-frame duration above half-maximum light; top panel), $L_{\rm bb}$ (peak blackbody luminosity; middle panel), and $R_{\rm bb}$ (blackbody radius at peak light; bottom panel), all as a function of black hole mass. For $t_{1/2}$, we adopt the values from the first, second, and fourth flares with complete temporal coverage, while the peak blackbody luminosity and radius are derived from blackbody fits to the most recent flare with multiband observations. We adopt a black hole mass of $10^{7.18}\,\rm M_\odot$, estimated from the total stellar mass via the relation in Reines \& Volonteri~\citep{Reines2015}. } 
\end{figure}

\begin{figure}[htb]
\centering
\begin{minipage}{1.0\textwidth}
\centering{\includegraphics[angle=0,width=1.0\textwidth]{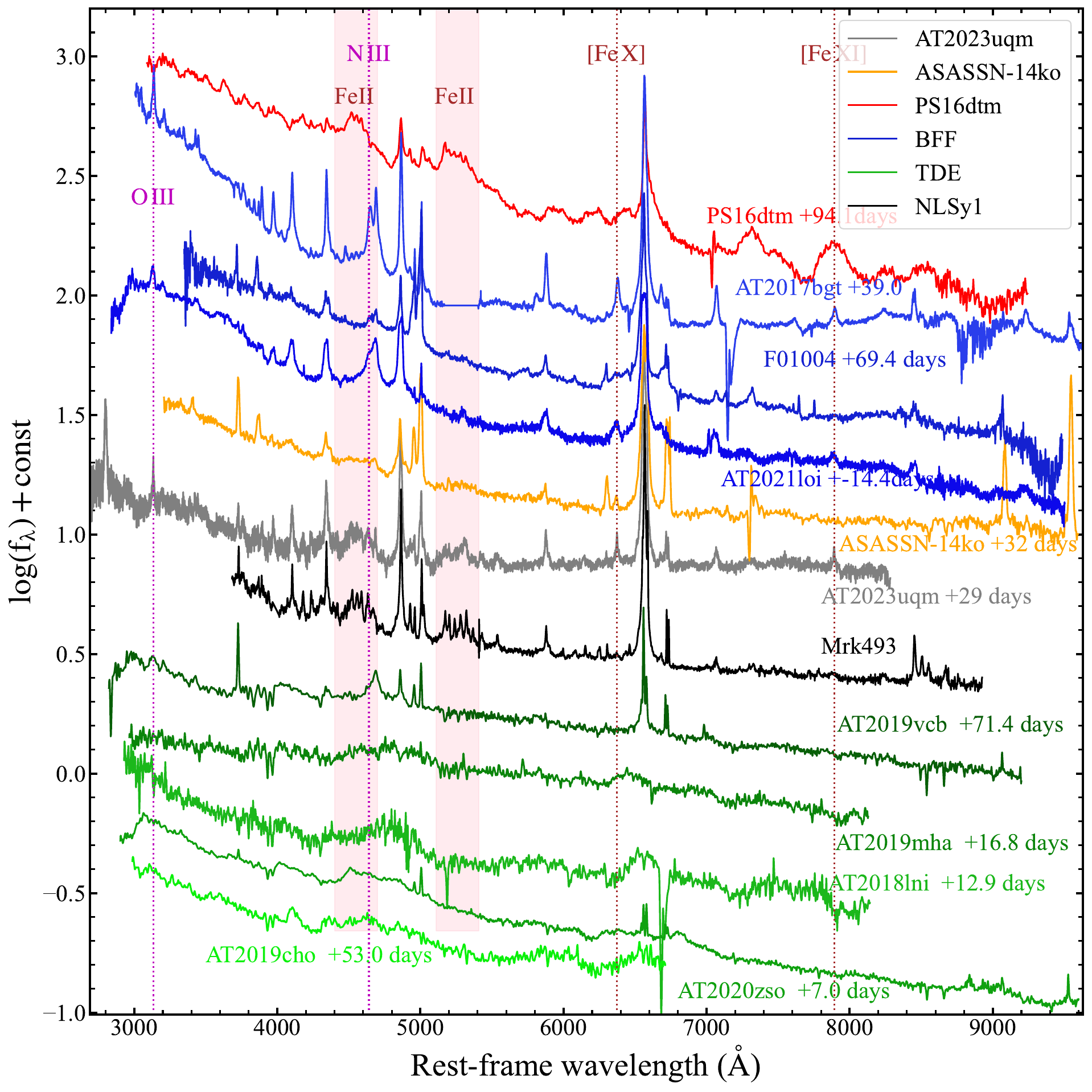}}
\end{minipage}
\caption{\label{fig:specCompare} Comparison of optical spectra of AT2023uqm (gray curve), typical TDEs (green curves), typical BFFs (blue curves), PS16dtm (red curve), NLSy1 Mrk493 (black curve), and the rpTDE candidate ASASSN-14ko (orange curve). The TDE spectra are from Hammerstein et al.~\citep{Hammerstein2023}, where telluric absorption has not been removed for AT2018lni. For the BFFs, the spectrum of AT2017bgt (telluric absorption not removed) is from Tadhunter et al.~\citep{Tadhunter2017}, the spectrum of F01004 is from Sun et al.~\citep{SunLuming2024}, and the spectrum of AT2021loi is retrieved from the Transient Name Server website. The spectra of ASASSN-14ko and PS16dtm are from Payne et al.~\citep{Payne2021} and Petrushevska et al.~\citep{Petrushevska2023}, respectively. The spectrum of Mrk493 is taken from the SDSS database. Vertical dashed lines mark the position of Bowen fluorescence emission lines and coronal lines. Pink shaded regions highlight the prominent $\rm Fe\,\textsc{ii}$ emission.} 
\end{figure}

\begin{figure}[htb]
\centering
\begin{minipage}{1.0\textwidth}
\includegraphics[angle=0,width=0.45\textwidth]{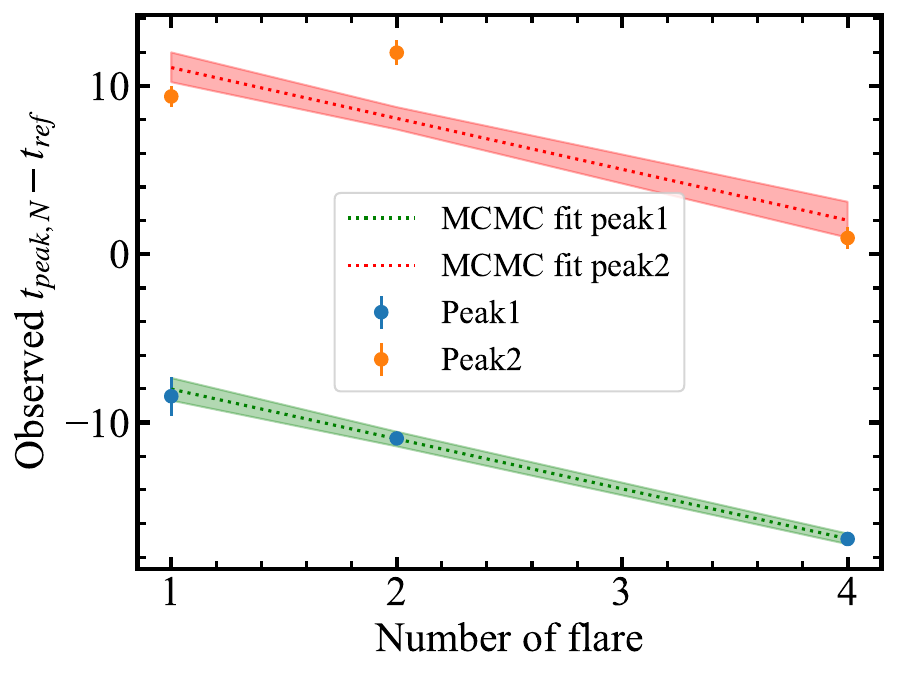}
\includegraphics[angle=0,width=0.45\textwidth]{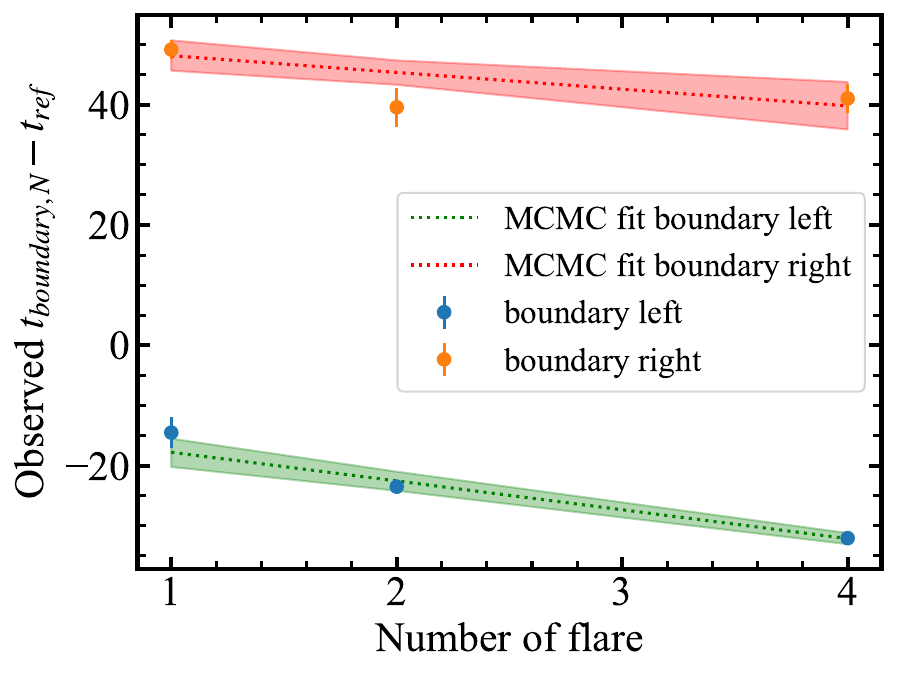}
\end{minipage}
\caption{\label{fig:mcmcfitting} The MCMC fitting results for the peaks (left panel) and boundaries (right panel) of the first, second, and fourth flares. The dashed lines indicate the median fits, while the shaded regions represent the 90\% credible intervals. The reference time $t_{\rm ref}$ is calculated as MJD = $58666+(n-1)\times527$, where $n$ is the flare number.  } 
\end{figure}

\begin{figure}[htb]
\centering
\begin{minipage}{1.0\textwidth}
\centering{\includegraphics[angle=0,width=1.0\textwidth]{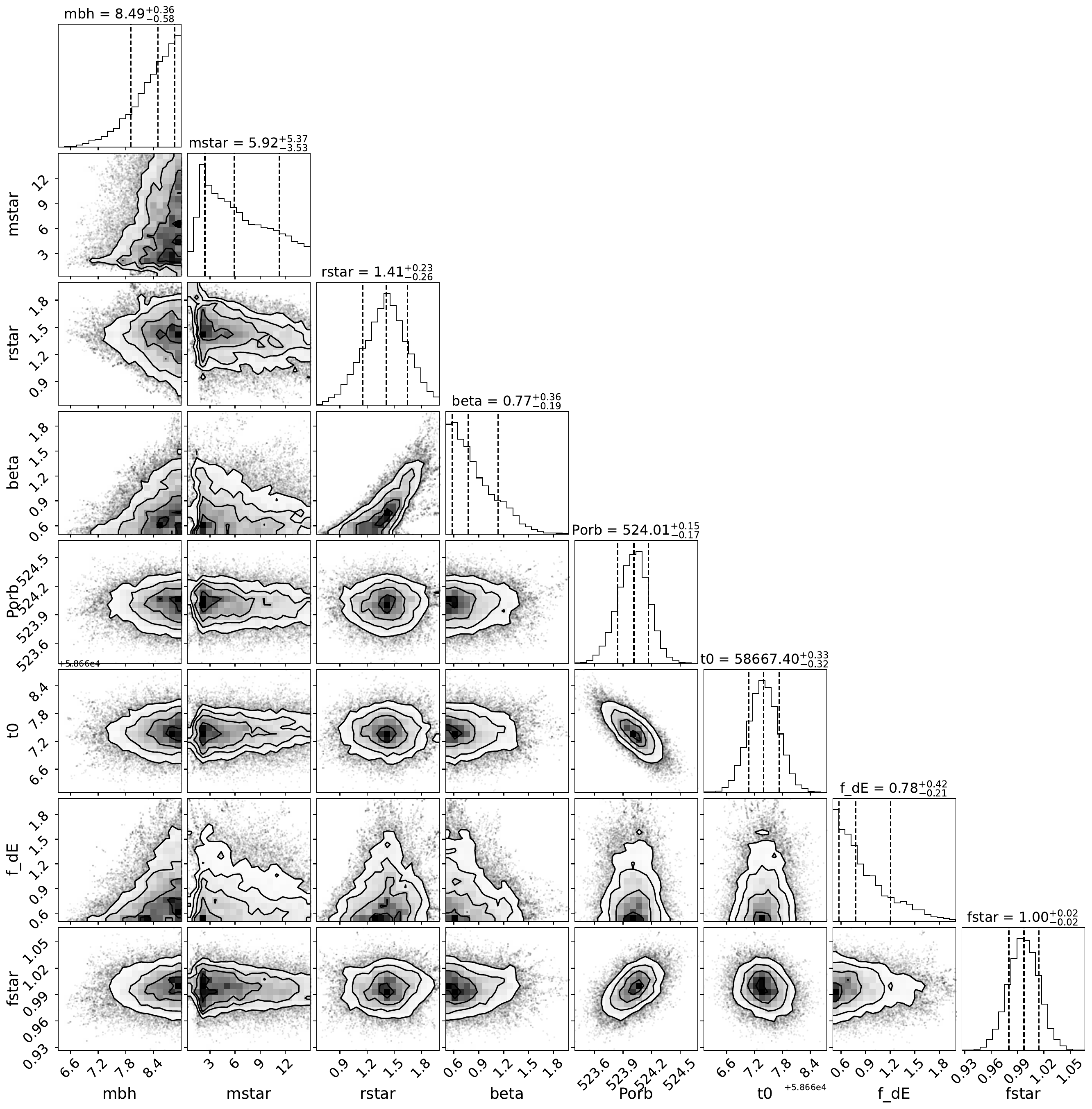}}
\end{minipage}
\caption{\label{fig:corner-peak} Corner plots of the MCMC fitting results for the flare peaks. The dashed lines in the parameter distributions and the values provided for each parameter correspond to 
the 16th, 50th, and 84th percentiles.} 
\end{figure}

\begin{figure}[htb]
\centering
\begin{minipage}{1.0\textwidth}
\centering{\includegraphics[angle=0,width=1.0\textwidth]{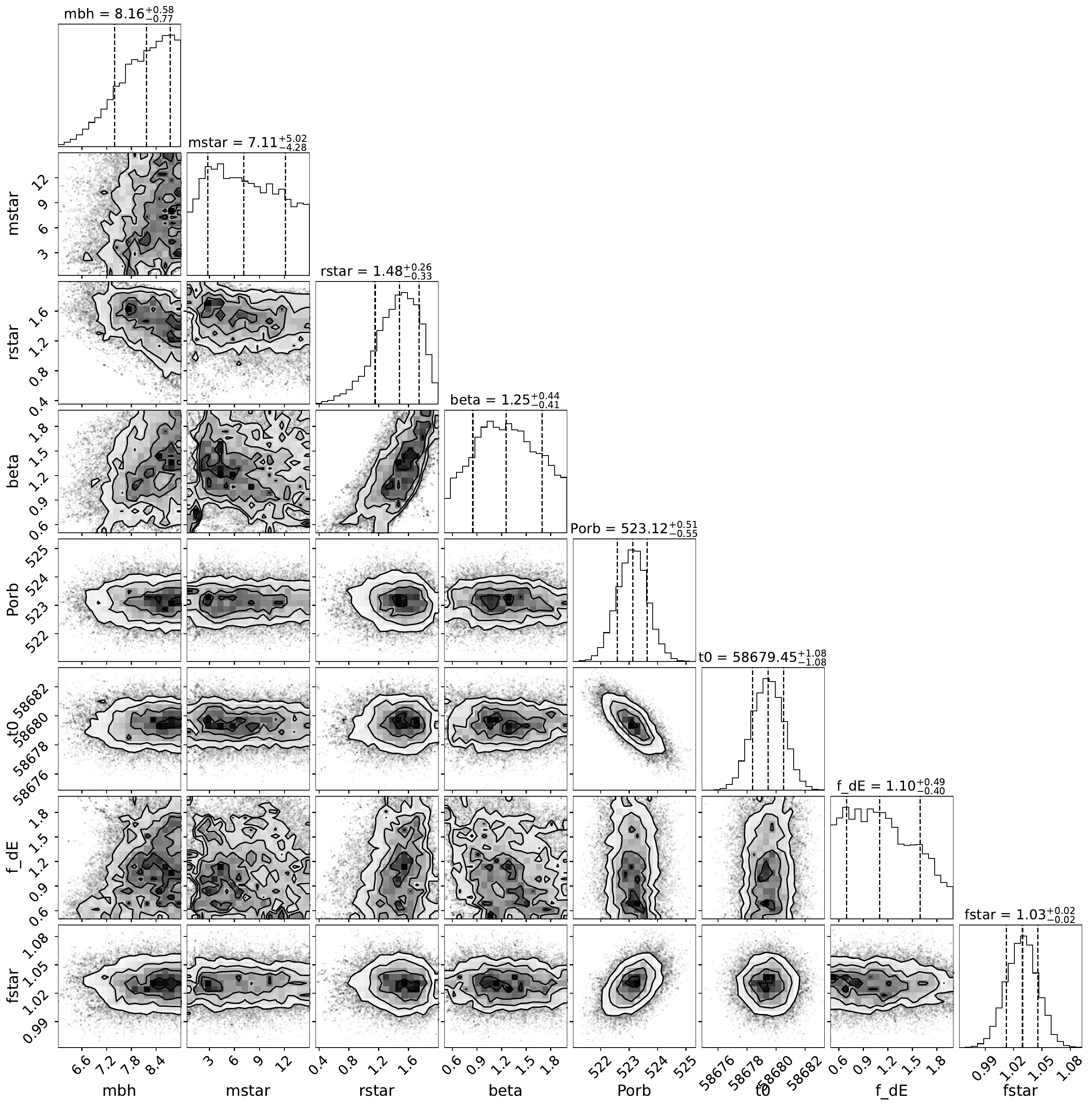}}
\end{minipage}
\caption{\label{fig:corner-boundary} Corner plots of the MCMC fitting results for the flare boundaries. The dashed lines in the parameter distributions and the values provided for each parameter correspond to 
the 16th, 50th, and 84th percentiles.} 
\end{figure}

\begin{figure}[htb]
\centering
\begin{minipage}{1.0\textwidth}
\centering{\includegraphics[angle=0,width=1.0\textwidth]{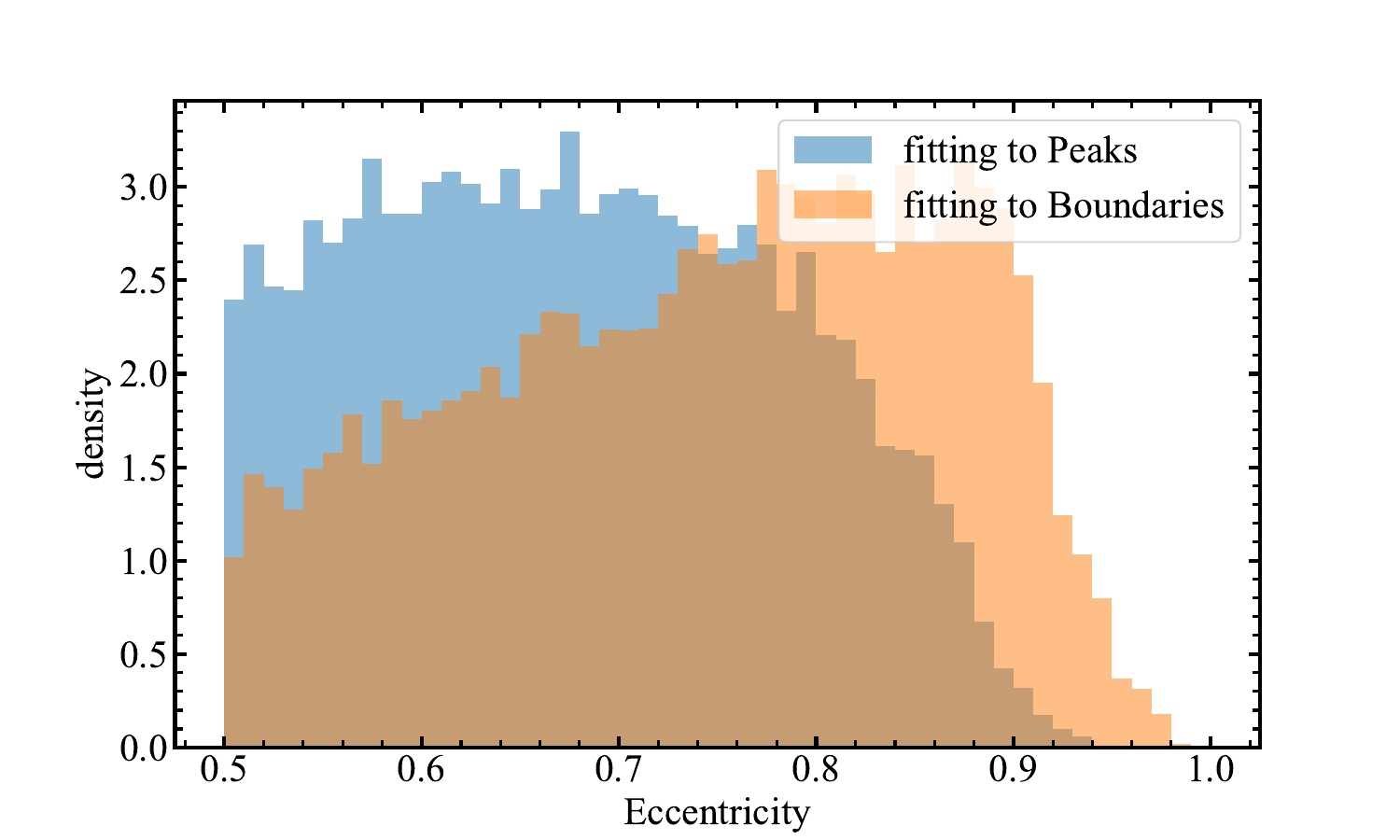}}
\end{minipage}
\caption{\label{fig:eccentricity} The eccentricity distributions derived from two MCMC fitting results. The fit to the flare peaks favors lower eccentricities ( $0.68^{+0.12}_{-0.11}$; $1 \sigma$ error) compared to the fit to the flare boundaries ( $0.76^{+0.12}_{-0.15}$) . This difference arises due to the same lower bound in the prior distribution for $f_E$. Note that in both fittings, we imposed a lower limit of 0.5 on the eccentricity.  } 
\end{figure}

\begin{figure}[htb]
\centering
\begin{minipage}{0.8\textwidth}
\centering{\includegraphics[angle=0,width=1.0\textwidth]{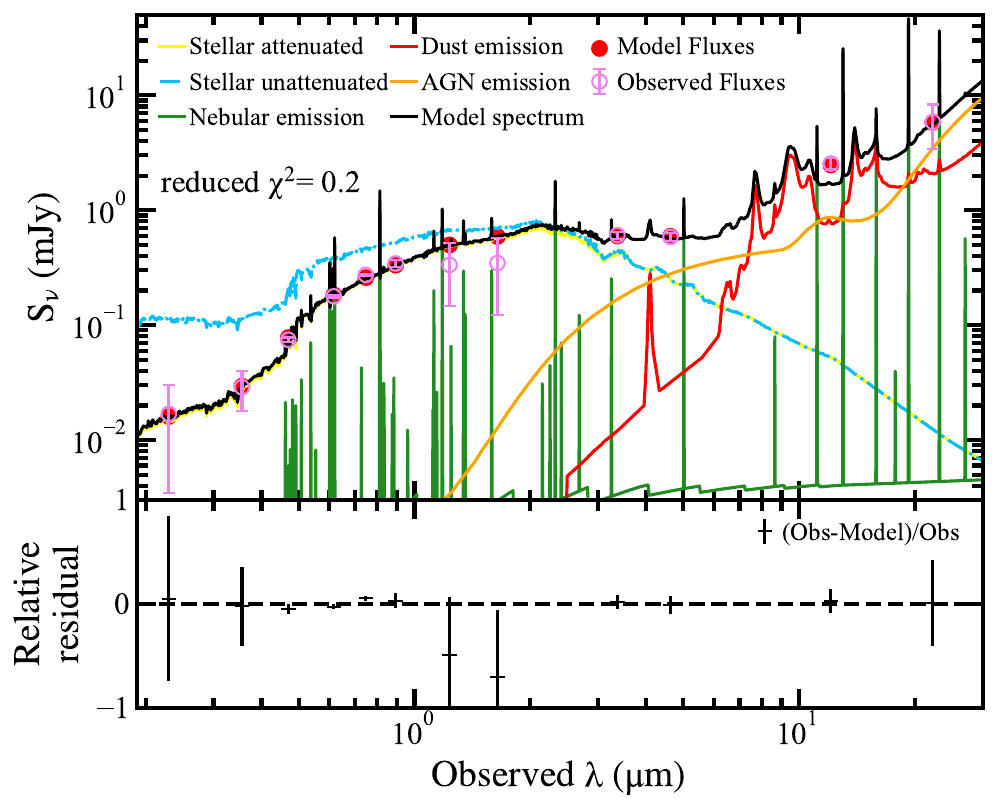}}
\end{minipage}
\caption{\label{fig:cigale} SED fitting results for the host galaxy using CIGALE. Magenta circles show the observed fluxes in each band, red points indicate the modeled fluxes, and residuals are displayed in the bottom panel. } 
\end{figure}

\begin{figure}[htb]
\centering
\centering{\includegraphics[width=0.45\textwidth]{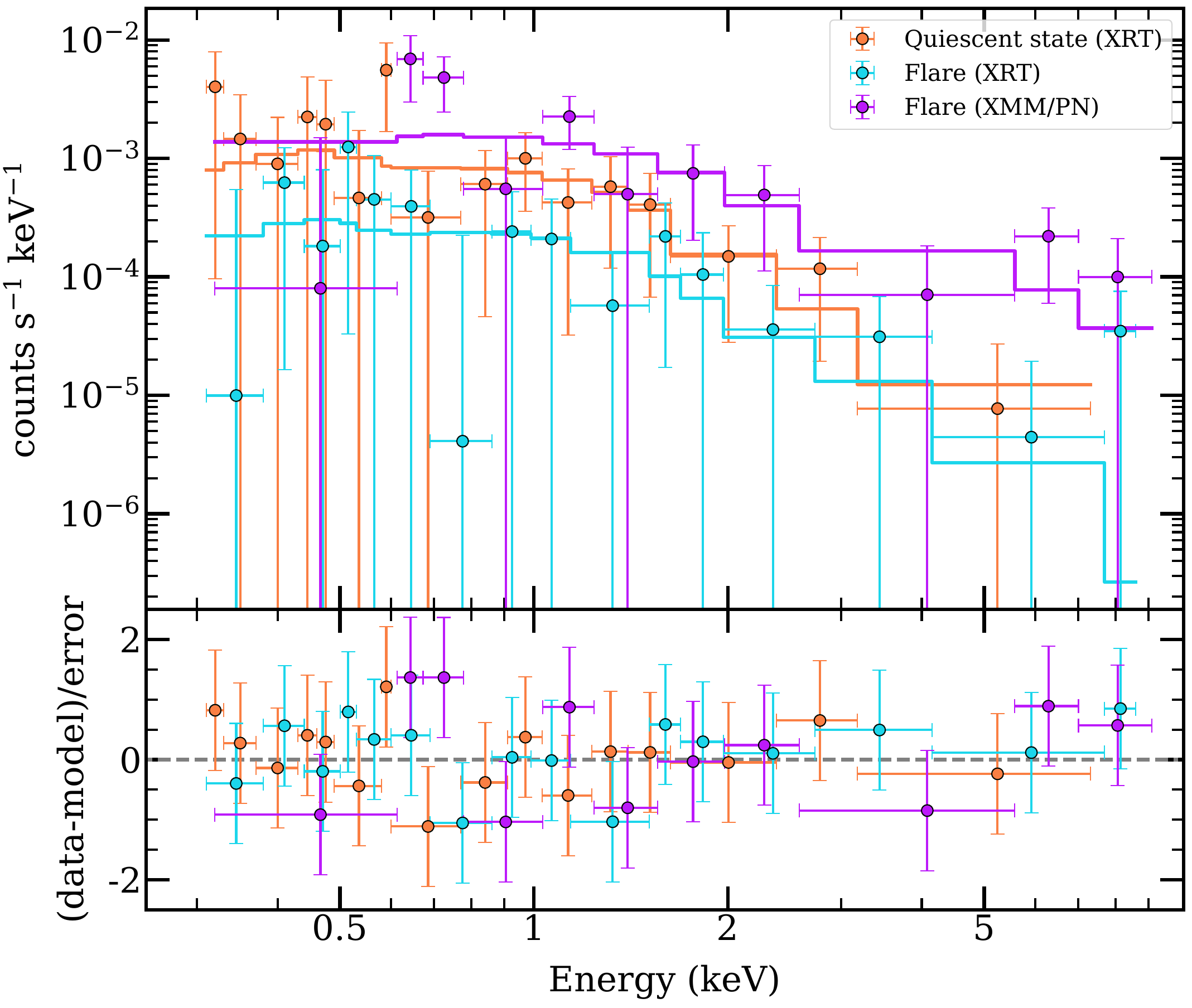}}
\label{fig:xspec} 
\caption{ The X-ray spectra for XRT and XMM-Newton. All the spectra were fitted by the \texttt{tbabs*zashift*powerlaw} model.} 
\end{figure}

\begin{figure}
\centering
\begin{minipage}{0.7\textwidth}
\centering{\includegraphics[width=1.0\textwidth]{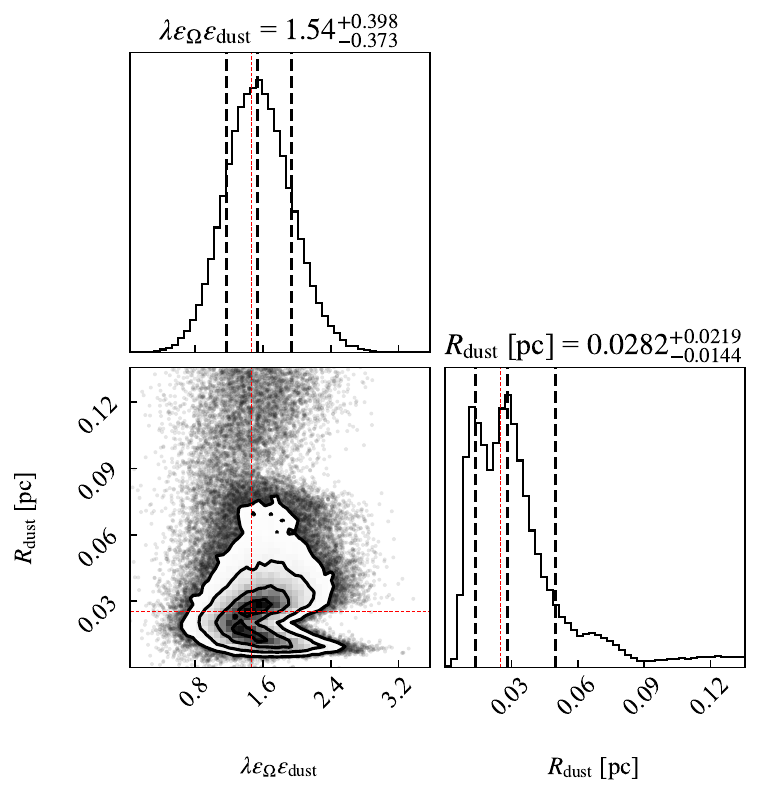}}
\end{minipage}
\caption{The distribution of dust-echo model parameters for AT2023uqm. The red dashed lines denote the location of maximum likelihood, corresponding to the parameter values of $\lambda \epsilon_\Omega \epsilon_{\text{dust}}=1.465$ and\ $R_{\text{dust}}=0.025$~pc. The values and errors above the plots correspond to the $15.87\%$, $50\%$ and $84.13\%$ 
values of the posterior samples of the parameters, and the black dashed lines show the locations of these values in the plots. The contours in the 2D plots reflect the relative numerical density, where darker color means a larger numerical density.}
\label{fig:IR-para}
\end{figure}

\begin{figure}
\centering
\begin{minipage}{0.8\textwidth}
\centering{\includegraphics[width=0.9\textwidth]{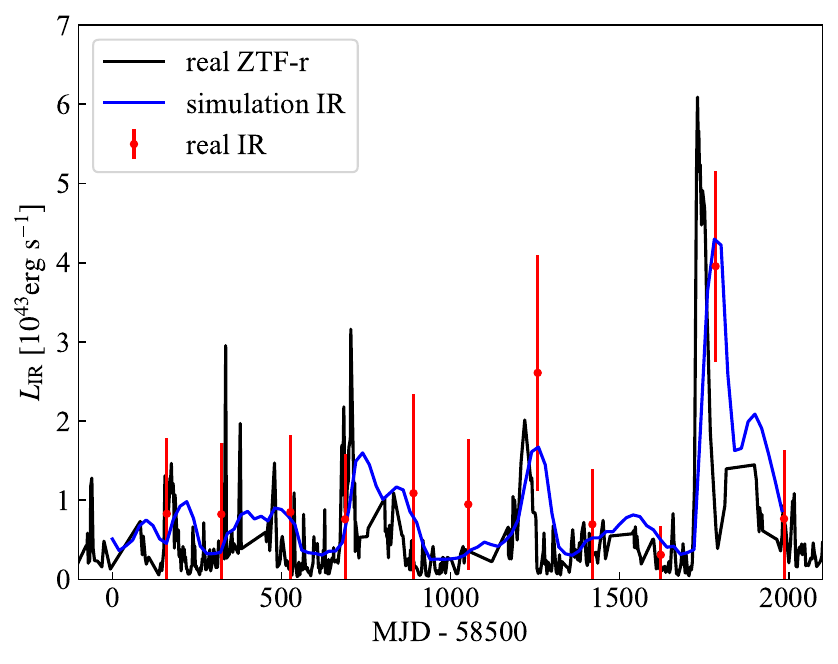}}
\end{minipage}
\caption{The best-fitting IR luminosity light curve. The black curve denotes the monochromatic luminosity of the ZTF-r band. The red error bars denote the real IR luminosity estimated from blackbody fits to the WISE data, while the blue curve denotes the simulated IR light curve.}
\label{fig:bestir}
\end{figure}

\begin{figure}[htb]
\centering
\begin{minipage}{0.9\textwidth}
\centering{\includegraphics[angle=0,width=1.0\textwidth]{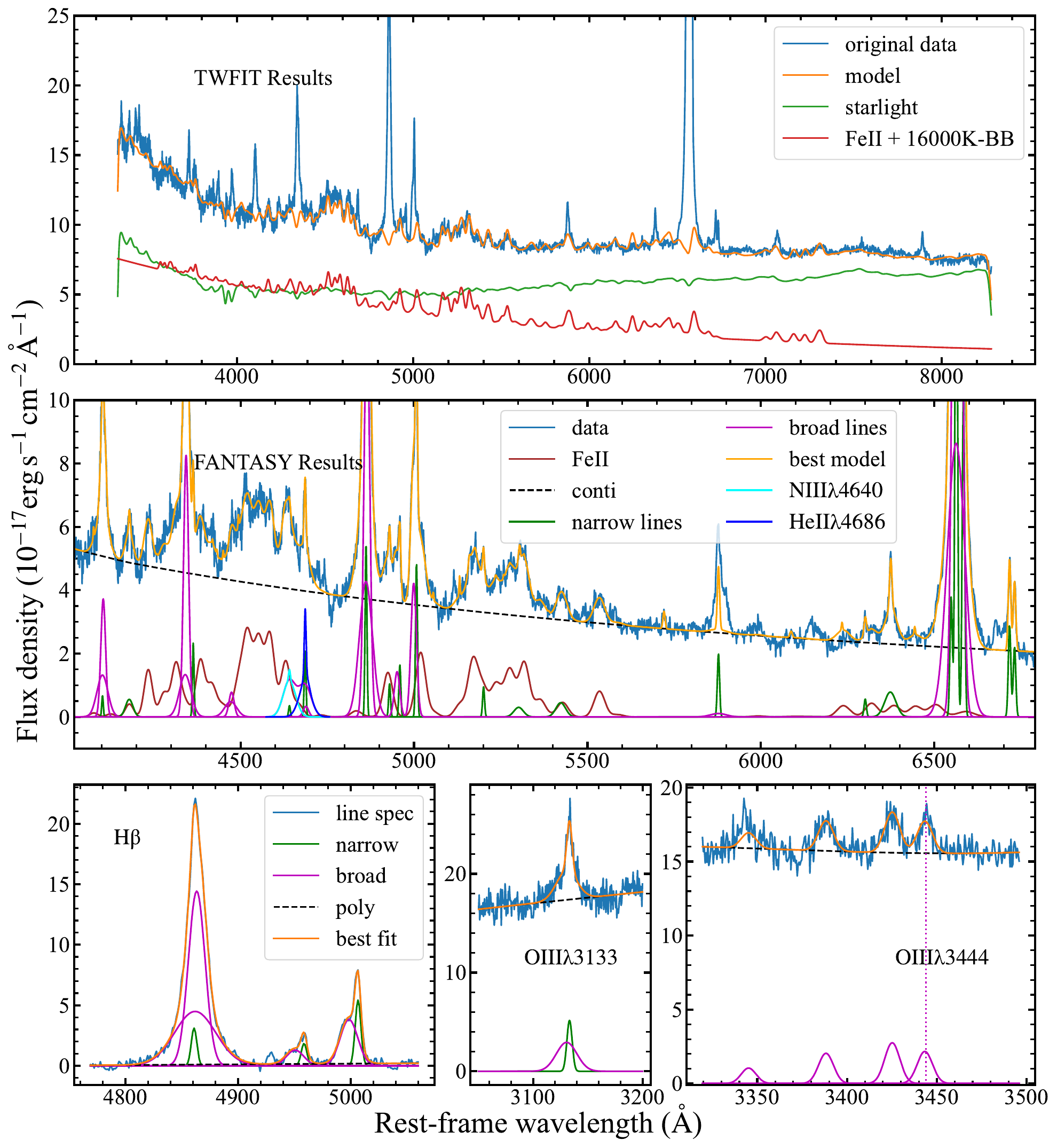}}
\end{minipage}
\caption{\label{fig:specfit}Spectroscopic decomposition and emission-line fitting of the Keck/LRIS spectrum.
(a) The top panel shows the continuum fitting results (orange curve), modeled as a combination of a starlight component (green curve) and a red component representing \feii\ emission and a blackbody continuum.
(b) The middle panel presents the spectroscopic fitting performed with FANSTASY. The best-fit result is shown in yellow, composed of a power-law continuum (black dashed line), narrow emission lines (green curve), broad emission lines (magenta curve), and \feii\ emission components. Notably, the $\rm He\,\textsc{ii}$ and $\rm N\,\textsc{iii}$ emission lines are highlighted in blue and cyan, respectively.
(c) The bottom three panels display examples of individual emission-line fittings. The dashed curves indicate polynomial fits to the residual continuum in regions where the continuum was subtracted, or to the continuum itself in regions where it was not.} 
\end{figure}

\begin{figure}[htb]
\centering
\begin{minipage}{0.8\textwidth}
\centering{\includegraphics[angle=0,width=1.0\textwidth]{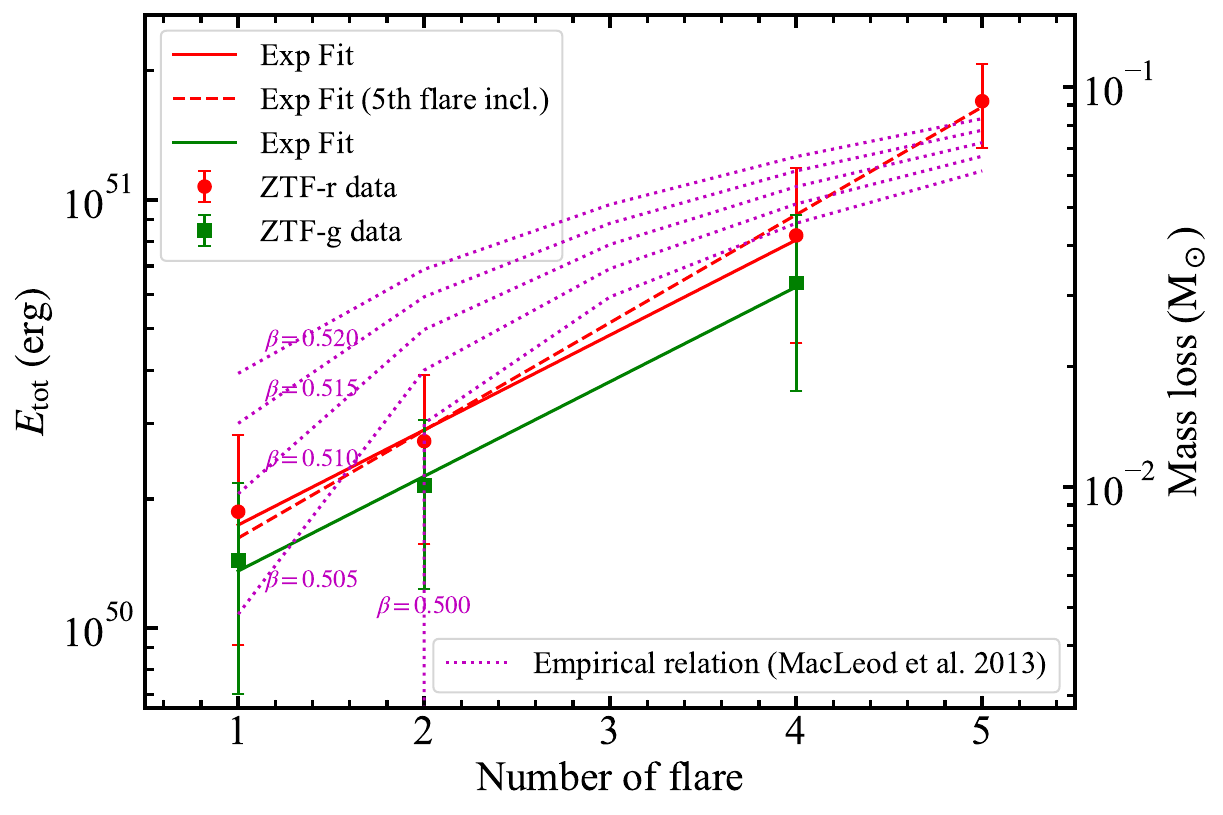}}
\end{minipage}
\caption{\label{fig:mloss} The evolution of integrated energy (left ordinate) and mass loss (right ordinate) as a function of flare number. Solid lines of the corresponding colors show exponential-function fits to the data. The red dashed line represents the fit to the ZTF-r data when the fifth flare is included. The magenta dashed lines indicate the mass-loss evolution for a $3\,\rm M_\odot$ giant star, based on the empirical relation given by MacLeod et al.~\citep{MacLeod2013}, with different values of $\beta$ labeled on each line. }  
\end{figure}

\begin{figure}
\centering
\begin{minipage}{0.95\textwidth}
\centering{\includegraphics[width=1.0\textwidth]{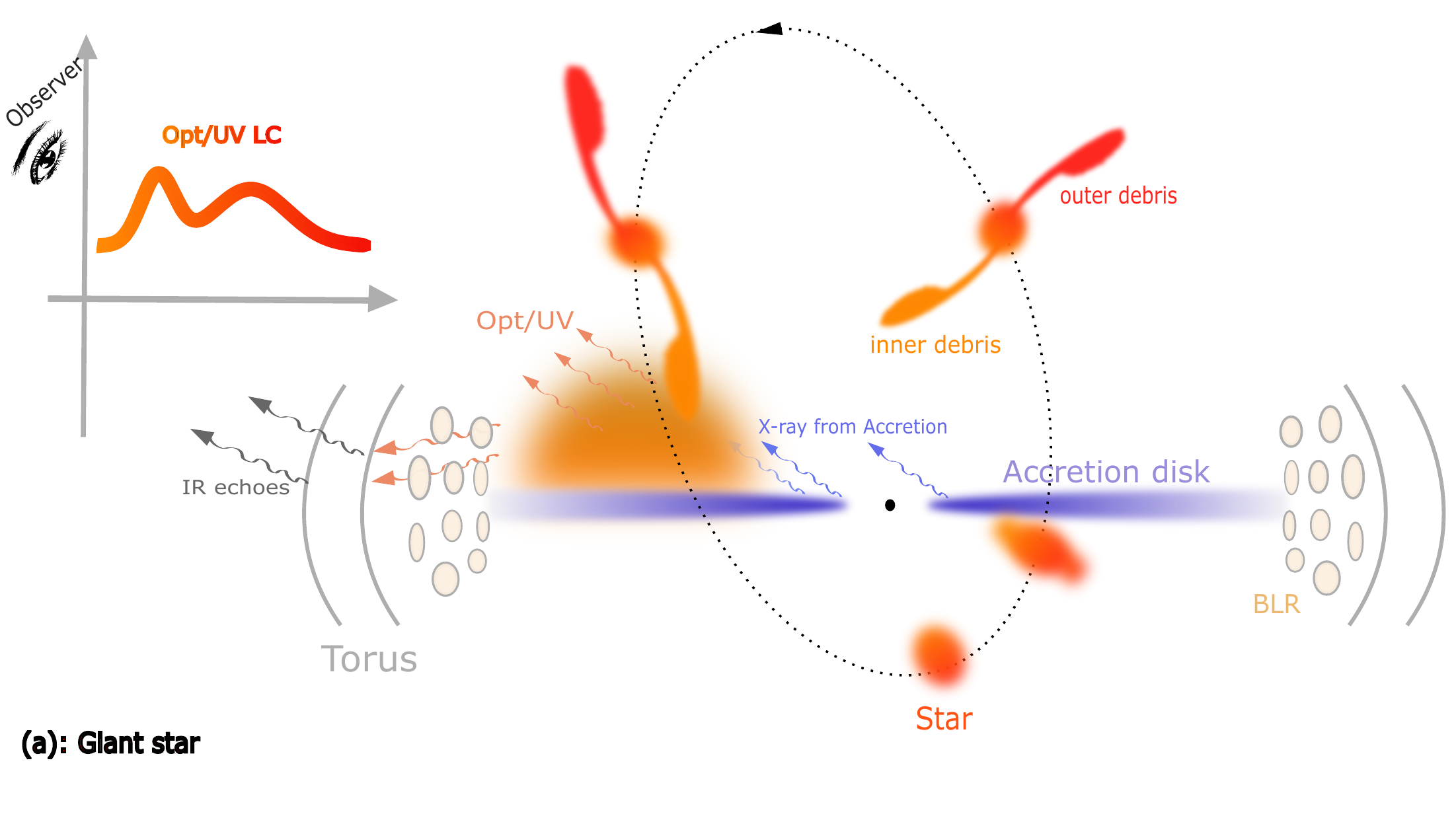}
           \includegraphics[width=1.0\textwidth]{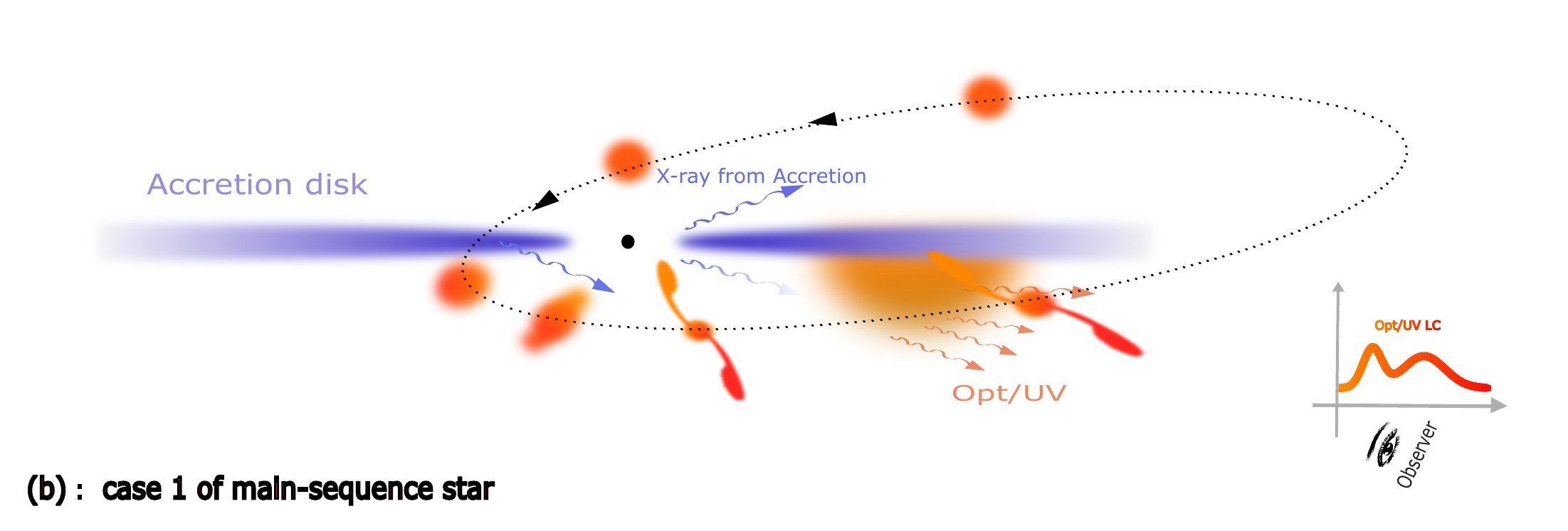}
           \includegraphics[width=1.0\textwidth]{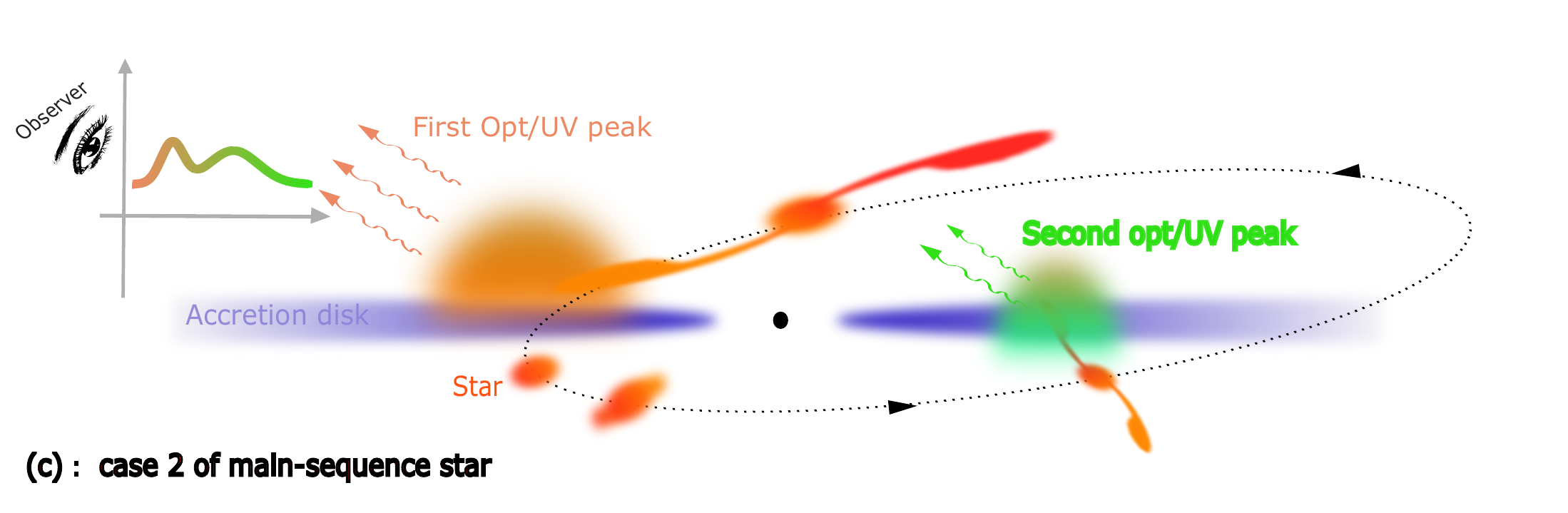}
           }
\end{minipage}
\caption{A schematic illustrating the different scenarios discussed in Section 5, accounting for the presence of an existing accretion disk. Each panel shows the proposed origin of radiation in different bands. These schematics assume the optical/UV radiation is primarily produced by debris-disk collisions. However, the pictures in (a) and (b) are also applicable if the collision merely injects material into the disk and the radiation (including optical/UV) is primarily produced by subsequent accretion. For clarity, the torus and BLR components are omitted from panels (b) and (c), as they would be similar to those shown in panel (a). \\
\scriptsize{
\textbf{Panel (a): Post-apocenter collision (giant-star scenario):} A scenario where the primary collision occurs after apocenter, especially for those close to pericenter. An initial, pre-apocenter collision is assumed to have a negligible effect on the debris and produces no significant radiation. This scenario requires a giant-star progenitor. The optical/UV outburst is produced by the post-apocenter debris-disk collision or subsequent accretion. The double-peaked light curve originates from successive fallback of the inner (orange) and outer (red) debris streams, as shown. X-ray radiation is produced by direct accretion. During the optical/UV outburst, an outflow induced by the debris-disk collision or accretion partially obscures the X-ray source, causing it to fade. IR emission arises from echoes of the optical/UV outburst by the dust torus.
\textbf{Panel (b): Pre-apocenter dissipation (main-sequence-star scenario)}:
A strong collision before apocenter dissipates most of the debris, and only the star itself crosses the disk. This scenario permits a main-sequence star on a highly eccentric orbit, with its semi-major axis nearly parallel to the disk plane. The double-peaked light curve still originates from the inner and outer parts of the remaining debris stream.
\textbf{Panel (c): Two discrete collisions (main-sequence-star scenario):} The double-peaked light curve originates from two distinct debris-disk collisions, as illustrated. This scenario also allows for a main-sequence star.} } 
\label{fig:schematic}
\end{figure}

\begin{figure}[htb]
\centering
\begin{minipage}{0.8\textwidth}
\centering{\includegraphics[angle=0,width=1.0\textwidth]{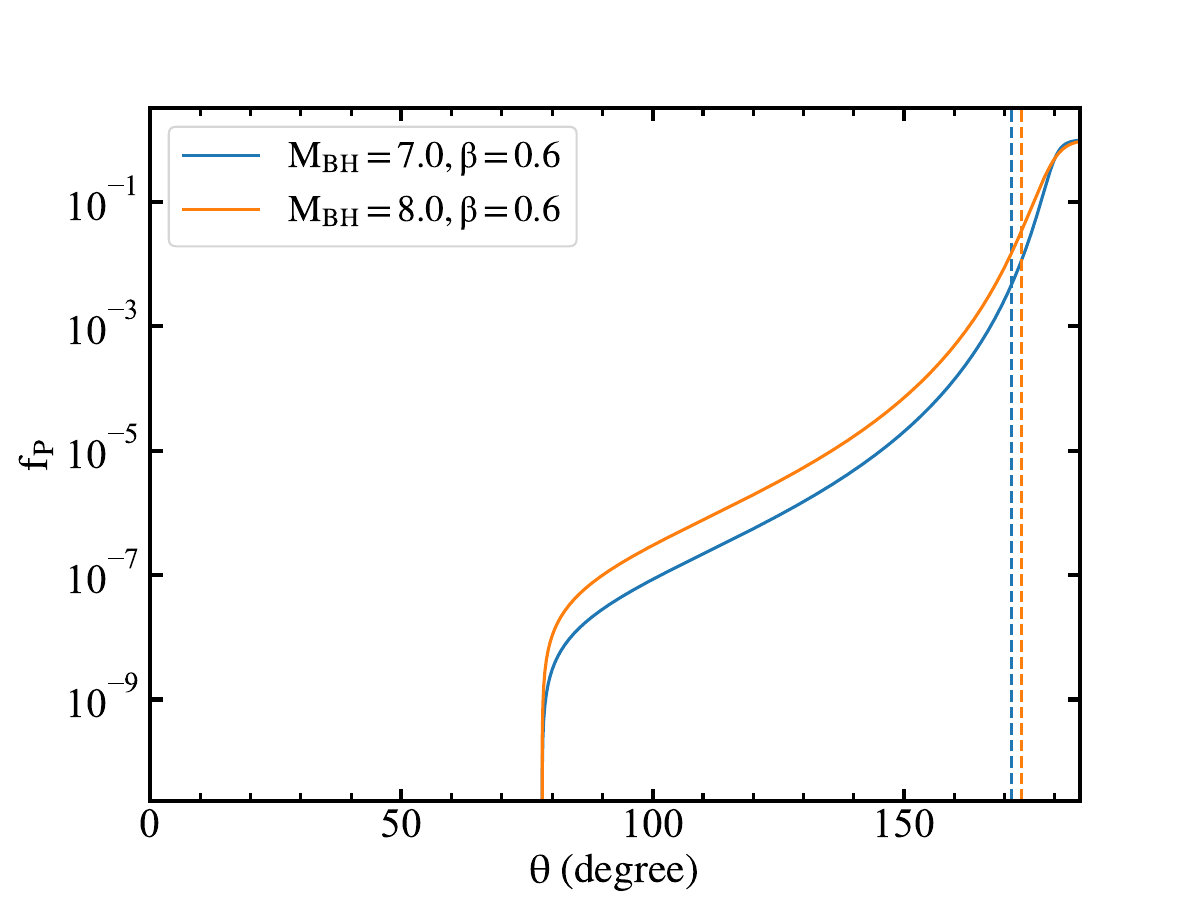}}
\end{minipage}
\caption{\label{fig:fP}  $f_P$ in Equation (10) as a function of the true anomaly $\theta$. The curves are calculated for a solar-like star orbiting black holes of two different masses with $\beta=0.6$. The vertical dashed lines mark the positions where the ratio $\Delta t_{\rm peak} /P$ aligns with the observational values. }  
\end{figure}

\begin{figure}[htb]
\centering
\begin{minipage}{0.8\textwidth}
\centering{\includegraphics[angle=0,width=1.0\textwidth]{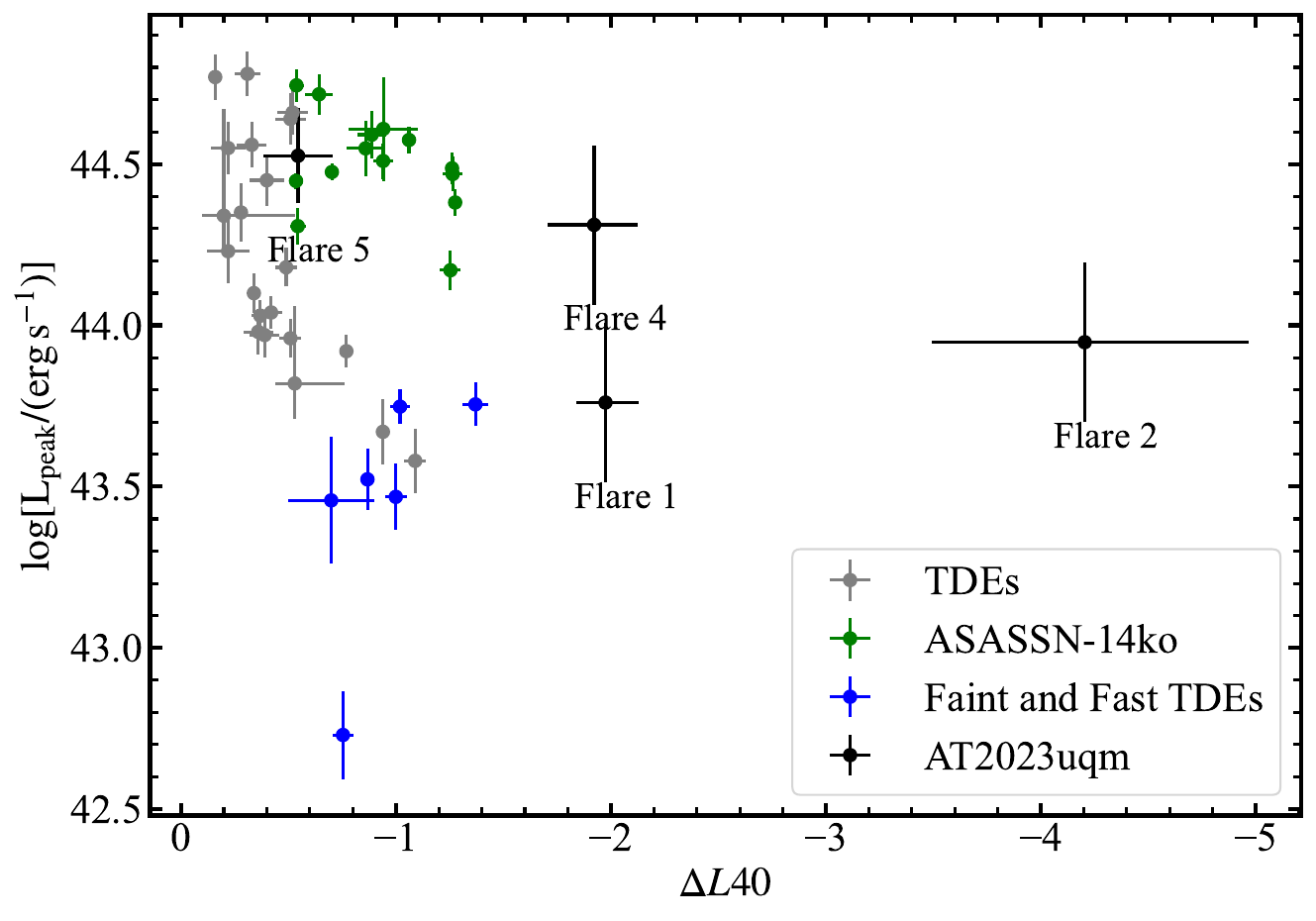}}
\end{minipage}
\caption{\label{fig:LpeakD40} Peak bolometric optical/UV luminosity versus decline rate, defined as $\Delta L_{40} = \log_{10}(L_{40}/L_{\rm peak})$, where $L_{40}$ is the luminosity 40 days after peak brightness. Gray points show typical TDEs \citep{Hinkle2020Declinerate}, blue points indicate fast and faint TDEs \citep{Hoogendam2024}, and green points denote ASASSN-14ko \citep{Payne2021}. Black points represent flares of AT2023uqm. For the fifth flare, both quantities are measured following Hinkle et al.~\citep{Hinkle2020Declinerate}; for the others, they are derived from two-Gaussian fitting results (see Section 3.2.1) after converting the ZTF-r light curve to bolometric luminosity assuming $T_{\rm bb}\approx 17,500\pm2000\,\rm K$. }  
\end{figure}

\clearpage 
\bibliographystyle{scibull}
\bibliography{Intro}